\documentclass[a4paper,11pt]{article}
\usepackage{amsmath,mathtools,bm,amssymb,color}

\usepackage{graphicx}
\usepackage{subcaption}
\usepackage{jheppub} 
\usepackage{xspace}
\usepackage{booktabs}
\usepackage{array}
\usepackage{multirow}
\usepackage{adjustbox}
\usepackage{xcolor}
\usepackage[utf8]{inputenc}

\newcommand{\NNLO}{\text{NNLO}\xspace}
\newcommand{\MF}{\text{MF}\xspace}
\DeclareRobustCommand{\NNLOJET}{NNLO\scalebox{.8}{JET}\xspace}

\def\l{\left}
\def\r{\right}
\def\be{\begin{equation}}
\def\ee{\end{equation}}
\def\ba{\begin{eqnarray}}
\def\ea{\end{eqnarray}}
\def\t{\tilde}

\def\bs{\boldsymbol}

\newcommand{\df}{\mathrm{d}}

\newcommand{\al}{\alpha}
\newcommand{\bt}{\beta}
\newcommand{\ga}{\gamma}
\newcommand{\Ga}{\Gamma}
\newcommand{\de}{\delta}

\newcommand{\ep}{\epsilon}
\newcommand{\la}{\lambda}
\newcommand{\si}{\sigma}
\newcommand{\nn}{\nonumber}

\newcommand{\qb}{\bar{q}} 
\newcommand{\Qb}{\bar{Q}} 

\newcommand{\eq}[1]{\eqref{eq:#1}}

\newcommand{\fig}[1]{Figure~\ref{fig:#1}}

\renewcommand{\sec}[1]{Section~\ref{sec:#1}}
\newcommand{\secs}[2]{Sections~\ref{sec:#1} and \ref{sec:#2}}
\newcommand{\msecs}[2]{Sections~\ref{sec:#1}--\ref{sec:#2}}
\newcommand{\app}[1]{Appendix~\ref{app:#1}}

\allowdisplaybreaks

\title{Antenna subtraction for jet production observables in full colour at NNLO}

\author{X.\ Chen$^{a,b}$, T.\ Gehrmann$^{c}$, E.W.N.\ Glover$^d$, J.~Mo$^c$}

\affiliation{
$^a$Institute for Theoretical Physics, Karlsruhe Institute of Technology, 76131 Karlsruhe, Germany\\
$^b$Institute for Astroparticle Physics, Karlsruhe Institute of Technology,\\ 76344 Eggenstein-Leopoldshafen, Germany\\
$^c$Physik-Institut, Universit\"at Z\"urich, Winterthurerstrasse 190, 8057 Z\"urich, Switzerland\\
$^d$Institute for Particle Physics Phenomenology, Department of Physics, University of Durham, Durham, DH1 3LE, UK}

\emailAdd{xuan.chen@kit.edu}
\emailAdd{thomas.gehrmann@uzh.ch}
\emailAdd{e.w.n.glover@durham.ac.uk}
\emailAdd{jmo@physik.uzh.ch}

\abstract{
We describe the details of the calculation of the full colour NNLO QCD corrections to jet production observables at the LHC with antenna subtraction. All relevant matrix elements for the process $pp \to jj$ at NNLO in full colour are colour-decomposed and given in a $N_c$ and $n_f$ expansion, making identification of leading and subleading colour contributions transparent. The colour-ordered antenna subtraction method has previously successfully been used to construct the NNLO subtraction terms for processes with up to five partons or in the leading colour approximation. However, it is challenged by the more involved subleading colour structure of the squared matrix elements in processes with six or more partons. Here, we describe the methods needed to successfully construct the NNLO subtraction terms for the subleading colour contributions to dijet production within the antenna subtraction formalism. 
}

\preprint{{\raggedleft%
		ZU-TH 27/22\\
		KA-TP-24-2022\\
		IPPP/22/60\\
		P3H-22-087\\
	}}

\begin{document}
\maketitle
\flushbottom

\section{Introduction}
Jet production observables at hadron colliders are among the most fundamental probes of quantum chromodynamics (QCD). High precision data from the Tevatron and the LHC provide important constraints on the parton content of the colliding hadrons and enable accurate determinations of the strong coupling constant. To interpret these data in precision studies of QCD requires theory predictions at a level of accuracy that is comparable to the quality of the experimental data. To attain this accuracy in the theory predictions requires the computation of higher-order perturbative corrections.

The most basic jet production processes are initiated at Born level by $2\to 2$ parton scattering, with single jet inclusive and dijet observables both deriving from the same Born-level reaction. For these hadron collider jet production processes, the established next-to-leading order (NLO) QCD 
predictions~\cite{Ellis:1992en,Giele:1994gf,Giele:1994xd,Nagy:2001fj,Gao:2012he}, supplemented with NLO electroweak 
corrections~\cite{Dittmaier:2012kx,Campbell:2016dks,Frederix:2016ost} and parton shower based resummation~\cite{Alioli:2010xa,Hoeche:2012fm}, have long been the baseline of comparisons between experiment and theory. With residual theory uncertainties of around ten percent at NLO QCD, these predictions were however no longer sufficient with the advent of high precision jet production data from the LHC~\cite{Gehrmann:2021qex}. 
 
Higher-order QCD calculations of jet observables require the handling of infrared singular configurations that arise from real and virtual corrections to the underlying Born-level process, and that cancel each other only once all corrections at a given order are added together. This cancellation of infrared divergences among different subprocess contributions is usually accomplished by a subtraction method. Several subtraction methods have been developed~\cite{Catani:2007vq,Gehrmann-DeRidder:2005btv,Currie:2013vh,Czakon:2010td,Boughezal:2011jf,Gaunt:2015pea,DelDuca:2016ily,Caola:2017dug} for next-to-next-to-leading order (NNLO) QCD calculations, and have been applied to a substantial number of hadron collider processes~\cite{Heinrich:2020ybq} in calculations performed on a process-by-process basis with up to now only limited automation. 
 
NNLO QCD corrections to jet production observables were initially computed~\cite{Currie:2016bfm,Currie:2017eqf,Gehrmann-DeRidder:2019ibf} using the antenna subtraction method~\cite{Gehrmann-DeRidder:2005btv,Currie:2013vh,Daleo:2006xa}, thereby retaining only the leading terms in an expansion in the number of colours and quark flavours in QCD. Since the antenna subtraction method is based on the colour-ordered decomposition of QCD amplitudes, it was particularly well-adapted to extract these leading-colour contributions. Subsequently, full-colour NNLO QCD results~\cite{Czakon:2019tmo} for jet production processes were obtained using a residue-subtraction scheme~\cite{Czakon:2010td}. Most recently, we completed the full-colour NNLO calculation of jet production processes~\cite{Chen:2022tpk} using the antenna subtraction method. In this paper, we would like to complement our numerical results and phenomenological interpretations that we presented in~\cite{Chen:2022tpk} by a description of the technical details of their calculation. 
  
The implementation of the antenna subtraction method is achieved on a process-by-process basis using the colour-ordered antenna approach. In this approach, single and double real emission patterns in colour-ordered matrix elements are identified and used to guide the construction of the subtraction terms. The subtraction for the double real emission contribution is usually constructed first and subsequently integrated with additional terms to form the real-virtual part, and finally the double virtual part. For low multiplicity processes or in leading colour approximations, this approach usually results in relatively compact expressions for the subtraction terms. However, the complexity of the implementation of the colour-ordered antenna subtraction method scales with the number of coloured particles involved, which for the process considered in this paper, $pp \to jj$, reaches up to six at NNLO. Going beyond the leading colour approximation and including all subleading colour contributions also comes with new complications. In this paper, we discuss these complications and the successful implementation of the colour-ordered antenna subtraction method for the full colour dijet production, thus enabling full colour NNLO QCD predictions for jet production observables at the LHC~\cite{Chen:2022tpk}. 

This paper is set up as follows. In \sec{MEs} we list all the relevant matrix elements in a colour-decomposed manner for dijet production from proton-proton collisions at NNLO QCD in full colour. One of the main obstacles is how to deal with interference terms and this is discussed in \sec{RRsub} for the construction of the double real antenna subtraction. The construction of the real-virtual subtraction terms is discussed in \sec{RVsub}, again focusing on the novel aspects that arise at subleading colour level. Lastly, the double virtual subtraction is discussed in \sec{VVsub}, thereby demonstrating the analytical cancellation of infrared pole terms among the different subprocess contributions and  establishing the generic structure of the integrated antenna subtraction terms, combined with 
the QCD mass factorization terms, in the form of single and double integrated dipole factors. We conclude with \sec{conc} and document the colour-ordered infrared pole operators and integrated dipole factors in appendices.

\section{Relevant matrix elements} \label{sec:MEs}
In this following, we discuss all the relevant matrix elements for the NNLO QCD corrections to dijet production at the LHC, aiming to establish the notation and to 
outline the colour-connections at all colour levels, which will subsequently become relevant for the construction of the antenna subtraction terms. 

At leading order (LO), the basic parton-level dijet production processes are given by all crossings of the two-to-two parton processes: $gg\to gg$, $qg\to qg$ and $qq' \to qq'$, where $q$ and $q'$ denote quarks of identical or different flavour. The cross section at NNLO is then assembled by combining these partonic subprocess contributions, integrated over their respective phase spaces, with all terms that result from the mass factorization (MF) of the parton distributions at NNLO. The NNLO QCD corrections can be divided into three types according to the number of loops and external legs of their Feynman diagrams:
\begin{itemize}
	\item RR: the real-real or double real corrections, which consist of the tree-level diagrams with six external partons.
	\item RV: the real-virtual corrections, which consist of the one-loop diagrams with five external partons.
	\item VV: the virtual-virtual or double virtual corrections, which consist of the two-loop corrections to the four parton processes.
\end{itemize}
The NNLO contribution to the cross section thus reads: 
\ba \label{eq:NNLOcs}
\df \si_{\NNLO} = \int_{\Phi_{n+2}} \df \si^{RR} + \int_{\Phi_{n+1}} \l( \df \si^{RV} + \df \si^{\MF,1} \r) +  \int_{\Phi_{n}} \left( \df \si^{VV} + \df \si^{\MF,2} \right),
\ea
where the $\Phi_{n}$ denote the $2\to (n-2)$-parton phase space integrations, and $n=4$ is the parton multiplicity of the underlying Born-level process. The $\df \si^{RR,RV,VV,\MF}$ contain the projection of all subprocesses onto 
$jj$ final states, involving the jet reconstruction algorithm and the application of kinematical cuts. 

The phase space integrals of different multiplicities in \eq{NNLOcs} cannot be computed in a direct manner, since each contribution on the right-hand side contains infrared singularities from real or virtual unresolved parton exchanges, which only cancel among each other once all contributions are added together. The singular real-radiation behaviour of the RR and RV contributions must be subtracted prior to integration, and the explicit infrared poles must be separated in the RV and VV contributions. We perform the cancellation of infrared-divergent terms among the different contributions using the antenna subtraction method~\cite{Gehrmann-DeRidder:2005btv,Daleo:2006xa,Currie:2013vh}, which introduces subtraction terms at the RR, RV and VV levels, resulting in:
\ba
\df \si_{\NNLO} = \int_{\Phi_{n+2}} \l( \df \si^{RR} - \df \si^{S} \r) + \int_{\Phi_{n+1}} \l( \df \si^{RV} - \df \si^{T} \r) + \int_{\Phi_{n}} \l( \df \si^{VV} - \df \si^{U} \r),
\ea
where each bracketed term on the right hand side is now free of infrared singularities and can be integrated numerically. They are implemented into the parton-level event generator \NNLOJET, which enables to compute any infrared-safe observable related to $pp\to jj$, such as multi-differential single-jet inclusive or dijet cross sections. Our numerical results for several LHC jet production observables are reported in a companion paper~\cite{Chen:2022tpk}. 

The newly introduced subtraction terms have the following structure:
\ba \label{eq:sigsub}
\df \si^{S} &=& \df \si^{S,1} + \df \si^{S,2}, \nn\\
\df \si^{T} &=& \df \si^{V,S} - \int_1 \df \si^{S,1} - \df \si^{\MF,1}, \nn\\
\df \si^{U} &=& -\int_1 {\rm d} \si^{V,S} - \int_2 \df \si^{S,2} - \df \si^{\MF,2}.
\ea
In here, $\df \si^{S,1,2}$ correspond to the parts contributing to the $(n-1)$ and $(n-2)$ parton final state respectively, and $\df \si^{V,S}$ subtracts all one-parton unresolved emissions from $\df \si^{RV}$. The integrations over the phase spaces relevant to one-parton or two-parton emissions (indicated by the respective suffixes) factorize from their original higher multiplicity phase spaces and are performed analytically~\cite{Gehrmann-DeRidder:2003pne,Gehrmann-DeRidder:2005svg,Gehrmann-DeRidder:2005alt,Gehrmann-DeRidder:2005btv,Daleo:2009yj,Boughezal:2010mc,Gehrmann:2011wi,Gehrmann-DeRidder:2012too}.

The construction of the antenna subtraction terms will be described in detail in \msecs{RRsub}{VVsub}, where we focus in particular on the novel features of the subtraction terms that appear for the first time in $pp\to jj$ at full colour. These considerations will in particular be important in view of a future automated generation~\cite{Chen:2022ktf} of antenna subtraction terms for generic processes.

As we sum over all possible allowed parton identities, we have to classify the processes contributing at NNLO according to their parton content. We define the following types of (squared) matrix elements:
\begin{itemize}
	\item A-type matrix elements: processes for which all the external partons are gluons.
	\item B-type matrix elements: processes which contain only one quark pair, plus any additional number of gluons.
	\item C-type matrix elements: processes which contain two quark pairs of different flavour, plus any additional number of gluons.
	\item D-type matrix elements: processes which contain two quark pairs of identical flavour, plus any additional number of gluons.
	\item E-type matrix elements: processes which contain three quark pairs, all of different flavour.
	\item F-type matrix elements: processes which contain two quark pairs of the same flavour and a third quark pair of different flavour.
	\item G-type matrix elements: processes which contain three quark pairs, all of the same flavour.
\end{itemize}
The distinction between the identical-flavour and different-flavour multi-quark processes is such that the identical flavour processes contain only those contributions that appear anew (interference terms from interchanging external quark states that are admitted only for identical flavours) if quark flavours are set equal. For example, non-identical flavour quark-quark scattering at leading order receives contributions from C-type squared matrix elements, while identical flavour quark-quark scattering receives contributions from both C-type and D-type squared matrix elements. 

The double real corrections to dijet production contain six external partons, so they can be of any A-, B-, C-, D-, E-, F- and G-type, while the real-virtual and double virtual corrections contain only five and four partons respectively, meaning they can only be of A-, B-, C- or D-type. We use the notation $\mathcal{M}_n^l(\{p\})$ with $\mathcal{M} \in \{\mathcal{A},\mathcal{B},\mathcal{C},\mathcal{D},\mathcal{E},\mathcal{F},\mathcal{G}\}$ for a generic amplitude containing $n$ external gluons and $l$ loops depending on the external parton momenta $\{p\}$. The helicity labels of the external partons are suppressed and when considering squared quantities $M_n^l(\{p\})$ we always assume a sum over the helicity configurations.  Squared matrix elements are denoted by capital italic letters, which are put in boldface when a summation over momentum
permutations is involved. 

In the following sections we define all the relevant NNLO matrix elements according to this convention. In view of the antenna subtraction which will be applied separately for each matrix element type per colour level, all the matrix elements are broken down in a tower of colour levels, each having their own powers of the colour factors $N_c$ and $n_f$. Each colour level may be further sub-divided in functions which display some desired behaviour, i.e. functions which have clear colour connections. Besides all the matrix elements which form the NNLO corrections to dijet production, we also look at several lower multiplicity matrix elements. These matrix elements are part of the NLO corrections to dijet production or can appear as reduced matrix elements in the NNLO subtraction. We list them along with several of their properties, such as their pole structures.

\subsection{A-type matrix elements}
The A-type amplitudes are defined as the all-gluonic amplitudes. We denote an A-type amplitude containing $n$ gluons and $l$ loops with $\mathcal{A}_n^l(g_1,g_2,\dots,g_n)$, where the ordering of the labels $g_1,g_2,\dots,g_n$ reflects the colour connections of the gluons, i.e. gluon $g_2$ is colour connected to gluon $g_1$ and gluon $g_3$, gluon $g_3$ colour connected to gluon $g_2$ and gluon $g_4$, and so on. The gluons $g_1$ and $g_n$ at the endpoint are also colour connected to each other.

\subsubsection{Amplitudes}
The full tree-level A-type amplitude with $n$ gluons is given by
\be \label{eq:AtypeAmp}
\mathcal{A}_n^0 = g_s^{n-2} \sum_{\si \in S_n/Z_n} (g_{\si(1)} g_{\si(2)} \dots g_{\si(n)}) \, a_n^0(g_{\si(1)}, g_{\si(2)}, \dots, g_{\si(n)}),
\ee
where $g_s$ is the QCD gauge coupling (related to the strong coupling constant by $4\pi \al_s = g_s^2$), $S_n$ the permutation group of $n$ objects, $Z_n$ its subgroup of cyclic permutations, $(g_1 g_2 \dots g_n)$ a shorthand notation for the colour structure $\mathrm{Tr}(T^{g_1}T^{g_2}\dots T^{g_n})$ and $a_n^0(g_1, g_2, \dots, g_n)$ the colour-ordered $n$-gluon partial amplitudes containing the kinematic information~\cite{Berends:1987me}. The partial amplitudes $a_n^0$ fulfill several relations~\cite{Mangano:1990by}:
\begin{itemize}
	\item Invariance under cyclic permutations: 
	\be
	a_n^0(g_1,g_2,\dots,g_n) = a_n^0(g_{\si(1)},g_{\si(2)},\dots,g_{\si(n)}),
	\ee
	with $\si \in Z_n$. 
	\item Line reversal symmetry:
	\be
	a_n^0(g_1,g_2,\dots,g_n) = (-1)^n a_n^0(g_n, g_{n-1}, \dots, g_1).
	\ee
	\item Dual Ward Identity:
	\ba
	a_n^0(g_1,g_2,g_3,\dots,g_n) + a_n^0(g_2,g_1,g_3,\dots,g_n) + a_n^0(g_2,g_3,g_1,\dots,g_n) \nn\\
	+ \dots + a_n^0(g_2,g_3,\dots,g_1, g_n) = 0,
	\ea
	i.e. the sum of amplitudes where one gluon ($g_1$) is placed in between all other gluons, while keeping the relative order of the other gluons the same, vanishes.
	\item Gauge invariance of $a_n^0(g_1,g_2,\dots,g_n)$. 
	\item Colour ordering:  the order of the arguments of $a_n^0(g_1,g_2,\dots,g_n)$ reflects their colour ordering, such that kinematical singularities can appear only 
	if adjacent gluons become unresolved with respect to each other. 
\end{itemize} 
The one-loop amplitude for four external gluons is given by~\cite{Ellis:1985er}
\ba \label{eq:FullA4g1amp}
\mathcal{A}_4^1 &=& g_s^2 \l( \frac{\al_s}{2\pi}\r) N_c \Big[ \sum_{\si \in S_4/Z_4} (g_{\si(1)} g_{\si(2)} g_{\si(3)} g_{\si(4)}) \, \mathcal{A}_{4;1}^1(g_{\si(1)}, g_{\si(2)}, g_{\si(3)}, g_{\si(4)}) \nn\\
&+& \frac{1}{N_c} 
\sum_{\rho \in S_4/Z_2 \times Z_2}  (g_{\rho(1)} g_{\rho(2)}) (g_{\rho(3)} g_{\rho(4)}) \, \mathcal{A}_{4;3}^1(g_{\rho(1)},g_{\rho(2)};g_{\rho(3)},g_{\rho(4)}) \Big],
\ea
where in the second sum the permutations only go over the orderings which are inequivalent under cyclic permutations of the two subsets $\{\rho(1),\rho(2)\}$ and $\{\rho(3),\rho(4)\}$ amongst themselves, and the interchange of these subsets. The subleading colour partial amplitudes $\mathcal{A}_{4;3}^1$ can be expressed in terms of the leading colour partial amplitudes $\mathcal{A}_{4;1}^1$. In the case of four gluons, the relation is simply
\ba \label{eq:A43decompose}
\mathcal{A}_{4;3}^1(g_1,g_2;g_3,g_4) = \mathcal{A}_{4;1}^1(g_2,g_1,g_3,g_4) + \mathcal{A}_{4;1}^1(g_2,g_3,g_1,g_4) + \mathcal{A}_{4;1}^1(g_3,g_2,g_1,g_4).
\ea
We note that the $\mathcal{A}_{4;3}^1$ amplitude is completely symmetric under the exchange of any two gluon indices, resulting in cancellations of the subleading colour partial amplitudes with each other at the squared matrix element level. The leading colour partial amplitude can be further colour-decomposed into more primitive amplitudes:
\ba
\mathcal{A}_{n;1}^1(g_1,g_2,\dots,g_n) = a_n^1(g_1,g_2,\dots,g_n) + \frac{n_f}{N_c} \hat{a}_n^1(g_1,g_2,\dots,g_n),
\ea
where the quark loop contribution proportional to $n_f$ has been separated. Adding one more gluon, we have for the five-gluon one-loop amplitude~\cite{Bern:1993mq}:
\ba \label{eq:FullA5g1amp}
\mathcal{A}_5^1 &=& g_s^3 \l( \frac{\al_s}{2\pi}\r) N_c \Big[ \sum_{\si \in S_5/Z_5} (g_{\si(1)} g_{\si(2)} g_{\si(3)} g_{\si(4)} g_{\si(5)}) \, \mathcal{A}_{5;1}^1(g_{\si(1)},g_{\si(2)},g_{\si(3)},g_{\si(4)} g_{\si(5)}) \nn\\
&+& \frac{1}{N_c} \sum_{\rho \in S_5/Z_2 \times Z_3} (g_{\rho(1)} g_{\rho(2)}) (g_{\rho(3)} g_{\rho(4)} g_{\rho(5)}) \, \mathcal{A}_{5;3}^1(g_{\si(1)},g_{\si(2)};g_{\si(3)},g_{\si(4)} g_{\si(5)}) \Big], \nn\\
\ea
where the notation follows by extension from the four-gluon one-loop amplitude. The subleading colour partial amplitude $\mathcal{A}_{5;3}^1$ can again be expressed in terms of the leading colour amplitudes:
\ba \label{eq:A53decompose}
\mathcal{A}_{5;3}^1(g_1,g_2,g_3,g_4,g_5) &=& \mathcal{A}_{5;1}^1(g_2,g_1,g_3,g_4,g_5) + \mathcal{A}_{5;1}^1(g_2,g_3,g_1,g_4,g_5) \nn\\
&+& \mathcal{A}_{5;1}^1(g_2,g_3,g_4,g_1,g_5) + \mathcal{A}_{5;1}^1(g_3,g_2,g_1,g_4,g_5) \nn\\
&+& \mathcal{A}_{5;1}^1(g_3,g_2,g_4,g_1,g_5) + \mathcal{A}_{5;1}^1(g_3,g_4,g_2,g_1,g_5) \nn\\
&+& \mathcal{A}_{5;1}^1(g_1,g_2,g_3,g_4,g_5) + \mathcal{A}_{5;1}^1(g_1,g_3,g_2,g_4,g_5) \nn\\
&+& \mathcal{A}_{5;1}^1(g_1,g_3,g_4,g_2,g_5) + \mathcal{A}_{5;1}^1(g_3,g_1,g_2,g_4,g_5) \nn\\
&+& \mathcal{A}_{5;1}^1(g_3,g_1,g_4,g_2,g_5) + \mathcal{A}_{5;1}^1(g_3,g_4,g_1,g_2,g_5).
\ea

\subsubsection{Pole structure of the one-loop amplitudes}
The poles of the one-loop primitive amplitudes can be expressed in terms of tree-level amplitudes multiplying the $I^{(1)}$ infrared singularity operators~\cite{Catani:1998bh}, which we express in a colour-ordered form~\cite{Gehrmann-DeRidder:2005btv}, as summarised in \app{Iop_colord}:
\ba \label{eq:AtypeAmpPoles}
\lefteqn{\mathrm{Poles} \big[ a_n^1(g_1,g_2,g_3,\dots,g_n) \big] = } \nn\\
&& 2 \, \big[I_{gg}^{(1)}(\ep,s_{12}) + I_{gg}^{(1)}(\ep,s_{23}) + \dots + I_{gg}^{(1)}(\ep,s_{n1}) \big] a_n^0(g_1,g_2,g_3,\dots,g_n), \nn\\
\lefteqn{\mathrm{Poles} \big[ \hat{a}_n^1(g_1,g_2,g_3,\dots,g_n) \big] =} \nn\\ 
&& 2 \, \big[I_{gg,F}^{(1)}(\ep,s_{12}) + I_{gg,F}^{(1)}(\ep,s_{23}) + \dots + I_{gg,F}^{(1)}(\ep,s_{n1}) \big] a_n^0(g_1,g_2,g_3,\dots,g_n). 
\ea

\subsubsection{Four-parton tree-level}
The leading order tree-level matrix element for four-gluon scattering is obtained by squaring \eq{AtypeAmp} with $n=4$. After factoring out an overall $(N_c^2-1)$, all the subleading colour levels vanish, resulting in
\ba \label{eq:FullA4g0}
\big|\mathcal{A}_4^0\big|^2 &=& g_s^4 (N_c^2-1) N_c^2 \bs{A}_4^0 \nn\\
&=& g_s^4 (N_c^2-1) N_c^2 \sum_{\si \in S_3} A_4^0(g_1,g_{\si(2)},g_{\si(3)},g_{\si(4)}) \nn\\
&=& g_s^4 (N_c^2-1) N_c^2 \cdot 2 \big[ A_4^0(g_1,g_2,g_3,g_4) + A_4^0(g_1,g_2,g_4,g_3) + A_4^0(g_1,g_3,g_2,g_4) \big], \nn\\
\ea
where $A_n^0$ is the squared partial amplitude 
\be
A_n^0(g_1,g_2,\dots,g_n) = a_n^{0\dagger}(g_1,g_2,\dots,g_n) \, a_n^0(g_1,g_2,\dots,g_n),
\ee
and the properties of the all-gluonic amplitudes have been used to simplify the result into only three independent orderings of $A_4^0$. Besides being a Born diagram for dijet production at the LHC, these functions will also appear as reduced matrix elements which are multiplied by antenna functions in the antenna subtraction. For this reason it is also convenient to define the interference term of a conjugated amplitude with colour ordering $(i_1,i_2,\dots,i_n)$ multiplied with an amplitude with colour ordering $(j_1,j_2,\dots,j_n)$, which do not need to be the same:
\ba
A_{n,\mathrm{int}}^0(i_1,i_2,\dots,i_n;j_1,j_2,\dots,j_n) = a_n^{0\dagger}(i_1,i_2,\dots,i_n) \, a_n^0(j_1,j_2,\dots,j_n).
\ea

\subsubsection{Five-parton tree-level}
For $n=5$ all subleading colour levels vanish as well, yielding
\ba \label{eq:FullA5g0}
\big|\mathcal{A}_5^0\big|^2 &=& g_s^6 (N_c^2-1) N_c^3 \bs{A_5^0} \nn\\
&=& g_s^6 (N_c^2-1) N_c^3 \sum_{\si \in S_4} A_5^0(g_1,g_{\si(2)},g_{\si(3)},g_{\si(4)},g_{\si(5)}).
\ea
This matrix element is part of the NLO correction, for which single unresolved limits need to be subtracted, and appears as a reduced matrix element in the NNLO subtraction. 

\subsubsection{Six-parton tree-level}
For $n=6$, we obtain the matrix element~\cite{Gunion:1986cb} which forms the all-gluonic double real correction for dijet production. Double real subtraction terms need to be constructed to mimic all the single and double unresolved limits of the matrix element, such that it can be numerically integrated over the whole phase space. Starting from six gluons, not all the subleading colour levels turn out to be zero after squaring \eq{AtypeAmp}. We find a leading colour level and a subleading colour level suppressed by $1/N_c^2$,
\ba
\big|\mathcal{A}_6^0\big|^2 = g_s^8 (N_c^2-1) N_c^4 \Big[ \bs{A}_6^0 + \frac{2}{N_c^2} \bs{\t{A}}_6^0 \Big],
\ea
where
\ba \label{eq:A6g0funcs}
\bs{A}_6^0 &=& \sum_{\si \in S_5} A_6^0(g_1,g_{\si(2)},g_{\si(3)},g_{\si(4)},g_{\si(5)},g_{\si(6)}), \nn\\
\bs{\t{A}}_6^0 &=& \sum_{\si \in S_5} a_6^{0\dagger}(g_1,g_{\si(2)},g_{\si(3)},g_{\si(4)},g_{\si(5)},g_{\si(6)}) \big[ a_6^0(g_1,g_{\si(3)},g_{\si(5)},g_{\si(2)},g_{\si(6)},g_{\si(4)}) \nn\\
&+& a_6^0(g_1,g_{\si(3)},g_{\si(6)},g_{\si(4)},g_{\si(2)},g_{\si(5)}) +
a_6^0(g_1,g_{\si(4)},g_{\si(2)},g_{\si(6)},g_{\si(3)},g_{\si(5)}) \big]. \nn\\		
\ea
In \eq{A6g0funcs} we see that the leading colour level consists of a sum of coherent squared orderings just like for the lower multiplicity matrix elements, while the subleading colour level consists of three incoherent interferences, summed over permutations.

\subsubsection{Four-parton one-loop}
The full four-gluon one-loop matrix element contribution is given by interfering the one-loop amplitude \eq{FullA4g1amp} with the tree-level amplitude \eq{AtypeAmp}:
\ba
M_4^1(g_1,g_2,g_3,g_4) = 2\, \mathrm{Re} \big[ \mathcal{A}_4^{0\dagger} \mathcal{A}_4^1 \big] &=& 
\l( \frac{\al_s}{2\pi}\r) g_s^4 (N_c^2-1) N_c^3 \Big[ \sum_{\si \in S_3} A_4^1(g_1,g_{\si(2)},g_{\si(3)},g_{\si(4)}) \nn\\
&+& \frac{n_f}{N_c} \sum_{\si \in S_3} \hat{A}_4^1(g_1,g_{\si(2)},g_{\si(3)},g_{\si(4)}) \Big],
\ea
where
\ba
A_n^1(g_1,g_2,\dots,g_n) &=& \mathrm{Re} \big[ a_n^{1\dagger}(g_1,g_2,\dots,g_n) \, a_n^0(g_1,g_2,\dots,g_n) \big], \nn\\
\hat{A}_n^1(g_1,g_2,\dots,g_n) &=& \mathrm{Re} \big[ \hat{a}_n^{1\dagger}(g_1,g_2,\dots,g_n) \, a_n^0(g_1,g_2,\dots,g_n) \big].
\ea
Besides the leading $N_c$ colour level, we also have the colour level proportional to $n_f/N_c$ coming from a closed quark loop and in general it is also possible to have additional subleading $n_f/N_c^3$, $n_f/N_c^5, \, \dots, n_f/N_c^{2n+1}$ colour levels. The four-parton A-type one-loop matrix element  has only a single $n_f/N_c$ colour level, which is referred to as the leading $n_f$ contribution and is considered part of the leading colour approximation. This matrix element is part of the virtual correction of dijet production, but like for the lower multiplicity tree-level matrix element, it will also appear as a reduced matrix element in the subtraction terms. The one-loop matrix element contains infrared poles which factorize onto the tree-level matrix elements. The pole structure of the functions $A_4^1$ and $\hat{A}_4^1$ for a given ordering follows from \eq{AtypeAmpPoles}:
\ba
\mathrm{Poles} \big[ A_4^1(a,b,c,d) \big] &=& 2 \, \big[ I_{gg}^{(1)}(\ep,s_{ab}) + I_{gg}^{(1)}(\ep,s_{bc}) + I_{gg}^{(1)}(\ep,s_{cd}) + I_{gg}^{(1)}(\ep,s_{da}) \big] \nn\\
&& \times A_4^0(a,b,c,d), \nn\\
\mathrm{Poles} \big[ \hat{A}_4^1(a,b,c,d) \big] &=& 2 \, \big[ I_{gg,F}^{(1)}(\ep,s_{ab}) + I_{gg,F}^{(1)}(\ep,s_{bc}) + I_{gg,F}^{(1)}(\ep,s_{cd}) + I_{gg,F}^{(1)}(\ep,s_{da}) \big] \nn\\
&& \times A_4^0(a,b,c,d).
\ea

\subsubsection{Five-parton one-loop}
The real-virtual A-type matrix element correction is obtained by interfering the five-gluon one-loop amplitude~\cite{Bern:1993mq} with the tree-level amplitude. Broken down in its colour levels it is given by
\ba
\lefteqn{M_5^1(g_1,g_2,g_3,g_4,g_5) = \l(\frac{\al_s}{2\pi}\r) g_s^6 (N_c^2-1) N_c^4 \Big[} \nn\\
&&  \sum_{\si \in S_4} A_5^1(g_1,g_{\si(2)},g_{\si(3)},g_{\si(4)},g_{\si(5)}) + \frac{n_f}{N_c} \sum_{\si \in S_4} \hat{A}_5^1(g_1,g_{\si(2)},g_{\si(3)},g_{\si(4)},g_{\si(5)}) \nn\\
&&+  \frac{1}{N_c^2} \sum_{\si \in S_4} \t{A}_5^1(g_1,g_{\si(2)},g_{\si(3)},g_{\si(4)},g_{\si(5)}) + \frac{n_f}{N_c^3} \sum_{\si \in S_4} \hat{\t{A}}_5^1(g_1,g_{\si(2)},g_{\si(3)},g_{\si(4)},g_{\si(5)}) \Big], \nn\\
\ea
where the subleading colour functions $\t{A}_5^1$ and $\hat{\t{A}}_5^1$ are obtained from the one independent interference ordering which contains no common neighbouring partons:
\ba
\t{A}_5^1(g_1,g_{\si(2)},g_{\si(3)},g_{\si(4)},g_{\si(5)}) = 2\, \mathrm{Re} \big[ a_5^{0\dagger}(g_1,g_{\si(2)},g_{\si(3)},g_{\si(4)},g_{\si(5)}) \nn\\
\times a_5^1(g_1,g_{\si(4)},g_{\si(2)},g_{\si(5)},g_{\si(3)}) \big], \nn\\
\hat{\t{A}}_5^1(g_1,g_{\si(2)},g_{\si(3)},g_{\si(4)},g_{\si(5)}) = 2\, \mathrm{Re} \big[ a_5^{0\dagger}(g_1,g_{\si(2)},g_{\si(3)},g_{\si(4)},g_{\si(5)}) \nn\\
\times \hat{a}_5^1(g_1,g_{\si(4)},g_{\si(2)},g_{\si(5)},g_{\si(3)}) \big].
\ea
The one-loop five-gluon matrix element contains both singularities in unresolved phase space regions and $\ep$-poles from the loop. The subtraction term for this matrix element will need to subtract these single unresolved limits as well as cancel the poles. The pole structures of the leading colour functions $A_5^1$ and $\hat{A}_5^1$ follow by extension from the four-gluon functions, while for the subleading colour functions they are
\ba
\mathrm{Poles} \big[ \t{A}_5^1(a,b,c,d,e) \big] &=& 2\, \big[ I_{gg}^{(1)}(\ep,s_{ad}) + I_{gg}^{(1)}(\ep,s_{db}) + I_{gg}^{(1)}(\ep,s_{be}) + I_{gg}^{(1)}(\ep,s_{ec}) \nn\\
&+& I_{gg}^{(1)}(\ep,s_{ca}) \big] \times \mathrm{Re} \big[ a_5^{0\dagger}(a,b,c,d,e) \, a_5^0(a,d,b,e,c) \big], \nn\\
\mathrm{Poles} \big[ \hat{\t{A}}_5^1(a,b,c,d,e) \big] &=& 2\, \big[ I_{gg,F}^{(1)}(\ep,s_{ad}) + I_{gg,F}^{(1)}(\ep,s_{db}) + I_{gg,F}^{(1)}(\ep,s_{be}) + I_{gg,F}^{(1)}(\ep,s_{ec}) \nn\\
&+& I_{gg,F}^{(1)}(\ep,s_{ca}) \big] \times \mathrm{Re} \big[ a_5^{0\dagger}(a,b,c,d,e) \, a_5^0(a,d,b,e,c) \big].
\ea

\subsubsection{Four-parton two-loop}
Lastly, we have the two-loop four-gluon matrix element which forms the double virtual correction. It consists of the two-loop amplitude interfered with the tree-level amplitude and the genuine square of the one-loop amplitude. Decomposed in all its colour levels it is given by~\cite{Glover:2001af,Glover:2001rd,Bern:2002tk}
\ba
M_4^2(g_1,g_2,g_3,g_4) &=& 2\, \mathrm{Re} \big[ \mathcal{A}_4^{0\dagger} \mathcal{A}_4^2 \big] + \mathcal{A}_4^{1\dagger} \mathcal{A}_4^1 = \l( \frac{\al_s}{2\pi}\r)^2 g_s^4 (N_c^2-1) N_c^4 \Big[ \nn\\
&& A_4^2(g_1,g_2,g_3,g_4) + \frac{n_f}{N_c} \hat{A}_4^2(g_1,g_2,g_3,g_4) + \frac{n_f^2}{N_c^2} \hat{\hat{A}}_4^2(g_1,g_2,g_3,g_4) \nn\\
&+& \frac{1}{N_c^2} \t{A}_4^2(g_1,g_2,g_3,g_4) + \frac{n_f}{N_c^3} \hat{\t{A}}_4^2(g_1,g_2,g_3,g_4) + \frac{n_f^2}{N_c^4} \hat{\t{A}}_4^2(g_1,g_2,g_3,g_4) \nn\\
&+& \frac{n_f^2}{N_c^6} \hat{\hat{\t{\t{A}}}}_4^2(g_1,g_2,g_3,g_4) \Big]. 
\ea
Its $\ep$-pole structure is documented in~\cite{Glover:2001af,Glover:2001rd,Bern:2002tk} and cancels with the integrated subtraction terms at the VV level. 

\subsection{B-type matrix elements}
The B-type matrix elements are matrix elements containing one quark pair plus any number of gluons. Thus, $\mathcal{B}_n^l$ denotes a B-type amplitude containing one quark pair, $n$ gluons and $l$ loops. For the functions defined in this section we will use the notation $1,2$ for the $q\qb$-pair and $i,j,k,\dots$ for the gluons. The convention for the ordering of the arguments of the B-type functions will be $B_n^l(1,i,j,k,\dots,2)$, where the quarks are in the first and last entries and the gluons in between. As for the A-type amplitudes, the ordering of the parton labels reflects the colour connections of the partons.

\subsubsection{Amplitudes}
Similar to the A-type tree-level amplitude, the full B-type tree-level amplitude for any number of gluons can also be written in a simple expression,  
\be \label{eq:BtypeAmp}
\mathcal{B}_n^0 = g_s^n \sum_{\si \in S_n} (g_{\si(1)},g_{\si(2)},\dots,g_{\si(n)})_{q\qb} \, b_n^0(q,g_{\si(1)},g_{\si(2)},\dots,g_{\si(n)},\qb),
\ee
where $(g_1,g_2,\dots,g_n)_{q\qb}$ abbreviates the colour matrix product $(T^{g_1} T^{g_2} \dots T^{g_n})_{q\qb}$ and the partial amplitude containing one quark pair and $n$ gluons is denoted by $b_n^0(q,g_1,g_2,\dots,g_n,\qb)$. At one loop, the amplitudes have a more elaborate colour structure. The one-loop amplitude with two gluons is given by
\ba \label{eq:FullB2g1amp}
\mathcal{B}_2^1 &=& g_s^2 \l( \frac{\al_s}{2\pi} \r) N_c \Big[ \sum_{\si \in S_2} (T^{g_{\si(1)}} T^{g_{\si(2)}})_{q\qb} \, \mathcal{B}_{2;1}^1(q,g_{\si(1)},g_{\si(2)},\qb) \nn\\
&& + \frac{1}{N_c} (T^{g_1} T^{g_2}) \de_{q\qb} \, \mathcal{B}_{2;3}^1(q,g_1,g_2,\qb) \Big],
\ea
and the five-parton one-loop B-type amplitude reads~\cite{Bern:1994fz,Signer:1995a}
\ba \label{eq:FullB3g1amp}
\mathcal{B}_3^1 &=& g_s^3 \l( \frac{\al_s}{2\pi} \r) N_c \Big[ \sum_{\si \in S_3} (T^{g_{\si(1)}} T^{g_{\si(2)}} T^{g_{\si(3)}})_{q\qb} \, \mathcal{B}_{3;1}^1(q,g_{\si(1)},g_{\si(2)},g_{\si(3)},\qb) \nn\\
&+& \frac{1}{N_c} \sum_{i=1,2,3} (T^{g_i})_{q\qb} \de_{g_j g_k} \mathcal{B}_{3;3}^1(q,g_i,\qb;g_j,g_k) \nn\\
&+& \frac{1}{N_c} (T^{g_1} T^{g_2} T^{g_3}) \de_{q\qb} \, \mathcal{B}_{3;4}^1(q,\qb;g_1,g_2,g_3) \nn\\
&+& \frac{1}{N_c} (T^{g_3} T^{g_2} T^{g_1}) \de_{q\qb} \, \mathcal{B}_{3;4}^1(q,\qb;g_3,g_2,g_1) \Big].
\ea
The leading partial amplitude $\mathcal{B}_{n;1}^1$ is the only one which contributes at leading colour and can be further colour-decomposed into primitive amplitudes:
\ba \label{eq:Bng1primamp}
\lefteqn{\mathcal{B}_{n;1}^1(q,g_1,\dots,g_n,\qb) =} \nn\\ &&
 b_n^1(q,g_1,\dots,g_n,\qb) + \frac{n_f}{N_c} \hat{b}_n^1(q,g_1,\dots,g_n,\qb) - \frac{1}{N_c^2} \t{b}_n^1(q,g_1,\dots,g_n,\qb). 
\ea
In contrast to the A-type amplitudes, the primitive amplitudes form the basic building blocks now, as the subleading colour partial amplitudes, $\mathcal{B}_{n;i}^1$ with $i>1$, cannot be written in terms of the leading colour partial amplitude $\mathcal{B}_{n;1}^1$, but are constituted as a sum over permutations of the same primitive amplitudes: 
\ba 
\lefteqn{\mathcal{B}_{n;i}^1(q,g_1,\dots,g_n,\qb) =} \nn\\ && (-1)^{i-1} \sum_{\si \in \mathrm{CY} \{\al\} \{\bt\}} \Big[ b_n^1(\si(q,g_1,\dots,g_n,\qb)) - \frac{n_f}{N_c} \hat{\t{b}}_n^1(\si(q,g_1,\dots,g_n,\qb)) \Big], 
\ea
where $\{\al\} = \{g_{i+1},g_i,\dots,g_1\}$, $\{\bt\} = \{q,\qb,g_{i+2},g_{i+3},\dots,g_n \}$, $\mathrm{CY}\{\al\}\{\bt\}$ are all the cyclically ordered permutations with $q$ held fixed and $\hat{\t{b}}_n^1$ a primitive amplitude which can be related to the amplitude $\hat{b}_n^1$ in \eq{Bng1primamp} by a reflection identity~\cite{Bern:1994fz}.  
 
\subsubsection{Pole structure of the one-loop amplitudes} \label{sec:Btype1LoopPoles}
The leading colour partial amplitude has the pole structure~\cite{Kunszt:1994np}:
\ba
\lefteqn{\mathrm{Poles} \big[ \mathcal{B}_{3;1}^1(q,g_1,g_2,g_3,\qb) \big] =  }\nn \\
&& \Big[ \l( I_{qg}^{(1)}(\ep,s_{q1}) + I_{gg}^{(1)}(\ep,s_{12}) + I_{gg}^{(1)}(\ep,s_{23}) + I_{qg}^{(1)}(\ep,s_{3\qb}) \r) \nn\\
&& +\frac{n_f}{N_c} \l( I_{qg,F}^{(1)}(\ep,s_{q1}) + I_{gg,F}^{(1)}(\ep,s_{12}) + I_{gg,F}^{(1)}(\ep,s_{23}) + I_{qg,F}^{(1)}(\ep,s_{3\qb}) \r) \nn\\
&& -\frac{1}{N_c^2} I_{q\qb}^{(1)}(\ep,s_{q\qb}) \Big] b_3^0(q,g_1,g_2,g_3,\qb),
\ea
while for the subleading partial amplitudes we have
\ba
\lefteqn{\mathrm{Poles} \big[ \mathcal{B}_{3;3}^1(q,g_1,\qb;g_2,g_3) \big] =} \nn\\
&& \big[ -I_{gg}^{(1)}(\ep,s_{13}) + I_{qg}^{(1)}(\ep,s_{1\qb}) + I_{gg}^{(1)}(\ep,s_{23}) - I_{qg}^{(1)}(\ep,s_{2\qb}) \big] b_3^0(q,g_1,g_2,g_3,\qb) \nn\\
&& +  \big[ 2 \leftrightarrow 3 \big] b_3^0(q,g_1,g_3,g_2,\qb) + \big[ q \leftrightarrow \qb \big] b_3^0(q,g_3,g_2,g_1,\qb) \nn\\
&& + \big[ 2 \leftrightarrow 3, q \leftrightarrow \qb \big] b_3^0(q,g_2,g_1,g_3,\qb), \nn\\ \nn\\
%
\lefteqn{\mathrm{Poles} \big[ \mathcal{B}_{3;4}^1(q,\qb;g_1,g_2,g_3) \big] = }\nn\\
&& \big[ - I_{qg}^{(1)}(\ep,s_{q3}) + I_{q\qb}^{(1)}(\ep,s_{q\qb}) + I_{gg}^{(1)}(\ep,s_{13}) - I_{qg}^{(1)}(\ep,s_{1\qb}) \big] b_3^0(q,g_1,g_2,g_3,\qb) \nn\\
&& + \big[ 1 \to 2, 2 \to 3, 3 \to 1 \big] b_3^0(q,g_2,g_3,g_1,\qb) \nn\\
&& + \big[ 1 \to 3, 3 \to 2, 2 \to 1 \big] b_3^0(q,g_3,g_1,g_2,\qb),
%
\ea
where the poles for $\mathcal{B}_{3;3}^1(q,g_2,\qb;g_3,g_1), \mathcal{B}_{3;3}^1(q,g_3,\qb;g_1,g_2)$ and $\mathcal{B}_{3;4}^1(q,\qb;g_3,g_2,g_1)$ can be obtained from appropriate permutations of the above expressions.

\subsubsection{Four-parton tree-level}
The B-type tree-level amplitude with $n=2$ gluons has two colour orderings:
\be
b_2^0(1,i,j,2) \quad \mathrm{and} \quad b_2^0(1,j,i,2).
\ee
Defining the following squared functions,
\ba
B_2^0(1,i,j,2) &=& \big|b_2^0(1,i,j,2)\big|^2, \nn\\
\bar{B}_2^0(1,i,j,2) &=& \big|b_2^0(1,i,j,2)+b_2^0(1,j,i,2)\big|^2,
\ea
the full squared four-parton B-type matrix element is then given by
\ba \label{eq:FullB2g0}
\big|\mathcal{B}_2^0\big|^2&=&g_s^4 (N_c^2-1) N_c \Big[ \bs{B}_2^0-\frac{1}{N_c^2}\bar{\bs{B}}_2^0 \Big],
\ea
with
\ba
\bs{B}_2^0 &=& \sum_{P(i,j)}B_2^0(1,i,j,2),\nn\\
\bar{\bs{B}}_2^0 &=& \bar{B}_2^0(1,i,j,2),
\ea
where $P(i,j)$ denotes all permutations of $i,j$. The leading colour function $B_2^0$ has colour connections as given by the ordering of its arguments, while the subleading colour function $\bar{B}_2^0$ contains abelian gluons: the gluons $i$ and $j$ are not colour connected to each other but both to the $q\qb$ pair. For later convenience, we also define the interference term
\ba \label{eq:BRt2g0}
\t{B}_{2,R}^0(1,i,j,2) = 2\, \mathrm{Re} \big[ b_2^{0\dagger}(1,i,j,2) \, b_2^0(1,j,i,2) \big],
\ea
which can be written in terms of the functions defined above
\ba
\t{B}_{2,R}^0(1,i,j,2) = \bar{B}_2^0(1,i,j,2) - B_2^0(1,i,j,2) - B_2^0(1,j,i,2).
\ea

\subsubsection{Five-parton tree-level}
For three gluons, there are $3! = 6$ colour-ordered amplitudes $b_3^0(1,i,j,k,2)$. We define the following squared matrix elements:
\ba
B_3^0(1,i,j,k,2) &=& \big|b_3^0(1,i,j,k,2)\big|^2, \nn\\
\t{B}_3^0(1,i,j,k,2) &=& \big|b_3^0(1,i,j,k,2) + b_3^0(1,j,i,k,2) + b_3^0(1,j,k,i,2)\big|^2, \nn\\
\bar{B}_3^0(1,i,j,k,2) &=& \big| \sum_{P(i,j,k)} b_3^0(1,i,j,k,2) \big|^2.
\ea
The leading colour function $B_3^0$ is a straightforward generalization of $B_2^0$ with an additional gluon. The other functions, which will appear at subleading colour levels, contain one or more abelian gluons. The function $\t{B}_3^0(1,i,j,k,2)$ has the colour structure $(1,i,2) \otimes (1,j,k,2)$, so it is understood as a matrix element with two non-abelian gluons $j,k$ and one abelian gluon $i$, indicated by the tilde above the function. By convention, the abelian gluons are placed before the non-abelian gluons in the argument list. When all the gluons in a matrix element are abelian, this will be indicated by a bar over the function. So $\bar{B}_3^0(1,i,j,k,2)$ is a matrix element where all three gluons are abelian and thus a generalization of $\bar{B}_2^0$. The full squared five-parton B-type matrix element can then be written as
\ba \label{eq:FullB3g0}
\big|\mathcal{B}_3^0\big|^{2} &=& g_s^6 (N_c^2-1)N_c^2 \Big[ \bs{B}_3^0-\frac{1}{N_c^2} \l( \t{\bs{B}}_3^0 - \bar{\bs{B}}_3^0 \r) + \frac{1}{N_c^4} \bar{\bs{B}}_3^0 \Big],
\ea
with
\ba
\bs{B}_{3}^{0}&=&\sum_{P(i,j,k)}B_{3}^{0}(1,i,j,k,2),\nn\\
\t{\bs{B}}_{3}^{0}&=&\sum_{P(i,j,k)}\t{B}_{3}^{0}(1,i,j,k,2),\nn\\
\bar{\bs{B}}_{3}^{0}&=&\bar{B}_{3}^{0}(1,i,j,k,2).
\ea
Note that the matrix element with all abelian gluons will always appear at the most subleading colour level, but can also contribute at the intermediate subleading colour levels.

\subsubsection{Six-parton tree-level} \label{sec:B4g0ME}
The double real B-type matrix element~\cite{Parke:1985fm} containing one quark pair and four gluons is given by
\ba \label{eq:FullB4g0}
\big|\mathcal{B}_{4}^{0}\big|^{2}=g_s^8 (N_c^2-1) N_c^{3}\Big[ \bs{B}_{4}^{0}-\frac{1}{N_c^2}\Big(\t{\bs{B}}_{4}^{0}-\bar{\bs{B}}_{4}^{0}+\bs{R}_{4}^{0}\Big)+\frac{1}{N_c^4}\Big(\t{\t{\bs{B}}}_{4}^{0}-3\bar{\bs{B}}_{4}^{0}\Big)-\frac{1}{N_c^6}\bar{\bs{B}}_{4}^{0}\Big], \nn\\
\ea
where
\ba \label{eq:6partBtypeFuncs}
\bs{B}_{4}^{0}&=&\sum_{P(i,j,k,l)}~B_{4}^{0}(1,i,j,k,l,2),\nn\\
\t{\bs{B}}_{4}^{0}&=&\sum_{P(i,j,k,l)}~\t{B}_{4}^{0}(1,i,j,k,l,2),\nn\\
\t{\t{\bs{B}}}_{4}^{0}&=&\sum_{P(i,j,k,l)/Z_{2}}\t{\t{B}}_{4}^{0}(1,i,j,k,l,2),\nn\\
\bar{\bs{B}}_{4}^{0}&=&\bar{B}_{4}^{0}(1,i,j,k,l,2),\nn\\
\bs{R}_{4}^{0}&=&\sum_{P(i,j,k,l)}~R_{4}^{0}(1,i,j,k,l,2).
\ea
Here $B_4^0, \t{B}_4^0, \t{\t{B}}_4^0$ and $\bar{B}_4^0$ are defined analogously as for the lower multiplicity B-type matrix elements, with the number of tildes above the function specifying the number of abelian gluons and the bar specifying that all gluons are abelian. In the definition of $\t{\t{\bs{B}}}_{4}^{0}$, the sum goes over all permutations of $i,j,k,l$ modulo the interchanges of the two abelian gluons. A new feature in the six-parton B-type matrix element is observed. It is no longer possible to group all the terms in terms of these squared (abelian gluonic) functions. Remainder interference terms remain which we group together in the R-term $\bs{R}_4^0$. Each single ordering of the R-term is an interference of eight amplitudes:
\ba \label{eq:soR4g0}
R_{4}^{0}(1,i,j,k,l,2)&=&\mathrm{Re}\big[ b_{4}^{0 \dagger}(1,i,j,k,l,2)\big( b_{4}^{0}(1,j,i,l,k,2)
+b_{4}^{0}(1,j,l,i,k,2)\nn\\
&&+b_{4}^{0}(1,j,l,k,i,2)
+b_{4}^{0}(1,k,i,l,j,2)
+b_{4}^{0}(1,k,j,l,i,2)\nn\\
&&+b_{4}^{0}(1,l,i,k,j,2)
+b_{4}^{0}(1,l,j,i,k,2)
+b_{4}^{0}(1,l,k,j,i,2)\big) \big].
\ea
These remainder interference terms also occur for $n > 6$ B-type matrix elements, with the consequence that it is no longer straightforward to identify the unresolved limits by just examining  the colour connections, which render the construction of the subtraction term more involved.

\subsubsection{Four-parton one-loop} \label{sec:B2g1Funcs}
For the following loop matrix elements, tildes and bars above the squared matrix elements will merely signify the degree in subleading colour, and not necessarily the number of abelian gluons, as was the convention for the tree-level matrix elements. Likewise, the number of hats above a functions signifies the power of its colour factor $n_f$. The full four-parton one-loop B-type matrix element~\cite{Ellis:1985er} is then written as
\ba
M_4^1(q,g_1,g_2,\qb) &=& \l(\frac{\al_s}{2\pi}\r) g_s^4 (N_c^2-1) N_c^2 \Big[ B_2^1(q,g_1,g_2,\qb) + B_2^1(q,g_2,g_1,\qb) \nn\\
&+& \frac{n_f}{N_c} \l( \hat{B}_2^1(q,g_1,g_2,\qb) + \hat{B}_2^1(q,g_2,g_1,\qb) \r) + \frac{1}{N_c^2} \t{B}_2^1(q,g_1,g_2,\qb) \nn\\
&+& \frac{n_f}{N_c^3} \hat{\t{B}}_2^1(q,g_1,g_2,\qb) + \frac{1}{N_c^4} \t{\t{B}}_2^1(q,g_1,g_2,\qb) \Big],
\ea
where each function appearing at the different colour levels is build up from the interferences of several tree-level amplitudes and one-loop amplitudes in \eq{Bng1primamp}:
\ba
B_2^1(q,g_1,g_2,\qb) &=& 2\, \mathrm{Re} \big[ b_2^{1\dagger}(q,g_1,g_2,\qb) \, b_2^0(q,g_1,g_2,\qb) \big], \nn\\
\hat{B}_2^1(q,g_1,g_2,\qb) &=& 2\, \mathrm{Re} \big[ \hat{b}_2^{1\dagger}(q,g_1,g_2,\qb) \, b_2^0(q,g_1,g_2,\qb) \big], \nn\\
\t{B}_2^1(\si(\{p\})) &=& \t{B}_{2,b}^1(\si(\{p\})) + \t{B}_{2,c}^1(\si(\{p\})) + \t{B}_{2,d}^1(\si(\{p\})) + \t{B}_{2,e}^1(\si(\{p\})) \nn\\
&+& \t{B}_{2,f}^1(\si(\{p\})), \nn\\ 
\hat{\t{B}}_2^1(q,g_1,g_2,\qb) &=& 2\, \mathrm{Re} \big[ \big( \hat{b}_2^{1\dagger}(q,g_1,g_2,\qb)+\hat{b}_2^{1\dagger}(q,g_2,g_1,\qb) \big) \big( b_2^0(q,g_1,g_2,\qb)+b_2^0(q,g_2,g_1,\qb) \big) \big], \nn\\
\t{\t{B}}_2^1(q,g_1,g_2,\qb) &=& 2\, \mathrm{Re} \big[ \big( b_2^{1\dagger}(q,g_1,g_2,\qb)+b_2^{1\dagger}(q,g_2,g_1,\qb) \big) \big( b_2^0(q,g_1,g_2,\qb)+b_2^0(q,g_2,g_1,\qb) \big) \big], \nn\\
\ea
where the first SLC matrix element $\t{B}_2^1$ has been partitioned into several subfunctions coming from the different contractions of the tree-level and one-loop colour structures:  
\ba
\t{B}_{2,b}^1(q,g_1,g_2,\qb) &=& 2\, \mathrm{Re} \big[ \mathcal{B}_{2;3}^{1\dagger}(q,g_1,g_2,\qb) \, b_2^0(q,g_1,g_2,\qb) \big] \big|_{\mathrm{SLC}}, \nn\\
\t{B}_{2,c}^1(q,g_1,g_2,\qb) &=& 2\, \mathrm{Re} \big[ \mathcal{B}_{2;3}^{1\dagger}(q,g_1,g_2,\qb) \, b_2^0(q,g_2,g_1,\qb) \big] \big|_{\mathrm{SLC}}, \nn\\
\t{B}_{2,d}^1(q,g_1,g_2,\qb) &=& 2\, \mathrm{Re} \big[ \mathcal{B}_{2;1}^{1\dagger}(q,g_1,g_2,\qb) \, b_2^0(q,g_1,g_2,\qb) \big] \big|_{\mathrm{SLC}}, \nn\\
\t{B}_{2,e}^1(q,g_1,g_2,\qb) &=& 2\, \mathrm{Re} \big[ \mathcal{B}_{2;1}^{1\dagger}(q,g_1,g_2,\qb) \, b_2^0(q,g_2,g_1,\qb) + \mathcal{B}_{2;1}^{1\dagger}(q,g_2,g_1,\qb) b_2^0(q,g_1,g_2,\qb)\big] \big|_{\mathrm{SLC}}, \nn\\
\t{B}_{2,f}^1(q,g_1,g_2,\qb) &=& 2\, \mathrm{Re} \big[ \mathcal{B}_{2;1}^{1\dagger}(q,g_2,g_1,\qb) \, b_2^0(q,g_2,g_1,\qb) \big] \big|_{\mathrm{SLC}}, 
\ea
where $\big|_{\mathrm{SLC}}$ means the operation of extracting the coefficient of the subleading $1/N_c^2$ contribution. The singular structure of the leading $N_c$ function $B_2^1(1,i,j,2)$ for a given ordering is
\be
\mathrm{Poles}\big[B_2^1(1,i,j,2)\big] = 2\, \big[ I_{qg}^{(1)}(\ep,s_{1i}) + I_{qg}^{(1)}(\ep,s_{2j}) + I_{gg}^{(1)}(\ep,s_{ij})\big] B_2^0(1,i,j,2),
\ee
and similar for the leading $n_f$ function $\hat{B}_2^1(1,i,j,2)$,
\be
\mathrm{Poles}\big[\hat{B}_2^1(1,i,j,2)\big] = 2\, \big[ I_{qg,F}^{(1)}(\ep,s_{1i}) + I_{qg,F}^{(1)}(\ep,s_{2j}) + I_{gg,F}^{(1)}(\ep,s_{ij})\big] B_2^0(1,i,j,2).
\ee
The subfunctions at the first subleading $N_c$ colour level have lengthy pole expressions:
\ba
\mathrm{Poles}\big[\t{B}_{2,b}^1(1,i,j,2)\big] &=& 
\big[I_{q\qb}^{(1)}(\ep,s_{12}) + I_{gg}^{(1)}(\ep,s_{ij}) + I_{qg}^{(1)}(\ep,s_{1i}) \nn\\
&-& 2 I_{qg}^{(1)}(\ep,s_{1j}) - 2 I_{qg}^{(1)}(\ep,s_{2i}) + I_{qg}^{(1)}(\ep,s_{2j})\big] B_2^0(1,i,j,2) \nn\\
&+& \big[- I_{q\qb}^{(1)}(\ep,s_{12}) - I_{gg}^{(1)}(\ep,s_{ij}) + I_{qg}^{(1)}(\ep,s_{1i}) + I_{qg}^{(1)}(\ep,s_{2j})\big] B_2^0(1,j,i,2) \nn\\
&+& \big[ I_{q\qb}^{(1)}(\ep,s_{12}) + I_{gg}^{(1)}(\ep,s_{ij}) - I_{qg}^{(1)}(\ep,s_{1i}) - I_{qg}^{(1)}(\ep,s_{2j})\big] \bar{B}_2^0(1,i,j,2), \nn\\
\mathrm{Poles}\big[\t{B}_{2,c}^1(1,i,j,2)\big] &=&
\big[- I_{q\qb}^{(1)}(\ep,s_{12}) - I_{gg}^{(1)}(\ep,s_{ij}) + I_{qg}^{(1)}(\ep,s_{1j}) + I_{qg}^{(1)}(\ep,s_{2i})\big] B_2^0(1,i,j,2) \nn\\
&+& \big[I_{q\qb}^{(1)}(\ep,s_{12}) + I_{gg}^{(1)}(\ep,s_{ij}) - 2 I_{qg}^{(1)}(\ep,s_{1i}) \nn\\
&+& I_{qg}^{(1)}(\ep,s_{1j}) + I_{qg}^{(1)}(\ep,s_{2i}) - 2 I_{qg}^{(1)}(\ep,s_{2j})\big] B_2^0(1,j,i,2) \nn\\
&+& \big[I_{q\qb}^{(1)}(\ep,s_{12}) + I_{gg}^{(1)}(\ep,s_{ij}) - I_{qg}^{(1)}(\ep,s_{1j}) - I_{qg}^{(1)}(\ep,s_{2i})\big] \bar{B}_2^0(1,i,j,2), \nn\\
\mathrm{Poles}\big[\t{B}_{2,d}^1(1,i,j,2)\big] &=& 2\,
\big[- I_{q\qb}^{(1)}(\ep,s_{12}) - I_{gg}^{(1)}(\ep,s_{ij}) \nn\\
&-& I_{qg}^{(1)}(\ep,s_{1i}) - I_{qg}^{(1)}(\ep,s_{2j})\big] B_2^0(1,i,j,2), \nn\\
\mathrm{Poles}\big[\t{B}_{2,e}^1(1,i,j,2)\big] &=& 
\big[2 I_{gg}^{(1)}(\ep,s_{ij}) + I_{qg}^{(1)}(\ep,s_{1i}) \nn\\
&+& I_{qg}^{(1)}(\ep,s_{1j}) + I_{qg}^{(1)}(\ep,s_{2i}) + I_{qg}^{(1)}(\ep,s_{2j})\big] B_2^0(1,i,j,2) \nn\\
&+& \big[2 I_{gg}^{(1)}(\ep,s_{ij}) + I_{qg}^{(1)}(\ep,s_{1i}) \nn\\
&+& I_{qg}^{(1)}(\ep,s_{1j}) + I_{qg}^{(1)}(\ep,s_{2i}) + I_{qg}^{(1)}(\ep,s_{2j})\big] B_2^0(1,j,i,2) \nn\\
&+& \big[- 2 I_{gg}^{(1)}(\ep,s_{ij}) - I_{qg}^{(1)}(\ep,s_{1i}) \nn\\
&-& I_{qg}^{(1)}(\ep,s_{1j}) - I_{qg}^{(1)}(\ep,s_{2i}) - I_{qg}^{(1)}(\ep,s_{2j})\big] \bar{B}_2^0(1,i,j,2), \nn\\
\mathrm{Poles}\big[\t{B}_{2,f}^1(1,i,j,2)\big] &=& 2\,
\big[- I_{q\qb}^{(1)}(\ep,s_{12}) - I_{gg}^{(1)}(\ep,s_{ij}) \nn\\
&-& I_{qg}^{(1)}(\ep,s_{1j}) - I_{qg}^{(1)}(\ep,s_{2i})\big] B_2^0(1,j,i,2),
\ea
whereas their total yields a relatively compact expression, with many poles cancelling against each other:
\ba
\lefteqn{\mathrm{Poles}\big[\t{B}_2^1(1,i,j,2)\big] = 
	- 2 I_{q\qb}^{(1)}(\ep,s_{12}) B_2^0(1,i,j,2) 
	- 2 I_{q\qb}^{(1)}(\ep,s_{12}) B_2^0(1,j,i,2)} \nn\\
&+& 2\, \big[ I_{q\qb}^{(1)}(\ep,s_{12}) - I_{qg}^{(1)}(\ep,s_{1i}) - I_{qg}^{(1)}(\ep,s_{1j}) - I_{qg}^{(1)}(\ep,s_{2i}) - I_{qg}^{(1)}(\ep,s_{2j})\big] \bar{B}_2^0(1,i,j,2). \nn\\
\ea
The subleading $n_f$ function $\hat{\t{B}}_2^1(1,i,j,2)$ has the pole structure
\ba
\mathrm{Poles}\big[\hat{\t{B}}_2^1(1,i,j,2)\big] &=& 2\, \big[- I_{qg,F}^{(1)}(\ep,s_{1i}) - I_{qg,F}^{(1)}(\ep,s_{1j}) \nn\\
&-& I_{qg,F}^{(1)}(\ep,s_{2i}) - I_{qg,F}^{(1)}(\ep,s_{2j})\big] \bar{B}_2^0(1,i,j,2),
\ea
and lastly, the pole structure of the subsubleading $N_c$ function $\t{\t{B}}_2^1(1,i,j,2)$ is given by
\be
\mathrm{Poles}\big[\t{\t{B}}_2^1(1,i,j,2)\big] = 2 I_{q\qb}^{(1)}(\ep,s_{12}) \bar{B}_2^0(1,i,j,2).
\ee

\subsubsection{Five-parton one-loop}
The full five-parton B-type squared matrix element~\cite{Bern:1994fz} is obtained by interfering the full one-loop amplitude \eq{FullB3g1amp} with its tree-level counterpart. It is decomposed into colour levels as follows:
\ba\label{eq:B51}
M_5^1(q,g_1,g_2,g_3,\qb) &=& \l(\frac{\al_s}{2\pi}\r) g_s^6 (N_c^2-1) N_c^3 \Big[ B_3^1(q,g_1,g_2,g_3,\qb) + \frac{n_f}{N_c} \hat{B}_3^1(q,g_1,g_2,g_3,\qb) \nn\\
&+& \frac{1}{N_c^2} \t{B}_3^1(q,g_1,g_2,g_3,\qb) + \frac{n_f}{N_c^3} \hat{\t{B}}_3^1(q,g_1,g_2,g_3,\qb) + \frac{1}{N_c^4} \t{\t{B}}_3^1(q,g_1,g_2,g_3,\qb) \nn\\
&+& \frac{n_f}{N_c^5} \hat{\t{\t{B}}}_3^1(q,g_1,g_2,g_3,\qb) + \frac{1}{N_c^6} \bar{B}_3^1(q,g_1,g_2,g_3,\qb) \Big]. 
\ea
The contributions at the leading $N_c$ and $n_f$ level simply consist of sums of coherent squares,
\ba
B_3^1(q,g_1,g_2,g_3,\qb) &=& \sum_{\si \in S_3} \mathrm{Re}\big[ b_3^{1\dagger}(q,g_{\si(1)},g_{\si(2)},g_{\si(3)},\qb) \, b_3^0(q,g_{\si(1)},g_{\si(2)},g_{\si(3)},\qb) \big], \nn\\
\hat{B}_3^1(q,g_1,g_2,g_3,\qb) &=& \sum_{\si \in S_3} \mathrm{Re}\big[ \hat{b}_3^{1\dagger}(q,g_{\si(1)},g_{\si(2)},g_{\si(3)},\qb) \, b_3^0(q,g_{\si(1)},g_{\si(2)},g_{\si(3)},\qb) \big], \nn\\
\ea
while the structure of the subleading colour contributions are more involved. The infrared pole structure of the functions at the different colour levels derives from the infrared $\ep$-pole structure of the underlying one-loop amplitudes that was described in \sec{Btype1LoopPoles} above.

\subsubsection{Four-parton two-loop}
The two-loop two-quark two-gluon matrix element yields the double virtual B-type contribution.  It consists of the two-loop amplitude interfered with the tree-level amplitude and the genuine square of the one-loop amplitude. Its colour level decomposition is given as~\cite{Anastasiou:2001sv,Anastasiou:2002zn,Bern:2003ck}
\ba
M_4^2(q,g_1,g_2,\qb) &=& \l(\frac{\al_s}{2\pi}\r)^2 g_s^4 (N_c^2-1) N_c^3 \Big[ B_2^2(q,g_1,g_2,\qb) + B_2^2(q,g_2,g_1,\qb) \nn\\
&+& \frac{n_f}{N_c} \l( \hat{B}_2^2(q,g_1,g_2,\qb) + \hat{B}_2^2(q,g_2,g_1,\qb) \r) \nn\\
&+& \frac{n_f^2}{N_c^2} \l( \hat{\hat{B}}_2^2(q,g_1,g_2,\qb) + \hat{\hat{B}}_2^2(q,g_2,g_1,\qb) \r) \nn\\
&+& \frac{1}{N_c^2} \t{B}_2^2(q,g_1,g_2,\qb) + \frac{n_f}{N_c^3} \hat{\t{B}}_2^2(q,g_1,g_2,\qb) + \frac{n_f^2}{N_c^4} \hat{\hat{\t{B}}}_2^2(q,g_1,g_2,\qb) \nn\\
&+& \frac{1}{N_c^4} \t{\t{B}}_2^2(q,g_1,g_2,\qb) + \frac{n_f}{N_c^5} \hat{\t{\t{B}}}_2^2(q,g_1,g_2,\qb) + \frac{1}{N_c^6} \t{\t{\t{B}}}_2^2(q,g_1,g_2,\qb) \Big]. \nn\\
\ea
Its $\ep$-pole structure is documented in~\cite{Anastasiou:2001sv,Bern:2003ck} and cancels with the integrated subtraction terms at the VV level. 

\subsection{C-type matrix elements}
We call matrix elements containing two quark pairs of non-identical flavour plus any number of gluons C-type matrix elements. We use the notation $C_n^l$ for a C-type matrix element containing $n$ gluons and $l$ loops.

\subsubsection{Amplitudes} \label{sec:CtypeAmps}
The full amplitude for two non-identical flavour quark pairs at tree-level is given by
\ba \label{eq:FullC0g0amp}
\mathcal{C}_0^0(q,\qb,Q,\Qb) = g_s^2 \Big( \de_{q\Qb}\de_{Q\qb} - \frac{1}{N_c} \de_{q\qb}\de_{Q\Qb} \Big) c_0^0(q,\Qb,Q,\qb),
\ea
where there are only the two colour structures of the quark lines: $\de_{q\Qb}\de_{Q\qb}$ for when the quarks of different flavour are connected and $\de_{q\qb}\de_{Q\Qb}$ for when quarks of the same flavour are connected, which is subleading by a factor of $1/N_c$. For the four-parton tree-level C-type amplitude, the partial amplitude $c_0^0$ multiplying the two colour structures is the same, and can thus be factorized. This is no longer the case for the one-loop amplitude
\ba \label{eq:FullC0g1amp}
\mathcal{C}_0^1(q,\qb,Q,\Qb) = g_s^2 \l(\frac{\al_s}{2\pi}\r) N_c \Big( \de_{q\Qb}\de_{Q\qb} \, c_{0,l}^1(q,\Qb,Q,\qb) - \frac{1}{N_c} \de_{q\qb}\de_{Q\Qb} \, c_{0,s}^1(q,\qb,Q,\Qb) \Big),
\ea
where the leading and subleading partial amplitudes are now different and can be further colour-decomposed as
\ba \label{eq:C0g1partamp}
c_{0,l}^1(q,\Qb,Q,\qb) &=& \mathcal{C}_{0,l}^1(q,\Qb,Q,\qb) + \frac{n_f}{N_c} \hat{\mathcal{C}}_{0,l}^1(q,\Qb,Q,\qb) + \frac{1}{N_c^2} \t{\mathcal{C}}_{0,l}^1(q,\Qb,Q,\qb), \nn\\
c_{0,s}^1(q,\qb,Q,\Qb) &=& \mathcal{C}_{0,s}^1(q,\qb,Q,\Qb) + \frac{n_f}{N_c} \hat{\mathcal{C}}_{0,s}^1(q,\qb,Q,\Qb) + \frac{1}{N_c^2} \t{\mathcal{C}}_{0,s}^1(q,\qb,Q,\Qb).
\ea
The five-parton C-type amplitude has the same structure at tree $i=0$ and one-loop $i=1$ level~\cite{Kunszt:1994nq,Signer:1995a}:
\ba \label{eq:C1giamp}
\mathcal{C}_1^i(q,\qb,Q,\Qb,g) &=& g_s^3 \l(\frac{\al_s}{2\pi}\r)^i \Big[ T_{q\Qb}^{g}\de_{Q\qb} \, c_{1,a}^{i}(q,g,\Qb;Q,\qb) + T_{Q\qb}^{g}\de_{q\Qb} \, c_{1,b}^{i}(q,\Qb;Q,g,\qb) \nn\\
&-& \frac{1}{N_c} T_{q\qb}^{g}\de_{Q\Qb} \, c_{1,c}^{i}(q,g,\qb;Q,\Qb) - \frac{1}{N_c} T_{Q\Qb}^{g}\de_{q\qb} \, c_{1,d}^{i}(q,\qb;Q,g,\Qb) \Big],
\ea
where we have four partial amplitudes $c_{1,j}^i$, with $j \in \{a,b,c,d\}$ corresponding to the four possible gluon positions in between the quarks. For the tree-level amplitudes, it is also useful to define
\ba
c_{1,\ga_1}^0(q,\qb,Q,\Qb,g) &=& c_{1,a}^0(q,g,\Qb;Q,\qb) + c_{1,b}^0(q,\Qb;Q,g,\qb), \nn\\
c_{1,\ga_2}^0(q,\qb,Q,\Qb,g) &=& c_{1,c}^0(q,g,\qb;Q,\Qb) + c_{1,d}^0(q,\qb;Q,g,\Qb),
\ea 
where the following relation holds:
\ba
c_{1,\ga}(q,\qb,Q,\Qb,g) = c_{1,\ga_1}(q,\qb,Q,\Qb,g) = c_{1,\ga_2}(q,\qb,Q,\Qb,g).
\ea
The one-loop amplitudes~\cite{Kunszt:1994nq} in \eq{C1giamp} can be further colour-decomposed into the more primitive sub-amplitudes:  
\ba
c_{1,a}^{1}(q,g,\Qb;Q,\qb)&=&N_c \, \mathcal{C}_{1,a}^{1}(q,g,\Qb;Q,\qb)+n_f \, \hat{\mathcal{C}}_{1,a}^{1}(q,g,\Qb;Q,\qb)+\frac{1}{N_c}\t{\mathcal{C}}_{1,a}^{1}(q,g,\Qb;Q,\qb),\nn\\
c_{1,b}^{1}(q,\Qb;Q,g,\qb)&=&N_c \, \mathcal{C}_{1,b}^{1}(q,\Qb;Q,g,\qb)+n_f \, \hat{\mathcal{C}}_{1,b}^{1}(q,\Qb;Q,g,\qb)+\frac{1}{N_c}\t{\mathcal{C}}_{1,b}^{1}(q,\Qb;Q,g,\qb),\nn\\
c_{1,c}^{1}(q,g,\qb;Q,\Qb)&=&N_c \, \mathcal{C}_{1,c}^{1}(q,g,\qb;Q,\Qb)+n_f \, \hat{\mathcal{C}}_{1,c}^{1}(q,g,\qb;Q,\Qb)+\frac{1}{N_c}\t{\mathcal{C}}_{1,c}^{1}(q,g,\qb;Q,\Qb),\nn\\
c_{1,d}^{1}(q,\qb;Q,g,\Qb)&=&N_c \, \mathcal{C}_{1,d}^{1}(q,\qb;Q,g,\Qb)+n_f \, \hat{\mathcal{C}}_{1,d}^{1}(q,\qb;Q,g,\Qb)+\frac{1}{N_c}\t{\mathcal{C}}_{1,d}^{1}(q,\qb;Q,g,\Qb). \nn\\
\ea
Like for the tree-level amplitudes, it is also useful to define
\ba 
\mathcal{C}_{1,\ga_{1}}^{1}(q,\qb,Q,\Qb,g)&=&\mathcal{C}_{1,a}^{1}(q,g,\Qb;Q,\qb)+{\mathcal{C}}_{1,b}^{1}(q,\Qb;Q,g,\qb),\nn\\
\mathcal{C}_{1,\ga_{2}}^{1}(q,\qb,Q,\Qb,g)&=&\mathcal{C}_{1,c}^{1}(q,g,\qb;Q,\Qb)+{\mathcal{C}}_{1,d}^{1}(q,\qb;Q,g,\Qb),\nn\\
\hat{\mathcal{C}}_{1,\ga_1}^{1}(q,\qb,Q,\Qb,g)&=&\hat{\mathcal{C}}_{1,a}^{1}(q,g,\Qb;Q,\qb)+\hat{\mathcal{C}}_{1,b}^{1}(q,\Qb;Q,g,\qb),\nn\\
\hat{\mathcal{C}}_{1,\ga_{2}}^{1}(q,\qb,Q,\Qb,g)&=&\hat{\mathcal{C}}_{1,c}^{1}(q,g,\qb;Q,\Qb)+\hat{\mathcal{C}}_{1,d}^{1}(q,\qb;Q,g,\Qb),\nn\\
\t{\mathcal{C}}_{1,\ga_{1}}^{1}(q,\qb,Q,\Qb,g)&=&\t{\mathcal{C}}_{1,a}^{1}(q,g,\Qb;Q,\qb)+\t{\mathcal{C}}_{1,b}^{1}(q,\Qb;Q,g,\qb),\nn\\
\t{\mathcal{C}}_{1,\ga_{2}}^{1}(q,\qb,Q,\Qb,g)&=&\t{\mathcal{C}}_{1,c}^{1}(q,g,\qb;Q,\Qb)+\t{\mathcal{C}}_{1,d}^{1}(q,\qb;Q,g,\Qb),
\ea
where only the hatted amplitude fulfills a similar identity as for the tree-level amplitude:
\ba
\hat{\mathcal{C}}_{1,\ga}^{1}(q,\qb,Q,\Qb,g)&=&\hat{\mathcal{C}}_{1,\ga_{1}}^{1}(q,\qb,Q,\Qb,g)=\hat{\mathcal{C}}_{1,\ga_{2}}^{1}(q,\qb,Q,\Qb,g).
\ea
The full expressions of the amplitudes can be found in~\cite{Kunszt:1994nq}. The six-parton C-type tree-level amplitude is given by
\ba \label{eq:C2g0amp}
&&\mathcal{C}_2^0(q,\qb,Q,\Qb,g_1,g_2) = g_s^4 \Big[ (T^{g_1} T^{g_2})_{q\Qb}\de_{Q\qb} \, c_{2,a}^0(q,g_1,g_2,\Qb;Q,\qb) \nn\\
&+& (T^{g_1} T^{g_2})_{Q\qb} \de_{q\Qb} \, c_{2,b}^0(q,\Qb;Q,g_1,g_2,\qb) + T^{g_1}_{q\Qb} T^{g_2}_{Q\qb} \, c_{2,c}^0(q,g_1,\Qb;Q,g_2,\qb) \nn\\
&+& \frac{1}{N_c} (T^{g_1} T^{g_2})_{q\qb}\de_{Q\Qb} \, \t{c}_{2,a}^0(q,g_1,g_2,\qb;Q,\Qb) + \frac{1}{N_c} (T^{g_1} T^{g_2})_{Q\Qb}\de_{q\qb} \, \t{c}_{2,b}^0(q,\qb;Q,g_1,g_2,\Qb) \nn\\
&+& \frac{1}{N_c} T^{g_1}_{q\qb} T^{g_2}_{Q\Qb} \, \t{c}_{2,c}^0(q,g_1,\qb;Q,g_2,\Qb) + (g_1 \leftrightarrow g_2) \Big],
\ea
where the six colour structures plus the additional six from interchanging $g_1$ and $g_2$ represent the different possibilities of the gluon placements in between the quark lines.  

\subsubsection{Pole structure of the one-loop amplitudes} \label{sec:Cpoles}
The pole structures of the five-parton one-loop partial amplitudes are given by~\cite{Kunszt:1994nq}
\ba
&&\mathrm{Poles}\big[c_{1,a}^1\big] = \nn\\
&& N_c \big[ I_{qg}^{(1)}(\ep,s_{\Qb g}) + I_{q\qb}^{(1)}(\ep,s_{\qb Q}) + I_{qg}^{(1)}(\ep,s_{qg}) \big] c_{1,a}^0(q,g,\Qb;Q,\qb) \nn\\
&-& \frac{1}{N_c} \big[ - I_{q\qb}^{(1)}(\ep,s_{\qb\Qb}) + I_{q\qb}^{(1)}(\ep,s_{q\Qb}) + I_{q\qb}^{(1)}(\ep,s_{Q\Qb}) 
\nn \\ &&+ I_{q\qb}^{(1)}(\ep,s_{q\qb}) + I_{q\qb}^{(1)}(\ep,s_{\qb Q}) - I_{q\qb}^{(1)}(\ep,s_{qQ})\big] c_{1,a}^0(q,g,\Qb;Q,\qb) \nn\\
&-& \frac{1}{N_c} \big[ - I_{q\qb}^{(1)}(\ep,s_{\qb\Qb}) + I_{qg}^{(1)}(\ep,s_{\Qb g}) + I_{q\qb}^{(1)}(\ep,s_{\qb Q}) - I_{qg}^{(1)}(\ep,s_{Qg}) \big] c_{1,c}^0(q,g,\qb;Q,\Qb) \nn\\
&-& \frac{1}{N_c} \big[ I_{q\qb}^{(1)}(\ep,s_{\qb Q}) - I_{qg}^{(1)}(\ep,s_{\qb g}) - I_{q\qb}^{(1)}(\ep,s_{qQ}) + I_{qg}^{(1)}(\ep,s_{qg}) \big] c_{1,d}^0(Q,g,\Qb;q,\qb), 
\ea
and
\ba
&&\mathrm{Poles}\big[c_{1,c}^1\big] = \nn\\
&& N_c \big[ I_{q\qb}^{(1)}(\ep,s_{Q\Qb}) + I_{qg}^{(1)}(\ep,s_{\qb g}) + I_{qg}^{(1)}(\ep,s_{qg}) \big] c_{1,c}^0(q,g,\qb;Q,\Qb) \nn\\
&-& \frac{1}{N_c} \big[ - I_{q\qb}^{(1)}(\ep,s_{\qb\Qb}) + I_{q\qb}^{(1)}(\ep,s_{q\Qb}) + I_{q\qb}^{(1)}(\ep,s_{Q\Qb})
\nn \\ && + I_{q\qb}^{(1)}(\ep,s_{q\qb}) + I_{q\qb}^{(1)}(\ep,s_{\qb Q}) - I_{q\qb}^{(1)}(\ep,s_{qQ})\big] c_{1,c}^0(q,g,\qb;Q,\Qb) \nn\\
&+& N_c \big[ I_{q\qb}^{(1)}(\ep,s_{\qb\Qb}) - I_{q\qb}^{(1)}(\ep,s_{Q\Qb}) - I_{qg}^{(1)}(\ep,s_{\qb g}) + I_{qg}^{(1)}(\ep,s_{Qg}) \big] c_{1,a}^0(q,g,\Qb;Q,\qb) \nn\\
&+& N_c \big[ - I_{q\qb}^{(1)}(\ep,s_{Q\Qb}) + I_{qg}^{(1)}(\ep,s_{\Qb g}) + I_{q\qb}^{(1)}(\ep,s_{qQ}) - I_{qg}^{(1)}(\ep,s_{qg}) \big] c_{1,b}^0(Q,g,\qb;q,\Qb). \nn\\
\ea
The pole structures of the remaining two partial amplitudes can be obtained by permuting the quark labels:
\ba
\mathrm{Poles}\big[c_{1,b}^1\big] = \mathrm{Poles}\big[c_{1,a}^1\big] \big|_{q \leftrightarrow Q, \qb \leftrightarrow \Qb}, \quad \mathrm{Poles}\big[c_{1,d}^1\big] = \mathrm{Poles}\big[c_{1,c}^1\big] \big|_{q \leftrightarrow Q, \qb \leftrightarrow \Qb}.
\ea

\subsubsection{Four-parton tree-level}
Squaring \eq{FullC0g0amp}, one obtains the full four-parton C-type matrix element
\be
\big|\mathcal{C}_0^0\big|^2 = g_s^4 (N_c^2-1) C_0^0(q,\Qb,Q,\qb), 
\ee
with $C_0^0 = |c_0^0|^2$.  

\subsubsection{Five-parton tree-level}
Squaring \eq{C1giamp} with $i=0$, we obtain the two quark-pairs of different flavour with one external gluon matrix element
\ba
\big|\mathcal{C}_1^0(q,\qb,Q,\Qb,g)\big|^{2} &=& g_s^6 (N_c^2-1) N_c \Big[ C_{1,a}^0(q,g,\Qb,Q,\qb) + C_{1,b}^0(q,\Qb,Q,g,\qb) \nn\\
&+&\frac{1}{N_c^2} \big( C_{1,c}^0(q,g,\qb,Q,\Qb) + C_{1,d}^0(q,\qb,Q,g,\Qb) - 2C_{1,\ga}^0(q,\qb,Q,\Qb,g)\big) \Big], \nn\\
\ea
where $C_{1,j}^0 = |c_{1,j}^0|^2$ for $j \in \{a,b,c,d,\ga\}$. For the construction of the subtraction terms, it is convenient to define the general interference of the five-parton C-type amplitudes. We define all six possibilities as
\ba
C_{1,ij}^{0}(q,\qb,Q,\Qb,g) = 2\, \mathrm{Re} \big[ c_{1,i}^0 \, c_{1,j}^{0\dagger} \big] \quad \mbox{with } i,j \in \{a,b,c,d\}.
\ea

\subsubsection{Six-parton tree-level}
The six-parton tree-level squared matrix element~\cite{Gunion:1985bp} is decomposed into colour levels as follows: 
\ba
\big|\mathcal{C}_2^0\big|^2 &=& g_s^8 (N_c^2-1) N_c^2 \Big[ C_2^0(q,\qb,Q,\Qb,g_1,g_2) + \frac{1}{N_c^2} \t{C}_2^0(q,\qb,Q,\Qb,g_1,g_2) \nn\\
&+& \frac{1}{N_c^4} \t{\t{C}}_2^0(q,\qb,Q,\Qb,g_1,g_2) \Big],
\ea
where the leading colour matrix element $C_2^0$ consists of coherent squares of the leading colour partial amplitudes in \eq{C2g0amp}, while the subleading matrix elements $\t{C}_2^0$ and $\t{\t{C}}_2^0$ receive contributions from interferences of both leading and subleading amplitudes.

\subsubsection{Four-parton one-loop} \label{sec:C0g1}
The full four-parton one-loop C-type matrix element is given by~\cite{Ellis:1985er}
\be
M_4^1(q,\Qb,Q,\qb) = \left(\frac{\al_s}{2\pi}\right) g_s^4 (N_c^2-1) N_c \Big[ C_0^1(q,\Qb,Q,\qb) + \frac{n_f}{N_c} \hat{C}_0^1(q,\Qb,Q,\qb) + \frac{1}{N_c^2} \t{C}_0^1(q,\Qb,Q,\qb) \Big],
\ee
where
\ba
C_0^1(q,\Qb,Q,\qb) &=& 2\, \mathrm{Re}\big[\mathcal{C}_{0,l}^{1\dagger}(q,\Qb,Q,\qb) \, c_0^0(q,\Qb,Q,\qb)\big], \nn\\
\hat{C}_0^1(q,\Qb,Q,\qb) &=& 2\, \mathrm{Re}\big[\hat{\mathcal{C}}_{0,l}^{1\dagger}(q,\Qb,Q,\qb) \, c_0^0(q,\Qb,Q,\qb)\big], \nn\\
\t{C}_0^1(q,\Qb,Q,\qb) &=& 2\, \mathrm{Re}\big[\t{\mathcal{C}}_{0,l}^{1\dagger}(q,\Qb,Q,\qb) \, c_0^0(q,\Qb,Q,\qb)\big].
\ea
Note that only the leading colour amplitude $c_{0,l}^1$ contributes at the squared matrix element with the subleading $c_{0,s}^1$ amplitude completely absent as it cancels out in the sum. Using the notation $(1,j,i,2) = (q,\Qb,Q,\qb)$, the singular structure of $C_0^1(1,j,i,2)$ is given by
\be 
\mathrm{Poles}[C_0^1(1,j,i,2)] = 2\, \big[ I_{q\qb}^{(1)}(\ep,s_{1j}) + I_{q\qb}^{(1)}(\ep,s_{2i})\big] C_0^0(1,j,i,2),
\ee
while the leading $n_f$ function $\hat{C}_0^1(1,j,i,2)$ displays no $\ep$-poles,
\be
\mathrm{Poles}\big[\hat{C}_0^1(1,j,i,2)\big] = 0.
\ee
At subleading colour we have for $\t{C}_0^1(1,j,i,2)$:
\ba \label{eq:Ct0g1poles}
\mathrm{Poles}\big[\t{C}_0^1(1,j,i,2)\big] &=& \big[-2 I_{q\qb}^{(1)}(\ep,s_{12}) - 2 I_{q\qb}^{(1)}(\ep,s_{ij}) - 4 I_{q\qb}^{(1)}(\ep,s_{1j}) - 4 I_{q\qb}^{(1)}(\ep,s_{2i}) \nn\\
&+& 4 I_{q\qb}^{(1)}(\ep,s_{1i}) + 4 I_{q\qb}^{(1)}(\ep,s_{2j})\big] C_0^0(1,j,i,2).
\ea
Here we can observe the different behaviour of $C_0^1$ and $\t{C}_0^1$ under an interchange of the secondary quark pair: $Q \leftrightarrow \Qb$. The pole structure of the leading colour matrix element $C_0^1$ is fully symmetric under this $i \leftrightarrow j$ interchange, with also $C_0^0(1,j,i,2) = C_0^0(1,i,j,2)$, while for the subleading colour matrix element $\t{C}_0^1(1,j,i,2)$, we note a symmetric part in the first two terms and an asymmetric part in the last four terms. This observation plays an important role in the construction of the subtraction term for the SLC real-virtual correction discussed in \sec{CtypeRVsub}.

\subsubsection{Five-parton one-loop}
Interfering the one-loop amplitude with the tree-level amplitude \eq{C1giamp} yields the five-parton one-loop C-type matrix element
\ba
&&M_5^1(q,\qb,Q,\Qb,g) = \left(\frac{\al_s}{2\pi}\right) g_s^6 (N_c^2-1) N_c^2 \Big[ \nn\\
&+&\Big(C_{1,a}^{1}(q,g,\Qb;Q,\qb)+C_{1,b}^{1}(q,\Qb;Q,g,\qb)\Big)\nn\\
&+&\frac{n_f}{N_c}\Big(\hat{C}_{1,a}^{1}(q,g,\Qb;Q,\qb)+\hat{C}_{1,b}^{1}(q,\Qb;Q,g,\qb)\Big)\nn\\
&+&\frac{1}{N_c^2}\Big(\t{C}_{1,a}^{1}(q,g,\Qb;Q,\qb)+\t{C}_{1,b}^{1}(q,\Qb;Q,g,\qb)+C_{1,c}^{1}(q,g,\qb;Q,\Qb)\nn\\
&&+C_{1,d}^{1}(q,\qb;Q,g,\Qb)-C_{1,\ga_{1}}^{1}(q,\qb,Q,\Qb,g)-C_{1,\ga_{2}}^{1}(q,\qb,Q,\Qb,g)\Big)\nn\\
&+&\frac{n_f}{N_c^3}\Big(\hat{C}_{1,c}^{1}(q,g,\qb;Q,\Qb)+\hat{C}_{1,d}^{1}(q,\qb;Q,g,\Qb)-2\hat{C}_{1,\ga}^{1}(q,\qb,Q,\Qb,g)\Big)\nn\\
&+&\frac{1}{N_c^4}\Big(\t{C}_{1,c}^{1}(q,g,\qb;Q,\Qb)+\t{C}_{1,d}^{1}(q,\qb;Q,g,\Qb)-\t{C}_{1,\ga_{1}}^{1}(q,\qb,Q,\Qb,g)-\t{C}_{1,\ga_{2}}^{1}(q,\qb,Q,\Qb,g)\Big) \Big], \nn\\
\ea
where the functions are given by
\ba 
C_{1,i}^{1} = 2\, \mathrm{Re}\big[\mathcal{C}_{1,i}^{0 \dagger} \, \mathcal{C}_{1,i}^{1}\big], \quad \hat{C}_{1,i}^{1} = 2\, \mathrm{Re}\big[\mathcal{C}_{1,i}^{0 \dagger} \, \hat{\mathcal{C}}_{1,i}^{1}\big],\quad \t{C}_{1,i}^{1} = 2\, \mathrm{Re}\big[\mathcal{C}_{1,i}^{0 \dagger} \, \t{\mathcal{C}}_{1,i}^{1}\big],
\ea
for $i \in \{a,b,c,d,\ga,\ga_1,\ga_2\}$. The above expression clearly shows where each partial amplitude ends up when squaring the full amplitude to obtain the squared matrix element, which is helpful in the construction of the antenna subtraction terms. However, due to possible cancellations of unresolved limits or $\ep$-poles between the several functions, it is sometimes simpler to construct the subtraction term for the whole colour level at once. For this reason and for later convenience in referencing this matrix element, we collect all the functions per colour level into one single function,
\ba \label{eq:FullC1g1MEsimple}
M_5^1(q,\qb,Q,\Qb,g) &=& \l(\frac{\al_s}{2\pi}\r) g_s^6 (N_c^2-1) N_c^2 \Big[ C_1^1(q,\qb,Q,\Qb,g) + \frac{n_f}{N_c} \hat{C}_1^1(q,\qb,Q,\Qb,g) \nn\\
&+& \frac{1}{N_c^2} \t{C}_1^1(q,\qb,Q,\Qb,g) + \frac{n_f}{N_c^3} \hat{\t{C}}_1^1(q,\qb,Q,\Qb,g) + \frac{1}{N_c^4} \t{\t{C}}_1^1(q,\qb,Q,\Qb,g) \Big]. \nn\\
\ea
The infrared pole structure derives from the pole structure of the underlying one-loop amplitudes documented in \sec{Cpoles} above. 

\subsubsection{Four-parton two-loop}
The two-loop non-identical flavour four-quark matrix element yields the double virtual C-type contribution.  It consists of the two-loop amplitude interfered with the tree-level amplitude and the genuine square of the one-loop amplitude. Its colour level decomposition is given as~\cite{Anastasiou:2000kg,DeFreitas:2004kmi}
\ba
M_4^2(q,\Qb,Q,\qb) &=& \l(\frac{\al_s}{2\pi}\r)^2 g_s^4 (N_c^2-1) N_c^2 \Big[ C_0^2(q,\Qb,Q,\qb) + \frac{n_f}{N_c} \hat{C}_0^2(q,\Qb,Q,\qb) \nn\\
&+& \frac{n_f^2}{N_c^2} \hat{\hat{C}}_0^2(q,\Qb,Q,\qb)
+ \frac{1}{N_c^2} \t{C}_0^2(q,\Qb,Q,\qb) + \frac{n_f}{N_c^3} \hat{\t{C}}_0^2(q,\Qb,Q,\qb) \nn\\
&+& \frac{1}{N_c^4} \t{\t{C}}_0^2(q,\Qb,Q,\qb) \Big].
\ea
Its $\ep$-pole structure is documented in~\cite{Anastasiou:2000kg,DeFreitas:2004kmi} and cancels with the integrated subtraction terms at the VV level. 

\subsection{D-type matrix elements}
For matrix elements containing two quark pairs of identical flavour, new interferences and consequently new colour connections independent of the C-type squared matrix elements appear because of the possibility to interchange external quark states. We label these new identical-flavour-only interferences as D-type matrix elements. We use the notation $D_n^l$ for a D-type squared matrix element containing $n$ gluons and $l$ loops.

At the level of the amplitudes, no new objects are introduced since all D-type squared matrix elements are constituted from interferences of C-type amplitudes that were not permitted for the non-identical flavour case.

\subsubsection{Four-parton tree-level}
The two identical flavour quark pair amplitude is obtained from the non-identical amplitude by addition of the same amplitude where the antiquark momenta have been interchanged: $\qb \leftrightarrow \Qb$. Denoting the identical two quark pair squared tree-level matrix element by $M_{4,\mathrm{id.}}^0$ and explicitly keeping the different quark pair labels for the identical flavour quarks to see which momenta exactly are interchanged, we obtain
\ba \label{eq:C0g0ID}
M_{4,\mathrm{id.}}^0(q,\qb,q,\qb) &=& \big|\mathcal{C}_0^0(q,\Qb,Q,\qb) - \mathcal{C}_0^0(q,\qb,Q,\Qb)\big|^2 \nn\\
&=& C_0^0(q,\Qb,Q,\qb) + C_0^0(q,\qb,Q,\Qb) + \frac{1}{N_c} D_0^0(q,\qb,Q,\Qb),
\ea
where we have two C-type matrix elements, with one related by $\qb \leftrightarrow \Qb$ from the other, and the interference term
\ba
D_0^0(q,\qb,Q,\Qb) = -2\, \mathrm{Re}\big[\mathcal{C}_0^{0\dagger}(q,\Qb,Q,\qb) \, \mathcal{C}_0^0(q,\qb,Q,\Qb)\big].
\ea
These type of interferences between the C-type amplitudes defined above with C-type amplitudes with interchanged antiquark momenta are precisely what defines our D-type squared matrix elements. By definition, they have symmetry properties under interchange of $\qb \leftrightarrow \Qb$ or $q \leftrightarrow Q$. It can be seen in \eq{C0g0ID} that the D-type contributions are suppressed by a factor of $1/N_c$ with respect to the C-type contributions, resulting in all D-type matrix elements belonging to the SLC. This means that all the D-type contributions appearing as double real, real-virtual or double virtual corrections, even their leading colour level, are beyond the LC-approximation at NNLO used for dijet production from proton-proton collisions~\cite{Currie:2018xkj,Currie:2017eqf,Gehrmann-DeRidder:2019ibf}. For the higher multiplicity or loop matrix elements in the remainder of this section, we only discuss the genuine D-type matrix elements, and not the full matrix element for identical quarks, as this can be obtained by addition of the C-type matrix elements defined above. 

\subsubsection{Five-parton tree-level}
The full D-type five-parton tree-level matrix element is given by
\ba
M_4^0(q,\qb,Q,\Qb,g) &=& -2\, \mathrm{Re}\big[\mathcal{C}_1^{0\dagger}(q,\qb,Q,\Qb,g) \, \mathcal{C}_1^0(q,\Qb,Q,\qb,g)\big] \nn\\
&=& g_s^6 (N_c^2-1) \Big[ D^0_1(q,\qb,Q,\Qb,g) + D^0_1(q,\Qb,Q,\qb,g) - D^0_{1,\ga}(q,\qb,Q,\Qb,g) \nn\\
&-& \frac{1}{N_c^2} D^0_{1,\ga}(q,\qb,Q,\Qb,g) \Big],
\ea
where the D-type subfunctions are defined by
\ba \label{eq:D-ME}
D^0_1(q,\qb,Q,\Qb,g) &=& 2\, \mathrm{Re} \big[ {\mathcal{C}}_{1,a}^0(q,g,\qb,Q,\Qb) \, {\mathcal{C}}_{1,c}^{0\dagger}(q,g,\qb,Q,\Qb) \nn\\
&+& \mathcal{C}_{1,b}^0(q,\qb,Q,g,\Qb) \, \mathcal{C}_{1,d}^{0\dagger}(q,\qb,Q,g,\Qb)\big], \nn\\
D^0_{1,\ga}(q,\qb,Q,\Qb,g) &=& 2\, \mathrm{Re} \big[ \l( {\mathcal{C}}_{1,a}^0(q,g,\Qb,Q,\qb) + {\mathcal{C}}_{1,b}^0(q,\Qb,Q,g,\qb)\r) \nn\\
&\times& \l( {\mathcal{C}}_{1,c}^0(q,g,\Qb,Q,\qb) + {\mathcal{C}}_{1,d}^0(q,\Qb,Q,g,\qb)\r)^{\dagger} \big].
\ea
Novel five-parton interference terms appear as the reduced matrix element in the NLO subtraction, so we conveniently define
\ba \label{eq:Dtypeint}
D_{1,i}^{0}(q,\qb,Q,\Qb,g) &=& 2\, \mathrm{Re} \big[ \mathcal{C}_{1,i}^0(q,g,\Qb,Q,\qb) \, \mathcal{C}_{1,i}^{0\dagger}(q,g,\qb,Q,\Qb) \big] \quad \mbox{with }
i \in \{a,b,c,d\}, \nn\\
D_{1,ij}^{0}(q,\qb,Q,\Qb,g) &=& 2\, \mathrm{Re} \big[ \mathcal{C}_{1,i}^0(q,g,\Qb,Q,\qb) \, \mathcal{C}_{1,j}^{0\dagger}(q,\qb,Q,g,\Qb) \big] \quad \mbox{with }
i,j \in \{a,b,c,d\}. \nn\\
\ea
The D-type matrix elements in \eq{D-ME} can also be expressed in terms of the interference terms
\ba \label{D-ME-byint}
D^0_1(q,\qb,Q,\Qb,g) &=& D_{1,ca}^{0}(q,\qb,Q,\Qb,g) + D_{1,db}^{0}(q,\qb,Q,\Qb,g), \nn\\
D^0_1(q,\Qb,Q,\qb,g) &=& D_{1,ac}^{0}(q,\qb,Q,\Qb,g) + D_{1,bd}^{0}(q,\qb,Q,\Qb,g), \nn\\
D^0_{1,\ga}(q,\qb,Q,\Qb,g) &=& D_{1,a}^{0}(q,\qb,Q,\Qb,g) + D_{1,b}^{0}(q,\qb,Q,\Qb,g)\nn\\
&+& D_{1,ab}^{0}(q,\qb,Q,\Qb,g) + D_{1,ba}^{0}(q,\qb,Q,\Qb,g).
\ea

\subsubsection{Six-parton tree-level}
Similar to the C-type, the D-type six-parton matrix element is decomposed into the following colour levels:
\ba \label{eq:D2g0full}
M_6^0(q,\qb,Q,\Qb,g_1,g_2) &=& -2\, \mathrm{Re}\big[\mathcal{C}_1^0(q,\qb,Q,\Qb,g_1,g_2)^{\dagger} \, \mathcal{C}_1^0(q,\Qb,Q,\qb,g_1,g_2)\big] \nn\\
&=& g_s^8 (N_c^2-1) N_c \Big[ D_2^0(q,\qb,Q,\Qb,g_1,g_2) \nn\\
&+& \frac{1}{N_c^2} \t{D}_2^0(q,\qb,Q,\Qb,g_1,g_2) + \frac{1}{N_c^4} \t{\t{D}}_2^0(q,\qb,Q,\Qb,g_1,g_2) \Big]. \nn\\ 
\ea

\subsubsection{Four-parton one-loop}
The full one-loop D-type squared matrix element reads as follows~\cite{Ellis:1985er}:
\ba
M_4^1(q,\qb,Q,\Qb) &=& \left(\frac{\al_s}{2\pi}\right) g_s^4 (N_c^2-1) \Big[ D_0^1(q,\Qb,Q,\qb) + \frac{n_f}{N_c} \hat{D}_0^1(q,\Qb,Q,\qb) \nn\\
&+& \frac{1}{N_c^2} \t{D}_0^1(q,\Qb,Q,\qb) \Big],
\ea
where
\ba
D_0^1(q,\Qb,Q,\qb) &=& 2\, \mathrm{Re}\big[\mathcal{C}_{0,s}^{1\dagger}(q,\qb,Q,\Qb) \, c_0^0(q,\qb,Q,\Qb) + (\qb \leftrightarrow \Qb)\big], \nn\\
\hat{D}_0^1(q,\Qb,Q,\qb) &=& 2\, \mathrm{Re}\big[\hat{\mathcal{C}}_{0,s}^{1\dagger}(q,\qb,Q,\Qb) \, c_0^0(q,\qb,Q,\Qb) + (\qb \leftrightarrow \Qb)\big], \nn\\
\t{D}_0^1(q,\Qb,Q,\qb) &=& 2\, \mathrm{Re}\big[\t{\mathcal{C}}_{0,s}^{1\dagger}(q,\qb,Q,\Qb) \, c_0^0(q,\qb,Q,\Qb) + (\qb \leftrightarrow \Qb)\big].
\ea
Due to the interchanged antiquarks, the roles of the leading and subleading partial amplitudes in \eq{C0g1partamp} are reversed: the four-parton one-loop D-type matrix elements consists of only the subleading amplitude $c_{0,s}^1$ in place of $c_{0,l}^1$. For later convenience in the construction of the subtraction terms we define the following subfunctions:
\ba \label{eq:D0g1subfuncs}
D_{0,a}^1(q,\Qb,Q,\qb) &=& \mathrm{Re}\big[\mathcal{C}_{0,l}^{1\dagger}(q,\Qb,Q,\qb) \, c_0^0(q,\qb,Q,\Qb)\big], \nn\\ 
D_{0,b}^1(q,\Qb,Q,\qb) &=& -\mathrm{Re}\big[\mathcal{C}_{0,s}^{1\dagger}(q,\qb,Q,\Qb) \, c_0^0(q,\qb,Q,\Qb)\big], \nn\\
D_{0,c}^1(q,\Qb,Q,\qb) &=& \mathrm{Re}\big[\mathcal{C}_{0,l}^{1\dagger}(q,\qb,Q,\Qb) \, c_0^0(q,\Qb,Q,\qb)\big], \nn\\
D_{0,d}^1(q,\Qb,Q,\qb) &=& -\mathrm{Re}\big[\mathcal{C}_{0,s}^{1\dagger}(q,\Qb,Q,\qb) \, c_0^0(q,\Qb,Q,\qb)\big],
\ea 
where the following relation holds:
\ba
D_{0}^1(q,\Qb,Q,\qb) = -2 \big( D_{0,b}^1(i,l,k,j) + D_{0,d}^1(i,l,k,j) \big).
\ea
Using the notation $(i,j,k,l) = (q,\qb,Q,\Qb)$, the infrared pole structure of the individual functions can be expressed as follows:
\ba
\mathrm{Poles}\big[D_0^1(i,l,k,j)\big] &=&  \big[ 2 I_{q\qb}^{(1)}(\ep,s_{ik}) + 2 I_{q\qb}^{(1)}(\ep,s_{jl}) \big] D_0^0(i,l,k,j), \nn\\
\mathrm{Poles}\big[\hat{D}_0^1(i,l,k,j)\big] &=& 0, \nn\\
\mathrm{Poles}\big[\t{D}_0^1(i,l,k,j)\big] &=& \big[ - 2 I_{q\qb}^{(1)}(\ep,s_{il}) - 2 I_{q\qb}^{(1)}(\ep,s_{kj}) -2 I_{q\qb}^{(1)}(\ep,s_{ij}) - 2 I_{q\qb}^{(1)}(\ep,s_{kl}) \nn\\
&+& 2 I_{q\qb}^{(1)}(\ep,s_{ik}) + 2 I_{q\qb}^{(1)}(\ep,s_{jl}) \big] D_0^0(i,l,k,j).
\ea
The pole structure of the D-type subfunctions are given by
\ba
\mathrm{Poles}\big[D_{0,a}^1(i,l,k,j)\big] &=& \frac{1}{2}\big[ I_{q\qb}^{(1)}(\ep,s_{il}) + I_{q\qb}^{(1)}(\ep,s_{kj}) \big] D_0^0(i,l,k,j), \nn \\
\mathrm{Poles}\big[D_{0,b}^1(i,l,k,j)\big] &=& -\frac{1}{2}\big[ I_{q\qb}^{(1)}(\ep,s_{ik}) + I_{q\qb}^{(1)}(\ep,s_{jl}) \big] D_0^0(i,l,k,j), \nn \\
\mathrm{Poles}\big[D_{0,c}^1(i,l,k,j)\big] &=& \frac{1}{2}\big[ I_{q\qb}^{(1)}(\ep,s_{ij}) + I_{q\qb}^{(1)}(\ep,s_{kl}) \big] D_0^0(i,l,k,j), \nn \\
\mathrm{Poles}\big[D_{0,d}^1(i,l,k,j)\big] &=& -\frac{1}{2}\big[ I_{q\qb}^{(1)}(\ep,s_{ik}) + I_{q\qb}^{(1)}(\ep,s_{jl}) \big] D_0^0(i,l,k,j).
\ea

\subsubsection{Five-parton one-loop}
The five-parton one-loop D-type contribution is obtained by interfering the C-type four-quark one-gluon amplitudes with their tree-level counterparts with a crossing of the external quarks. It reads as follows: 
\ba \label{eq:D1g1full}
&& M_5^1(q,\qb,Q,\Qb,g) = -2\, \mathrm{Re} \big[ \mathcal{C}_1^{1\dagger}(q,\qb,Q,\Qb,g_1) \, \mathcal{C}_1^0(q,\Qb,Q,\qb,g_1) + (\qb \leftrightarrow \Qb) \big] \nn\\
&& = \left(\frac{\al_s}{2\pi}\right) g_s^6 (N_c^2-1) N_c \Big[ D_1^1(q,\qb,Q,\Qb,g) - D_{1,\ga_1}^1(q,\qb,Q,\Qb,g) \nn\\
&&+ \frac{n_f}{N_c} \big( \hat{D}_1^1(q,\qb,Q,\Qb,g) - \hat{D}_{1,\ga}^1(q,\qb,Q,\Qb,g) \big) \nn\\
&&+ \frac{1}{N_c^2} \big( \t{D}_1^1(q,\qb,Q,\Qb,g) - \t{D}_{1,\ga_1}^1(q,\qb,Q,\Qb,g) - D_{1,\ga_2}^1(q,\qb,Q,\Qb,g) \big) \nn\\
&&- \frac{n_f}{N_c^3} \hat{\t{D}}_{1,\ga}^1(q,\qb,Q,\Qb,g) - \frac{1}{N_c^4} \t{D}_{1,\ga_2}^1(q,\qb,Q,\Qb,g) + (\qb \leftrightarrow \Qb) \Big],
\ea
where the functions are defined in terms of the C-type amplitudes in \sec{CtypeAmps},
\ba
\lefteqn{D_{1}^{1}(q,\qb,Q,\Qb,g) =} \nn\\ 
&&2\,\mathrm{Re}\Big[\mathcal{C}_{1,c}^{0 \dagger}(q,g,\qb,Q,\Qb)\,\mathcal{C}_{1,a}^{1}(q,g,\qb,Q,\Qb)
+\mathcal{C}_{1,d}^{0 \dagger}(q,\qb,Q,g,\Qb)\,\mathcal{C}_{1,b}^{1}(q,\qb,Q,g,\Qb)\nn\\
&&+\mathcal{C}_{1,a}^{0 \dagger}(q,g,\qb,Q,\Qb)\,\mathcal{C}_{1,c}^{1}(q,g,\qb,Q,\Qb)
+\mathcal{C}_{1,b}^{0 \dagger}(q,\qb,Q,g,\Qb)\,\mathcal{C}_{1,d}^{1}(q,\qb,Q,g,\Qb)\Big],\nn\\
\lefteqn{\hat{D}_{1}^{1}(q,\qb,Q,\Qb,g)=} \nn\\ 
&&2\,\mathrm{Re}\Big[\mathcal{C}_{1,c}^{0 \dagger}(q,g,\qb,Q,\Qb)\,\hat{\mathcal{C}}_{1,a}^{1}(q,g,\qb,Q,\Qb)
+\mathcal{C}_{1,d}^{0 \dagger}(q,\qb,Q,g,\Qb)\,\hat{\mathcal{C}}_{1,b}^{1}(q,\qb,Q,g,\Qb)\nn\\
&&+\mathcal{C}_{1,a}^{0 \dagger}(q,g,\qb,Q,\Qb)\,\hat{\mathcal{C}}_{1,c}^{1}(q,g,\qb,Q,\Qb)
+\mathcal{C}_{1,b}^{0 \dagger}(q,\qb,Q,g,\Qb)\,\hat{\mathcal{C}}_{1,d}^{1}(q,\qb,Q,g,\Qb)\Big],\nn\\
\lefteqn{\t{D}_{1}^{1}(q,\qb,Q,\Qb,g)=} \nn\\ 
&&2\,\mathrm{Re}\Big[\mathcal{C}_{1,c}^{0 \dagger}(q,g,\qb,Q,\Qb)\,\t{\mathcal{C}}_{1,a}^{1}(q,g,\qb,Q,\Qb)
+\mathcal{C}_{1,d}^{0 \dagger}(q,\qb,Q,g,\Qb)\,\t{\mathcal{C}}_{1,b}^{1}(q,\qb,Q,g,\Qb)\nn\\
&&+\mathcal{C}_{1,a}^{0 \dagger}(q,g,\qb,Q,\Qb)\,\t{\mathcal{C}}_{1,c}^{1}(q,g,\qb,Q,\Qb)
+\mathcal{C}_{1,b}^{0 \dagger}(q,\qb,Q,g,\Qb)\,\t{\mathcal{C}}_{1,d}^{1}(q,\qb,Q,g,\Qb)\Big],\nn\\
\lefteqn{D_{1,\ga_{1}}^{1}(q,\qb,Q,\Qb,g)=2\,\mathrm{Re}\big[\mathcal{C}_{1,\ga}^{0 \dagger}(q,\qb,Q,\Qb,g)\,\mathcal{C}_{1,\ga_{1}}^{1}(q,\Qb,Q,\qb,g)\big],}\nn\\
\lefteqn{D_{1,\ga_{2}}^{1}(q,\qb,Q,\Qb,g)=2\,\mathrm{Re}\big[\mathcal{C}_{1,\ga}^{0 \dagger}(q,\qb,Q,\Qb,g)\,\mathcal{C}_{1,\ga_{2}}^{1}(q,\Qb,Q,\qb,g)\big],}\nn\\
\lefteqn{\hat{D}_{1,\ga}^{1}(q,\qb,Q,\Qb,g) = \hat{D}_{1,\ga_{1}}^{1}(q,\qb,Q,\Qb,g) = \hat{D}_{1,\ga_{2}}^{1}(q,\qb,Q,\Qb,g),}\nn\\
\lefteqn{\hat{D}_{1,\ga_{1}}^{1}(q,\qb,Q,\Qb,g)=2\,\mathrm{Re}\big[\mathcal{C}_{1,\ga}^{0 \dagger}(q,\qb,Q,\Qb,g)\,\hat{\mathcal{C}}_{1,\ga_{1}}^{1}(q,\Qb,Q,\qb,g)\big],}\nn\\
\lefteqn{\hat{D}_{1,\ga_{2}}^{1}(q,\qb,Q,\Qb,g)=2\,\mathrm{Re}\big[\mathcal{C}_{1,\ga}^{0 \dagger}(q,\qb,Q,\Qb,g)\,\hat{\mathcal{C}}_{1,\ga_{2}}^{1}(q,\Qb,Q,\qb,g)\big],}\nn\\
\lefteqn{\t{D}_{1,\ga_{1}}^{1}(q,\qb,Q,\Qb,g)=2\,\mathrm{Re}\big[\mathcal{C}_{1,\ga}^{0 \dagger}(q,\qb,Q,\Qb,g)\,\t{\mathcal{C}}_{1,\ga_{1}}^{1}(q,\Qb,Q,\qb,g)\big],}\nn\\
\lefteqn{\t{D}_{1,\ga_{2}}^{1}(q,\qb,Q,\Qb,g)=2\,\mathrm{Re}\big[\mathcal{C}_{1,\ga}^{0 \dagger}(q,\qb,Q,\Qb,g)\,\t{\mathcal{C}}_{1,\ga_{2}}^{1}(q,\Qb,Q,\qb,g)\big].}
\ea
The infrared pole structure results from the pole structure of the underlying one-loop amplitudes documented in \sec{Cpoles} above. 

\subsubsection{Four-parton two-loop}
The two-loop identical-flavour-only four-quark matrix element yields the double virtual D-type contribution, obtained from the same one-loop and two-loop four-quark amplitudes as the C-type contribution, but interfered in a different manner~\cite{Anastasiou:2000ue}: 
\ba
M_4^2(q,\Qb,Q,\qb) &=& \l(\frac{\al_s}{2\pi}\r)^2 g_s^4 (N_c^2-1) N_c \Big[ D_0^2(q,\Qb,Q,\qb) + \frac{n_f}{N_c} \hat{D}_0^2(q,\Qb,Q,\qb) \nn\\
&+& \frac{n_f^2}{N_c^2} \hat{\hat{D}}_0^2(q,\Qb,Q,\qb) + \frac{1}{N_c^2} \t{D}_0^2(q,\Qb,Q,\qb) + \frac{n_f}{N_c^3} \hat{\t{D}}_0^2(q,\Qb,Q,\qb) \nn\\
&+& \frac{1}{N_c^4} \t{\t{D}}_0^2(q,\Qb,Q,\qb) \Big].
\ea
Its $\ep$-pole structure is documented in~\cite{Anastasiou:2000ue}. 

\subsection{E-type matrix elements}
We call matrix elements containing three quark pairs of non-identical flavour E-type matrix elements~\cite{Gunion:1986zh,Su:2004ym}. We use the notation $E_n^l$ for a E-type matrix element containing $n$ gluons and $l$ loops and use the notation $(q,\qb,Q,\Qb,R,\bar{R})$ for the three quark pairs. As they contain at least six external partons by definition, they are only present at the double real level for dijet production at NNLO. The full tree-level amplitude with zero gluons is given by
\ba \label{eq:EfullME}
\mathcal{E}_0^{0,\mathrm{full}} &=& g_s^4 \Big[ \de_{q\Qb}\de_{Q\bar{R}}\de_{R\qb} \, \mathcal{E}_0^{0}(q,\Qb,Q,\bar{R},R,\qb) + \de_{q\bar{R}}\de_{R\Qb}\de_{Q\qb} \, \mathcal{E}_0^{0}(q,\bar{R},R,\Qb,Q,\qb) \nn\\
&-& \frac{1}{N_c} \Big( \de_{q\qb}\de_{Q\bar{R}}\de_{R\Qb} \, \t{\mathcal{E}}_0^{0}(Q,\bar{R},R,\Qb,q,\qb) + \de_{Q\Qb}\de_{q\bar{R}}\de_{R\qb} \, \t{\mathcal{E}}_0^{0}(q,\bar{R},R,\qb,Q,\Qb) \nn\\
&+& \de_{R\bar{R}}\de_{q\Qb}\de_{Q\qb} \, \t{\mathcal{E}}_0^{0}(q,\Qb,Q,\qb,R,\bar{R}) \Big) + \frac{1}{N_c^2}\de_{q\qb}\de_{Q\Qb}\de_{R\bar{R}} \, \t{\t{\mathcal{E}}}_0^{0}(q,\qb,Q,\Qb,R,\bar{R}) \Big], \nn\\
\ea
where the following relations among the colour-ordered amplitudes hold:
\ba
\t{\t{\mathcal{E}}}_0^{0}(q,\qb,Q,\Qb,R,\bar{R}) &=& \mathcal{E}_0^{0}(q,\Qb,Q,\bar{R},R,\qb) + \mathcal{E}_0^{0}(q,\bar{R},R,\Qb,Q,\qb), \nn\\
2{\t{\t{\mathcal{E}}}}_0^{0}(q,\qb,Q,\Qb,R,\bar{R}) &=& \t{\mathcal{E}}_0^{0}(Q,\bar{R},R,\Qb,q,\qb) + \t{\mathcal{E}}_0^{0}(q,\bar{R},R,\qb,Q,\Qb) + \t{\mathcal{E}}_0^{0}(q,\Qb,Q,\qb,R,\bar{R}). \nn\\
\ea
Squaring \eq{EfullME} we obtain
\ba
\lefteqn{E_0^{0,\mathrm{full}}(q,\qb,Q,\Qb,R,\bar{R}) = \big|\mathcal{E}_0^{0,\mathrm{full}}(q,\qb,Q,\Qb,R,\bar{R}) \big|^{2}} \nn\\
&=& g_s^8 (N_c^2-1) N_c \Big[ E_0^0(q,\Qb,Q,\bar{R},R,\qb) + E_0^0(q,\bar{R},R,\Qb,Q,\qb) \nn\\
&+&\frac{1}{N_c^2} \Big( \t{E}_0^0(q,\Qb,Q,\qb,R,\bar{R}) + \t{E}_0^0(R,\Qb,Q,\bar{R},q,\qb) + \t{E}_0^0(q,\bar{R},R,\qb,Q,\Qb)\nn\\
&&-3\t{\t{E}}_0^0(q,\qb,Q,\Qb,R,\bar{R})  \Big)  \Big],
\ea
where 
\ba
E_0^0 = |\mathcal{E}_0^{0}|^2, \quad \t{E}_0^0 = |\t{\mathcal{E}}_0^{0}|^2, \quad \t{\t{E}}_0^0 = |\t{\t{\mathcal{E}}}_0^{0}|^2.
\ea

\subsection{F-type matrix elements}
We define the F-type matrix elements containing two quark pairs of identical flavour and one quark pair of distinct flavour in terms of the E-type amplitudes in the same way as we defined the D-type matrix elements related to the C-type amplitudes. Using the notation $F_n^l$ for an F-type matrix element containing $n$ gluons and $l$ loops and taking $(q,\qb) = (Q,\Qb)$, we have for the full two-identical-flavour six-quarks matrix element:
\ba
M_{0,\text{two-id.}}^0(q,\qb,q,\qb,R,\bar{R}) &=& \big|\mathcal{E}_0^{0,\mathrm{full}}(q,\qb,Q,\Qb,R,\bar{R})-\mathcal{E}_0^{0,\mathrm{full}}(q,\Qb,Q,\qb,R,\bar{R})\big|^{2} \nn\\
&=& E_0^{0,\mathrm{full}}(q,\qb,Q,\Qb,R,\bar{R}) + E_0^{0,\mathrm{full}}(q,\Qb,Q,\qb,R,\bar{R}) \nn\\
&+& F_0^{0,\mathrm{full}}(q,\qb,Q,\Qb,R,\bar{R}).
\ea
The full F-type matrix element is suppressed by $1/N_c$ with respect to the full E-type matrix element and is given by
\ba
\lefteqn{F_0^{0,\mathrm{full}}(q,\qb,Q,\Qb,R,\bar{R}) = g_s^8 (N_c^2-1) \Big[} \nn\\
&-&F_0^0(q,\qb,Q,\bar{R},R,\Qb) - F_0^0(q,\Qb,Q,\bar{R},R,\qb) \nn\\
&-& F_0^0(Q,\Qb,q,\bar{R},R,\qb) - F_0^0(Q,\qb,q,\bar{R},R,\Qb) \nn\\
&-& \t{F}_0^0(q,\qb,Q,\Qb,R,\bar{R}) -  \t{F}_0^0(q,\Qb,Q,\qb,R,\bar{R}) + 3\t{\t{F}}_0^0(q,\qb,Q,\Qb,R,\bar{R}) \nn\\
&+&\frac{1}{N_c^2} \big( \t{F}_0^0(q,\qb,Q,\Qb,R,\bar{R}) +  \t{F}_0^0(q,\Qb,Q,\qb,R,\bar{R}) - 3\t{\t{F}}_0^0(q,\qb,Q,\Qb,R,\bar{R}) \big) \Big], \nn\\
\ea
where
\ba
F_0^0(q,\qb,Q,\bar{R},R,\Qb) &=& 2\, \mathrm{Re} \big[ \mathcal{E}_{0}^0(q,\qb,Q,\bar{R},R,\Qb) \, \t{\mathcal{E}}_{0}^{0\dagger}(q,\bar{R},R,\qb,Q,\Qb) \big],\nn\\
\t{F}_0^0(q,\qb,Q,\Qb,R,\bar{R}) &=& 2\, \mathrm{Re} \big[ \t{\mathcal{E}}_{0}^0(q,\qb,Q,\Qb,R,\bar{R}) \, \t{\t{\mathcal{E}}}_{0}^{0\dagger}(q,\qb,Q,\Qb,R,\bar{R}) \big],\nn\\
\t{\t{F}}_0^0(q,\qb,Q,\Qb,R,\bar{R}) &=& 2\, \mathrm{Re} \big[ \t{\t{\mathcal{E}}}_{0}^0(q,\Qb,Q,\qb,R,\bar{R}) \, \t{\t{\mathcal{E}}}_{0}^{0\dagger}(q,\qb,Q,\Qb,R,\bar{R}) \big].
\ea

\subsection{G-type matrix elements}
Lastly, we have the G-type matrix elements containing three-identical-flavour quark pairs. We use the notation $G_n^l$ for a G-type matrix element containing $n$ gluons at $l$ loops and have $(q,\qb)=(Q,\Qb)=(R,\bar{R})$. The full identical matrix element consists of the six antisymmetrized permutations:
\ba
M_{0,\mathrm{id.}}^0(q,\qb,q,\qb,q,\qb) &=& 
\big|\mathcal{E}_0^{0,\mathrm{full}}(q,\qb,Q,\Qb,R,\bar{R}) - \mathcal{E}_0^{0,\mathrm{full}}(q,\Qb,Q,\qb,R,\bar{R}) \nn\\
&+& \mathcal{E}_0^{0,\mathrm{full}}(q,\Qb,Q,\bar{R},R,\qb) - \mathcal{E}_0^{0,\mathrm{full}}(q,\bar{R},Q,\Qb,R,\qb) \nn\\
&+& \mathcal{E}_0^{0,\mathrm{full}}(q,\bar{R},Q,\qb,R,\Qb) - \mathcal{E}_0^{0,\mathrm{full}}(q,\qb,Q,\bar{R},R,\Qb) \big|^{2}.  
\ea
The colour connections new to the E- and F-type matrix elements are collected in the G-type matrix element, which broken down in its colour levels reads
\ba
G_0^{0,\mathrm{full}}(q,\qb,Q,\Qb,R,\bar{R}) &=& g_s^8 (N_c^2-1) N_c \Big[ G_0^0(q,\qb,Q,\Qb,R,\bar{R}) \nn\\
&+& \frac{1}{N_c^2} \Big( \t{G}_0^0(q,\qb,Q,\Qb,R,\bar{R}) - \t{\t{G}}_0^0(q,\qb,Q,\Qb,R,\bar{R}) \Big) \nn\\
&-& \frac{1}{N_c^4} \t{\t{G}}_0^0(q,\qb,Q,\Qb,R,\bar{R}) \Big].
\ea

\section{Double real subtraction} \label{sec:RRsub}
All six-parton matrix elements that were introduced in the previous section enter the double-real contribution to the dijet cross section. Their numerical implementation into a parton-level event generator requires the construction of subtraction terms, which we discuss in the following. 

The double real subtraction, denoted by $\df \si^S$, needs to reproduce single and double unresolved limits of all squared matrix elements in $\df\sigma^{RR}$. We partition $\df \si^S$ into the following parts~\cite{Currie:2013vh}:
\ba
\df \si^S = \df \si^{S,a} + \df \si^{S,b} + \df \si^{S,c} + \df \si^{S,d} + \df \si^{S,e},
\ea
where each part has a specific purpose and general form:
\begin{itemize}
	\item $\df \si^{S,a}$: this part is responsible for subtracting the single unresolved limits of the RR squared matrix elements and has the general form of three-parton antennae times a reduced $(n+1)$-parton matrix element:
	\ba
	\df \si^{S,a} \supset \mathcal{N}_{RR} \, \df \Phi_{n+2} \, X_3^0(\{p_{n+2}\}) M_{n+1}^0(\{\t{p}_{n+1}\}) J_n^{(n+1)}(\{\t{p}_{n+1}\}),
	\ea
	where overall factors are collected in $\mathcal{N}_{RR}$, $\df \Phi_{n+2}$ is the $(2\to n)$-particle phase space measure and the tilde in $\{\t{p}_{n+1}\}$ denotes that the $(n+1)$-momenta set is a single mapped momenta set. The $X_3^0$ denotes a three-parton antenna function~\cite{Gehrmann-DeRidder:2004ttg,Gehrmann-DeRidder:2005svg,Gehrmann-DeRidder:2005alt} which mimics the behaviour of the matrix elements in single unresolved limits. The reduced matrix element contains $(n+1)$ particles and at NNLO the jet function $J_n^{(n+1)}$ allows one parton to become unresolved. These limits have to be accounted for by subtraction terms in the other parts.
	\item $\df \si^{S,b}$: this part consists of $X_4^0$ antenna functions~\cite{Gehrmann-DeRidder:2005btv}, which reproduce the behaviour of the RR matrix element in genuine double unresolved limits, and its associated iterated $X_3^0 X_3^0$ antennae.  We further partition $\df \si^{S,b}$ into $\df \si^{S,b_1}$ and $\df \si^{S,b_2}$ accordingly: 
	\ba
	\df \si^{S,b_1} &\supset& \mathcal{N}_{RR} \, \df \Phi_{n+2} \, X_4^0(\{p_{n+2}\}) M_n^0(\{\t{p}_{n}\}) J_n^{(n)}(\{\t{p}_{n}\}), \nn\\
	\df \si^{S,b_2} &\supset& \mathcal{N}_{RR} \, \df \Phi_{n+2} \, X_3^0(\{p_{n+2}\}) X_3^0(\{\t{p}_{n+1}\}) M_n^0(\{\t{\t{p}}_{n}\}) J_n^{(n)}(\{\t{\t{p}}_{n}\}).
	\ea
	Here the double tilde denotes a doubly mapped momenta set arising from the iterated $X_3^0$ antennae. The terms in $\df \si^{S,b_1}$ subtract colour-connected double unresolved limits. The $X_4^0$ antenna functions themselves contain single unresolved limits, which do not correspond to the single unresolved limits of the RR matrix element. In these limits, the $X_4^0$ antennae collapse onto $X_3^0$ antennae, so these limits are removed by the iterated $X_3^0$ antennae in $\df \si^{S,b_2}$. 
	\item $\df \si^{S,c}$: these terms have the general form:
	\ba
	\df \si^{S,c} \supset \mathcal{N}_{RR} \, \df \Phi_{n+2} \, X_3^0(\{p_{n+2}\}) X_3^0(\{\t{p}_{n+1}\}) M_n^0(\{\t{\t{p}}_{n}\}) J_n^{(n)}(\{\t{\t{p}}_{n}\}),
	\ea
	where the two $X_3^0$ antennae have one or more common momenta (related by a mapping). These terms appear when subtracting almost colour-connected limits and other spurious limits.
	\item $\df \si^{S,d}$: this part subtracts colour-disconnected limits and has the form
	\ba
	\df \si^{S,d} \supset \mathcal{N}_{RR} \, \df \Phi_{n+2} \, X_3^0(\{p_{n+2}\}) X_3^0(\{\t{p}_{n+1}\}) M_n^0(\{\t{\t{p}}_{n}\}) J_n^{(n)}(\{\t{\t{p}}_{n}\}),
	\ea
	where there are no common momenta in the two $X_3^0$ antenna functions. Consequently, although the arguments of the second $X_3^0$ are taken from 
	the $\{\t{p}_{n+1}\}$ momentum set (to assure phase space factorization), they are not affected by the first phase space mapping 
	and are taken unmodified  out of the $\{p_{n+2}\}$ momenta set. 
	\item $\df \si^{S,e}$: after all the previous parts have been added together, there is still the possibility of a mismatch in single soft limits between the subtraction terms and the squared matrix element, which is caused by the mapped momenta in the antennae. These mismatches are resolved by adding large angle soft (LAS) terms~\cite{Currie:2013vh}:
	\ba
	\df \si^{S,e} \supset \mathcal{N}_{RR} \, \df \Phi_{n+2} \, \l(S^{FF}(\{p_{n+2}\}) - S^{FF}(\{\t{p}_{n+1}\}) \r) \nn\\
	\times X_3^0(\{\t{p}_{n+1}\}) M_n^0(\{\t{\t{p}}_{n}\}) J_n^{(n)}(\{\t{\t{p}}_{n}\}).
	\ea
	In general, the mapping in the soft functions can be any of the initial-initial, initial-final or final-final type, but for our dijet processes, there are sufficient partons present to always use a final-final mapping. As the FF mapping has some advantages compared to the other mappings in terms of convergence in the unresolved limits, we choose to always employ the FF mappings for the LAS terms. 
\end{itemize}

The above subtraction terms are integrated analytically over the unresolved phase space associated to the antenna functions contained in them by assigning them to either $\df\sigma^{S,1}$ or $\df\sigma^{S,2}$ defined in \eq{sigsub} above: $\df \si^{S,a}$, $\df \si^{S,b_2}$ and $\df \si^{S,e}$ belong to $\df\sigma^{S,1}$, while $\df \si^{S,b_1}$ and $\df \si^{S,d}$ belong to $\df\sigma^{S,2}$. $\df\sigma^{S,c}$ is split between both. 

In the colour-ordered antenna subtraction method, we generally construct the double real subtraction in the order above, guided by the behaviour of the matrix element in unresolved limits. For most of the RR parton-level subprocesses described in \sec{MEs} above, it is straightforward to identify the hard radiator partons and the unresolved partons for each colour-ordered squared matrix element containing quarks, gluons and abelian gluons. 
The construction of the subtraction terms then follows in a straightforward, albeit tedious and lengthy manner, as described in detail in~\cite{Currie:2013vh}. We will not repeat this description here, but instead focus on the new type of antenna subtraction terms that are required for genuine interference-type squared matrix elements, which first appear for processes with four coloured partons at Born level. In the subleading colour all-gluon contribution to dijet production, the required antenna subtraction terms were already described elsewhere~\cite{Currie:2013dwa}. We thus focus in the following on the more general case involving quarks and gluons. 

The full squared six-parton B-type matrix element was shown in \sec{B4g0ME} to be constructed from functions containing quarks, gluons and abelian gluons, with some remainder interference terms at the $1/N_c^2$ colour level grouped together in $\bs{R}_4^0$. For the R-term $R_4^0$, there is no clear colour connection defined at the squared amplitude level, even after including the notion of abelian gluons, thus its unresolved limits are not immediately clear. In the following, we discuss how to find all the unresolved limits of $R_4^0$ and subsequently construct the double real subtraction for these terms. 

\subsection{Single unresolved limits of genuine interference B-type contributions} \label{sec:Rtermsub1}
We can exploit our knowledge of the unresolved limits of all the other B-type squared matrix elements besides the R-term to find the unresolved limits of $R_4^0$. 
This is done by knowing how the full B-type matrix element behaves in certain limits and comparing it with the terms coming from the well-defined B-type squared functions in that limit. Any missing term from the full B-type matrix element not produced by the B-type functions must come from the R-term. We do these comparisons in a colour-ordered way, such that we only need to consider the colour level relevant to the R-term.

\subsubsection{Single gluon-gluon collinear $gg\to G$}
We consider the full matrix element $|\mathcal{B}_4^0|$ in the gluon-gluon collinear limit. It factorizes into a splitting function and a reduced multiplicity matrix element
(omitting here and in the following the tensor indices related to angular-dependent terms in the gluon-to-gluon splitting function):
\ba\label{eq:B40gglimit}
\big|\mathcal{B}_4^0\big|^{2} &\xrightarrow{g\|g}& \bs{P}_{gg\to G} \big|\mathcal{B}_3^0\big|^2.
\ea
The full splitting function carries a factor of $g_s^{2}$ and $C_{A}=N_c$ and the full reduced matrix element is given by \eq{FullB3g0}. This yields the result
\ba
\big|\mathcal{B}_{4}^{0}\big|^{2} &\xrightarrow{g\|g}& g_s^{8}(N_c^2-1)N_c^{3} P_{gg\to G} \Big[ \bs{B}_{3}^{0}-\frac{1}{N_c^2}\Big(\t{\bs{B}}_{3}^{0}-\bar{\bs{B}}_{3}^{0}\Big)+\frac{1}{N_c^4}\bar{\bs{B}}_{3}^{0} \Big].
\ea
We can also obtain this result by taking the collinear limit of the B-type functions in \eq{FullB4g0}. Since abelian gluons have no collinear limit with each other, we immediately see why the $1/N_c^6$ colour level disappears in this limit, as $\bar{\bs{B}}_{4}^{0}$ contains no non-abelian gluons. Focusing on the $1/N_c^2$ colour level where the R-term resides, we thus note that $\bar{\bs{B}}_{4}^{0}$ has no contribution in the collinear limit and the $\t{\bs{B}}_{4}^{0}$ term can only contribute when the collinear gluons are non-abelian. We also know that the sum of all terms must factor onto a full $\t{\bs{B}}_{3}^{0}$ function minus a full $\bar{\bs{B}}_{3}^{0}$ function. The $\t{\bs{B}}_{4}^{0}$ function can never factor onto the all abelian-gluon $\bar{\bs{B}}_{3}^{0}$ function, leaving the R-term as its only source. Keeping track of all the orderings we see that in the gluon-gluon collinear limit $i\|j$
\ba
\t{\bs{B}}_{4}^{0} &\xrightarrow{i\|j}& P_{gg\to G} \Big[ \t{B}_{3}^{0}(1,k,(ij),l,2)+\t{B}_{3}^{0}(1,k,l,(ij),2)\nn\\
&&+\t{B}_{3}^{0}(1,l,(ij),k,2)+\t{B}_{3}^{0}(1,l,k,(ij),2) \Big],
\ea
which lacks the two orderings
\ba
\t{B}_{3}^{0}(1,(ij),k,l,2), \quad \t{B}_{3}^{0}(1,(ij),l,k,2),
\ea
needed to form the full $\t{\bs{B}}_{3}^{0}$ function at the reduced matrix element level. This means that the R-term must exactly factorize in such a way so as to provide these missing orderings plus the $\bar{\bs{B}}_{3}^{0}$ term:
\ba
\bs{R}_{4}^{0} \xrightarrow{i\|j} P_{gg\to G} \Big[ \t{B}_{3}^{0}(1,(ij),k,l,2)+\t{B}_{3}^{0}(1,(ij),l,k,2)-\bar{B}_{3}^{0}(1,(ij),k,l,2) \Big].
\ea
The R-term in the gluon-gluon collinear limit thus factorizes into functions where the composite gluon is abelian.

\subsubsection{Single quark-antiquark collinear $\qb q\to G$}
In the quark-antiquark collinear limit, the full B-type  squared matrix element factorizes into a splitting function and a reduced all-gluon squared matrix element:
\ba
\big|\mathcal{B}_{4}^{0}\big|^{2} &\xrightarrow{q\|\qb}& \bs{P}_{q\qb\to G} \big|\mathcal{A}_{5}^{0}\big|^{2}.
\ea
The splitting function carries a factor of $g_s^{2}$ and $T_R$ and the full reduced matrix element is given by \eq{FullA5g0}. The full five-gluon squared matrix element has no subleading colour terms, since these vanish by the Dual Ward Identity for all-gluonic amplitudes. This yields the result
\ba
\big|\mathcal{B}_{4}^{0}\big|^{2} &\xrightarrow{q\|\qb}& g_s^{8}(N_c^2-1)N_c^{3} (T_R {P}_{q\qb\to G}) \bs{A}_{5}^{0}.
\ea
Noting that both $\t{\bs{B}}_{4}^{0}$ and $\bar{\bs{B}}_{4}^{0}$ independently have no singular quark-antiquark collinear limit since they would reduce to a gluonic matrix element containing abelian gluon(s), we can conclude that the R-term also has no singular quark-antiquark collinear limit.

\subsubsection{Single quark-gluon collinear $qg\to Q$}
To determine the behaviour of the R-term in the single-collinear quark-gluon limits, we use the known collinear factorization for colour-ordered amplitudes and note that to get a sufficiently singular collinear limit at the squared amplitude level we must have the partons colour adjacent in both the amplitude and conjugated amplitude. From the definition of a single ordering of the R-term in \eq{soR4g0}, we can immediately see that there are no single quark-gluon collinear limits: quark $1$ is adjacent to gluon $i$ in the conjugated amplitude, but this is never the case in the amplitudes this particular conjugated amplitude multiplies; the same holds for quark $2$ with gluon $l$. 

\subsubsection{Single soft limit} \label{sec:SingleSoft}
The collinear limits of the R-term could easily be found because of the factorization onto the five-parton squared matrix elements, but unfortunately this is not the case for the soft gluon limit. As the R-term does not have clear colour connections at the squared matrix element level, its soft limit is not immediately clear. Thus, it is now necessary to look at how the matrix element behaves in unresolved limits at the amplitude level. The factorization of a colour-ordered amplitude in a single soft limit is given by
\ba
\mathcal{M}_n^0(\dots,a,i,b,\dots)&\xrightarrow{i\to 0}&\mathcal{S}_{\mu}(a,i,b)\ep^{\mu}\mathcal{M}_{n-1}^{0}(\dots,a,b,\dots),
\ea
where $\ep^\mu$ is the gluon polarization vector and $\mathcal{S}_{\mu}$ the single soft current~\cite{Sudakov:1954sw} 
\ba\label{eq:ssc}
\mathcal{S}^{\mu}(a,i,b)&=&2\bigg(\frac{a^{\mu}}{s_{ai}}-\frac{b^{\mu}}{s_{bi}}\bigg).
\ea
Squaring the single soft current and averaging over the unresolved gluon polarizations using the polarization tensor yields the familiar soft eikonal factor, 
\ba
S_{aib} = \frac{1}{2} d^{\mu \nu} S_\mu(a,i,b) S_\nu(a,i,b) = \frac{2 s_{ab}}{s_{ai}s_{ib}},
\ea
but when we have an interference term of an amplitude where $i$ becomes soft in between $a,b$ multiplied by a conjugated amplitude where $i$ becomes soft in between $c,d$ we get
\ba
&&\mathcal{M}_n^{0 \dagger}(\dots,a,i,b,\dots) \, \mathcal{M}_n^{0}(\dots,c,i,d,\dots) \xrightarrow{i\to 0} \nn\\
&&\frac{1}{2}\Big[S_{aid}+S_{bic}-S_{aic}-S_{bid}\Big]\mathcal{M}_{n-1}^{0 \dagger}(\dots,a,b,\dots) \,  \mathcal{M}_{n-1}^{0}(\dots,c,d,\dots),
\ea
which can simply be shown by interfering two incoherent colour orderings of the soft current:
\ba
\frac{1}{2}d^{\mu\nu}\mathcal{S}_{\mu}(a,i,b)\mathcal{S}_{\nu}(c,i,d)&=&2\Big(\frac{a^{\mu}}{s_{ai}}-\frac{b^{\mu}}{s_{bi}}\Big)\Big(\frac{c_{\mu}}{s_{ci}}-\frac{d_{\mu}}{s_{di}}\Big),\nn\\
&=&\frac{1}{2}\Big[ S_{aid}+S_{bic}-S_{aic}-S_{bid} \Big].
\ea
Alternatively, preparing for the double soft current below, the above relation can also be shown by using the following properties of the single soft current:
\begin{enumerate}
	\item Reflection: $\mathcal{S}_{\mu}(b,i,a)=-\mathcal{S}_{\mu}(a,i,b)$,
	\item Left shuffle: $\mathcal{S}_{\mu}(a,i,b)=\mathcal{S}_{\mu}(c,i,b)-\mathcal{S}_{\mu}(c,i,a)$,
	\item Right shuffle: $\mathcal{S}_{\mu}(a,i,b)=\mathcal{S}_{\mu}(a,i,c)+\mathcal{S}_{\mu}(c,i,b)$,
\end{enumerate}
where the first and second properties can simply be inferred from looking at the explicit expression of the soft current and the third property follows from the other two.

Using these identities for the soft limit of a general interference term, the soft limit of the R-term can be expressed by combinations of eikonal factors multiplying an interference of five-parton amplitudes. For example, for the single ordering of the R-term $R_4^0(1,i,j,k,l,2)$ we find in the $i$ soft limit:
\ba
R_{4}^{0}(1,i,j,k,l,2) &\xrightarrow{i \to 0}& \frac{1}{2}\big[S_{1il}+S_{jij}-S_{1ij}-S_{jil}\big] b_{3}^{0 \dagger}(1,j,k,l,2) \, b_{3}^{0}(1,j,l,k,2)\nn\\
&+&\frac{1}{2}\big[S_{1ik}+S_{jil}-S_{1il}-S_{jik}\big] b_{3}^{0 \dagger}(1,j,k,l,2) \, b_{3}^{0}(1,j,l,k,2)\nn\\
&+&\frac{1}{2}\big[S_{1i2}+S_{jik}-S_{1ik}-S_{ji2}\big] b_{3}^{0 \dagger}(1,j,k,l,2) \, b_{3}^{0}(1,j,l,k,2)\nn\\\nn\\
&+&\frac{1}{2}\big[S_{1il}+S_{jik}-S_{1ik}-S_{jil}\big] b_{3}^{0 \dagger}(1,j,k,l,2) \, b_{3}^{0}(1,k,l,j,2)\nn\\
&+&\frac{1}{2}\big[S_{1i2}+S_{jil}-S_{1il}-S_{ji2}\big] b_{3}^{0 \dagger}(1,j,k,l,2) \, b_{3}^{0}(1,k,j,l,2)\nn\\\nn\\
&+&\frac{1}{2}\big[S_{1ik}+S_{jil}-S_{1il}-S_{jik}\big] b_{3}^{0 \dagger}(1,j,k,l,2) \, b_{3}^{0}(1,l,k,j,2)\nn\\
&+&\frac{1}{2}\big[S_{1ik}+S_{jij}-S_{1ij}-S_{jik}\big] b_{3}^{0 \dagger}(1,j,k,l,2) \, b_{3}^{0}(1,l,j,k,2)\nn\\
&+&\frac{1}{2}\big[S_{1i2}+S_{jij}-S_{1ij}-S_{ji2}\big] b_{3}^{0 \dagger}(1,j,k,l,2) \, b_{3}^{0}(1,l,k,j,2). \nn\\
\ea
These reduced interference terms cannot be written into a squared function or sum of squared functions, even when summed over all 24 orderings. Thus, we define
\be
B_{3,\mathrm{int}}^0(i_1,i_2,i_3,i_4,i_5,j_1,j_2,j_3,j_4,j_5) = 2\, \mathrm{Re}\big[ b_{3}^{0 \dagger}(i_1,i_2,i_3,i_4,i_5) \, b_{3}^{0}(j_1,j_2,j_3,j_4,j_5)\big],
\ee
as the reduced matrix element when it consists of a general interference.

Having identified all the single unresolved limits of the R-term, the construction of the $\df \si^{S,a}$ subtraction follows as usual. Eikonal factors are promoted to $X_3^0$ antenna functions which capture all the soft as well as the collinear gluon-gluon limits. Any quark-gluon collinear limit produced by the $X_3^0$ antennae cancels out in the total sum of all terms, which is consistent with what we found before. This approach of taking the soft limit at amplitude level to construct the $\df \si^{S,a}$ subtraction term also works for the well-defined squared B-type functions containing (abelian) gluons. Exactly the same subtraction terms are generated in this way as in the standard approach using the colour connections of the squared matrix elements, meaning that although helpful to understand the matrix element better, it is not strictly necessary to organise the amplitudes into squared functions. As even higher multiplicity B-type matrix elements (e.g.\ $|\mathcal{B}_5^0|$) will also contain remaining interference terms like the R-term $\bs{R}_4^0$ at subleading colour levels, the approach at amplitude level works the same way as for the six-parton B-type matrix element to construct the $\df \si^{S,a}$ subtraction, and would be the preferred method, if one seeks to automate the construction of the subtraction terms~\cite{Chen:2022ktf}. 

\subsection{Double unresolved limits of genuine interference B-type contributions} \label{sec:Rtermsub2}
In the following, we examine the double unresolved limits of the R-term for the construction of the double real subtraction terms. The NNLO collinear limits can be determined in the same way as was done for the single collinear limits, because they factorize nicely onto the reduced squared matrix elements. For the double soft limit however, this is again not the case and we will need to determine how the R-term behaves in a double soft limit at the amplitude level.

\subsubsection{Triple collinear $ggg\to G$}
First we consider the gluon-gluon-gluon triple collinear limit, in which the full matrix element factorizes as
\ba
\big|\mathcal{B}_{4}^{0}\big|^{2} &\xrightarrow{g\|g\|g}& \bs{P}_{ggg\to G} \big|\mathcal{B}_{2}^{0}\big|^{2}.
\ea
The full splitting function carries a factor of $g_s^4$ and $C_A^2 = N_c^2$ as well as being summed over six gluon orderings. The reduced matrix element is given by \eq{FullB2g0}, resulting in
\ba
\big|\mathcal{B}_{4}^{0}\big|^{2} &\xrightarrow{g\|g\|g}& g_s^8 (N_c^2-1) N_c^3 {P}_{ggg\to G} \Big[ \bs{B}_{2}^{0} - \frac{1}{N_c^2} \bar{\bs{B}}_2^0 \Big].
\ea
The lack of non-abelian gluons at the $1/N_c^4$ and $1/N_c^6$ colour levels means that these terms do not contribute to the limit. At the $1/N_c^2$ colour level, the $\t{\bs{B}}_{4}^{0}$ term fully reproduces the $\bar{\bs{B}}_2^0$ term, and so there is no limit left over for $\bs{R}_4^0$ to contribute to.

\subsubsection{Triple collinear $qgg\to Q$}
The $qgg\to Q$ splitting function has the decomposition into the colour-ordered splitting functions:
\ba
\bs{P}_{qgg\to Q}&=&N_c^{2}P_{qgg\to Q}- (P_{qgg\to Q} - \t{P}_{qgg\to Q}) +\frac{1}{N_c^2}\t{P}_{qgg\to Q}.
\ea
The full B-type matrix element in this limit then factorizes according to
\ba
\big|\mathcal{B}_{4}^{0}\big|^{2} &\xrightarrow{q\|g\|g}& g_s^{8}(N_c^2-1)N_c \Big[N_c^{2}P_{qgg\to Q}- (P_{qgg\to Q} - \t{P}_{qgg\to Q}) +\frac{1}{N_c^2}\t{P}_{qgg\to Q}\Big] \nn\\
&& \times \Big[ \bs{B}_{2}^{0}-\frac{1}{N_c^2}\bar{\bs{B}}_{2}^{0}\Big] \nn\\
&=&g_s^{8}(N_c^2-1)N_c^{3} \Big[P_{qgg\to Q}\bs{B}_{2}^{0}\nn\\
&&-\frac{1}{N_c^2}\Big( (P_{qgg\to Q}-\t{P}_{qgg\to Q}){\bs{B}}_{2}^{0}+P_{qgg\to Q}\bar{\bs{B}}_{2}^{0} \Big)\nn\\
&&+\frac{1}{N_c^4}\Big( (P_{qgg\to Q}-\t{P}_{qgg\to Q})\bar{\bs{B}}_{2}^{0}+\t{P}_{qgg\to Q}{\bs{B}}_{2}^{0} \Big)\nn\\
&&-\frac{1}{N_c^6}\Big( \t{P}_{qgg\to Q}\bar{\bs{B}}_{2}^{0} \Big) \Big].
\ea
At the $1/N_c^2$ colour level we observe that we can receive contributions to this limit from the $\t{\bs{B}}_{4}^{0}$, $\bar{\bs{B}}_{4}^{0}$ and $\bs{R}_{4}^{0}$ terms. The $\t{\bs{B}}_{4}^{0}$ terms have two types of limits: a splitting factoring onto the subleading colour matrix element,
\ba
\t{B}_{4}^{0}(1,i,j,k,l,2) \xrightarrow{1\|j\|k} {P}_{qgg\to Q} \bar{B}_{2}^{0}((1jk),i,l,2),
\ea
and an abelian splitting factoring onto a leading colour matrix element,
\ba
\t{B}_{4}^{0}(1,i,j,k,l,2) \xrightarrow{1\|i\|j} \t{P}_{qgg\to Q} B_{2}^{0}((1ij),k,l,2).
\ea
The $\bar{\bs{B}}_{4}^{0}$ term only has the latter type, but factorizes onto the subleading colour matrix element instead. Keeping track of all the orderings, their full factorization is given by
\ba
\t{\bs{B}}_{4}^{0}&\xrightarrow{q\|g\|g}&2(\t{P}_{qgg\to Q}\bs{B}_{2}^{0}+P_{qgg\to Q}\bar{\bs{B}}_{2}^{0}),\nn\\
\bar{\bs{B}}_{4}^{0}&\xrightarrow{q\|g\|g}&\t{P}_{qgg\to Q}\bar{\bs{B}}_{2}^{0},
\ea
where we note that here $P_{qgg\to Q}$ contains both gluon orderings. From this we can infer the factorization of the R-term:
\ba
\bs{R}_{4}^{0}&\xrightarrow{q\|g\|g} &(\t{P}_{qgg\to Q} - P_{qgg\to Q})(\bar{\bs{B}}_{2}^{0}-\bs{B}_{2}^{0}),
\ea
where the combination of reduced matrix elements is exactly the interference matrix element $\t{B}_{2,R}^0$ defined in \eq{BRt2g0}. Note that the combination of splitting functions $(\t{P}_{qgg\to Q} - P_{qgg\to Q})$ can also be written as an interference of splitting amplitudes~\cite{Catani:1999ss} in the form,
\ba
\t{P}_{qgg\to Q} - P_{qgg\to Q} = 2\, \mathrm{Re}\Big[\sum_{\la}\mathrm{Split}^{\dagger}(1,i,j;\{\la\}) \, \mathrm{Split}(1,j,i;\{\la\})\Big].
\ea

\subsubsection{Triple collinear $g\qb q\to G$}
In the $g\|\qb\|q$ limit we have the factorization
\ba
\big|\mathcal{B}_{4}^{0}\big|^{2} &\xrightarrow{g\|\qb\|q}& \bs{P}_{g\qb q\to G} \big|\mathcal{A}_{4}^{0}\big|^{2},
\ea
where the full gluon matrix element is given by \eq{FullA4g0}. The full $g\qb q\to G$ splitting function has the decomposition
\ba
\bs{P}_{g\qb q\to G} = N_c T_R P_{g\qb q} - \frac{1}{N_c} T_R \t{P}_{g\qb q},
\ea
which gives the following factorization:
\ba
\big|\mathcal{B}_{4}^{0}\big|^{2} &\xrightarrow{g\|\qb\|q}& g_s^{8}(N_c^2-1)N_c^3 \Big[ T_R P_{g\qb q} \bs{A}_{4}^{0}-\frac{1}{N_c^2} T_R \t{P}_{g\qb q} \bs{A}_{4}^{0} \Big],
\ea
where $P_{g\qb q}$ is summed over the two orderings $g\qb q$ and $\qb qg$. In this limit $\t{\t{\bs{B}}}_{4}^{0}$ and $\bar{\bs{B}}_{4}^{0}$ reduce to all gluon matrix elements containing abelian gluon(s), which vanish by the Dual Ward Identity. At the $1/N_c^2$ colour level, the $\t{\bs{B}}_{4}^{0}$ does have some non-vanishing limits when the collinear gluon is abelian. In this limit it reduces exactly to $\bs{A}_{4}^{0}$, leaving no limit left over for the R-term.

\subsubsection{Double single collinear limits}
In the limit where distinct pairs of gluons become collinear, say $i\|j$ and $k\|l$, the full matrix element factorizes as
\ba
\big|\mathcal{B}_{4}^{0}\big|^{2} &\xrightarrow{i\|j,\, k\|l}& \bs{P}_{ij\to G}\bs{P}_{kl\to G} \big|\mathcal{B}_{2}^{0}\big|^{2}\nn\\
&=& g_s^8 (N_c^2-1) N_c^3 P_{ij\to G}P_{kl\to G} \Big[ \bs{B}_{2}^{0}-\frac{1}{N_c^2}\bar{\bs{B}}_{2}^{0} \Big].
\ea
Due to the lack of sufficient non-abelian gluons in $\t{\bs{B}}_{4}^{0}$ and $\bar{\bs{B}}_{4}^{0}$, we can immediately see the factorization of the R-term in this limit:
\ba
\bs{R}_4^0 &\xrightarrow{i\|j,\, k\|l}& P_{ij \to G} P_{kl \to G} \bar{\bs{B}}_2^0.
\ea
Alternatively, we could take the two $gg$-collinear limits in a sequential manner, using the result of the single $gg$-collinear factorization of the R-term, which gives the same result. Since the R-term does not have any singular $qg$-collinear limits, even after one $gg$-collinear limit, we can also conclude that it has no double $gg$,$qg$-collinear or double $qg$,$qg$-collinear limit. In the same way, the double $\qb q, gg$-collinear limit is absent because the R-term has no $\qb q$ limit.

\subsubsection{Double soft limit} \label{sec:DoubleSoft}
For the double soft limit of the R-term it is necessary to look into the double soft limit at the amplitude level. The double soft limit of a colour-ordered amplitude where the two soft gluons are not colour connected simply results in two single soft currents,
\ba
\mathcal{M}_n^{0}(\dots,a,i,b,\dots,c,j,d,\dots)&\xrightarrow{i,j\to 0}&\mathcal{S}_\mu(a,i,b)\mathcal{S}_\nu(c,j,d)\ep^{\mu}\ep^{\nu}\mathcal{M}_{n-2}^{0}(\dots,a,b\dots,c,d,\dots). \nn\\
\ea
In the colour-connected double soft limit we have a factorization into the double soft current~\cite{Catani:1999ss,Berends:1988zn,Weinzierl:2003fx},
\ba
\mathcal{M}_n^{0}(\dots,a,i,j,b,\dots)&\xrightarrow{i,j\to 0}&\mathcal{S}_{\mu\nu}(a,i,j,b)\ep^{\mu}\ep^{\nu}\mathcal{M}_{n-2}^{0}(\dots,a,b,\dots),
\ea
where
\ba \label{eq:Smunu}
\mathcal{S}_{\mu\nu}(a,i,j,b)&=&4\bigg( \frac{1}{s_{ij}(s_{ai}+s_{aj})}\big(a_{\mu}i_{\nu}-a_{\nu}j_{\mu}\big)+\frac{1}{s_{ij}(s_{bi}+s_{bj})}\big(b_{\nu}j_{\mu}-b_{\mu}i_{\nu}\big) \nn\\
&+& g_{\mu\nu}\frac{(s_{aj}s_{bi}-s_{ai}s_{bj})}{2s_{ij}(s_{ai}+s_{aj})(s_{bi}+s_{bj})}\nn\\
&+& a_{\mu}a_{\nu}\frac{1}{s_{ai}(s_{ai}+s_{aj})}+b_{\mu}b_{\nu}\frac{1}{s_{bj}(s_{bi}+s_{bj})}-a_{\mu}b_{\nu}\frac{1}{s_{ai}s_{bj}}\bigg).
\ea
Squaring the double soft amplitude and averaging over the soft gluon polarizations for a coherent ordering gives by definition the double soft function
\ba
S_{aijb} = \frac{1}{4}d^{\mu\rho}d^{\nu\si}\mathcal{S}_{\mu\nu}(a,i,j,b)\mathcal{S}_{\rho\si}(a,i,j,b).
\ea

The R-term consists of many interference terms, and for a generic interference amplitude, we now have several possibilities in the double soft limit:
\begin{enumerate}
	\item Coherent colour-connected ordering: 
	\ba
	\mathcal{M}_n^{\dagger}(\dots,a,i,j,b,\dots) \mathcal{M}_n(\dots,a,i,j,b,\dots) \xrightarrow{i,j\to 0}  S_{\mu\nu}(a,i,j,b) S^{\mu\nu}(a,i,j,b) M_{n-2}. \nn\\
	\ea
	\item Colour-connected gluons between different partons:
	\ba
	\mathcal{M}_n^{\dagger}(\dots,a,i,j,b,\dots) \mathcal{M}_n(\dots,c,i,j,d,\dots) \xrightarrow{i,j\to 0}  S_{\mu\nu}(a,i,j,b) S^{\mu\nu}(c,i,j,d) M_{n-2}. \nn\\
	\ea
	\item Colour-disconnected times colour-connected: 
	\ba
	\mathcal{M}_n^{\dagger}(\dots,a,i,b,\dots,c,j,d,\dots) \mathcal{M}_n(\dots,e,i,j,f,\dots) \xrightarrow{i,j\to 0} \nn\\
	S_\mu(a,i,b) S_\nu(c,j,d) S^{\mu\nu}(e,i,j,f) M_{n-2}.
	\ea
	\item All colour-disconnected: 
	\ba
	\mathcal{M}_n^{\dagger}(\dots,a,i,b,\dots,c,j,d,\dots) \mathcal{M}_n(\dots,e,i,f,\dots,g,j,h,\dots) \xrightarrow{i,j\to 0} \nn\\
	S_\mu(a,i,b)S_\nu(c,j,d) S^\mu(e,i,f) S^\nu(g,j,h) M_{n-2}.
	\ea
\end{enumerate}
Note that in the first and second possibilities the gluons in one of the amplitudes could also be in reverse order with respect to the other amplitude. We will touch upon these cases later on and see that they can be converted to the other possibilities. For possibility 1, the double soft amplitude can be squared and in \sec{SingleSoft} it was shown how the interference of single soft currents in possibility 4 can be expressed in terms of soft eikonal factors. This is necessary for the antenna subtraction method, because the antenna functions are only able to capture the soft singular limits in the form of squared soft factors. In contrast to the single soft current, it is now not straightforward to express the type of soft interferences from possibility 2 and 3 in terms of known squared factors, so we cannot easily identify which antenna functions are able to capture all the double soft limits. Therefore, we seek properties for the double soft current as was done for the single soft current to help simplify all the generic interferences of single and double soft amplitudes of the R-term into products of soft eikonal factors and double soft factors.

The double soft current satisfies a decoupling identity when symmetrized over soft gluons:
\ba
\mathcal{S}_{\mu\nu}(a,i,j,b)+\mathcal{S}_{\nu\mu}(a,j,i,b)&=&\mathcal{S}_{\mu}(a,i,b)\mathcal{S}_{\nu}(a,j,b),
\ea
which can easily be seen by the fact that the first three terms in \eq{Smunu} are antisymmetric under $(i\leftrightarrow j,\mu\leftrightarrow\nu)$ and by using the partial fractioning in the fourth and fifth terms to eliminate the three-particle denominators.

The decoupling identity can be used to derive the square of the interference of two double soft currents which differ only by soft gluon orderings,
\ba
\sum_{(i,j)}\frac{1}{4}d^{\mu\rho}d^{\nu\si}\mathcal{S}_{\mu\nu}(a,i,j,b)\mathcal{S}_{\si\rho}(a,j,i,b)&=&S_{aib}S_{ajb}-S_{aijb}-S_{ajib},
\ea
where $\sum_{(i,j)}$ denotes the sum over the $i \leftrightarrow j$ swapped term with also the indices of the double soft currents swapped ($\mu \leftrightarrow \nu$, $\rho \leftrightarrow \si$). Examining the double soft current further, it is obvious that it also possesses a reflection property,
\ba
\mathcal{S}_{\mu\nu}(a,i,j,b)&=&\mathcal{S}_{\nu\mu}(b,j,i,a),
\ea
Using the decoupling identity we can then obtain the expression for the double soft current with swapped hard partons, analogous to the reflection property of the single soft current,
\ba
\mathcal{S}_{\mu\nu}(b,i,j,a)&=&\mathcal{S}_{\nu\mu}(a,j,i,b)\nn\\
&=&-\mathcal{S}_{\mu\nu}(a,i,j,b)+\mathcal{S}_{\mu}(a,i,b)\mathcal{S}_{\nu}(a,j,b).
\ea
The double soft current also satisfies a left shuffle identity,
\ba \label{eq:lshuf}
\mathcal{S}_{\mu\nu}(a,i,j,b)&=&\mathcal{S}_{\mu\nu}(c,i,j,b)-\mathcal{S}_{\mu\nu}(c,i,j,a)-\mathcal{S}_{\mu}(c,i,a)\mathcal{S}_{\nu}(a,j,b),
\ea
and a right shuffle identity follows by using the reflection, decoupling and left shuffle properties:
\ba \label{eq:rshuf}
\mathcal{S}_{\mu\nu}(a,i,j,b)&=&\mathcal{S}_{\mu\nu}(a,i,j,c)+\mathcal{S}_{\mu\nu}(c,i,j,b)+\mathcal{S}_{\mu}(a,i,c)\mathcal{S}_{\nu}(c,j,b).
\ea
Using the decoupling identity (or by relabelling), we can also write down shuffle identities for the currents with reversed soft gluons,
\ba \label{eq:shufrev}
\mathcal{S}_{\nu\mu}(a,j,i,b)&=&\mathcal{S}_{\nu\mu}(c,j,i,b)-\mathcal{S}_{\nu\mu}(c,j,i,a)-\mathcal{S}_{\mu}(b,i,a)\mathcal{S}_{\nu}(a,j,c),\nn\\
\mathcal{S}_{\nu\mu}(a,j,i,b)&=&\mathcal{S}_{\nu\mu}(a,j,i,c)+\mathcal{S}_{\nu\mu}(c,j,i,b)+\mathcal{S}_{\mu}(b,i,c)\mathcal{S}_{\nu}(c,j,a).
\ea 

There are two relations corresponding to the two possible orderings of the symmetrized soft gluons,
\ba
\mathcal{S}_{\mu\nu}(a,i,j,b)\mathcal{S}^{\mu\nu}(c,i,j,d)+\mathcal{S}_{\nu\mu}(a,j,i,b)\mathcal{S}^{\nu\mu}(c,j,i,d), \nn\\
\mathcal{S}_{\mu\nu}(a,i,j,b)\mathcal{S}^{\nu\mu}(c,j,i,d)+\mathcal{S}_{\nu\mu}(a,j,i,b)\mathcal{S}^{\mu\nu}(c,i,j,d).
\ea
The second configuration, where the soft gluon ordering is different in the two interfered currents, can be related to the first configuration using the decoupling identity,
\ba
\sum_{(i,j)}\mathcal{S}_{\mu\nu}(a,i,j,b)\mathcal{S}^{\nu\mu}(c,j,i,d)&=&\sum_{(i,j)}\mathcal{S}_{\mu\nu}(a,i,j,b)\Big[-\mathcal{S}^{\mu\nu}(c,i,j,d)+\mathcal{S}^{\mu}(c,i,d)\mathcal{S}^{\nu}(c,j,d)\Big]\nn\\
&=&\sum_{(i,j)}-\mathcal{S}_{\mu\nu}(a,i,j,b)\mathcal{S}^{\mu\nu}(c,i,j,d)\nn\\
&&+\mathcal{S}^{\mu}(c,i,d)\mathcal{S}^{\nu}(c,j,d)\sum_{(i,j)}\mathcal{S}_{\mu\nu}(a,i,j,b)\nn\\
&=&\sum_{(i,j)}-\mathcal{S}_{\mu\nu}(a,i,j,b)\mathcal{S}^{\mu\nu}(c,i,j,d)\nn\\
&&+\mathcal{S}_{\mu}(a,i,b)\mathcal{S}_{\nu}(a,j,b)\mathcal{S}^{\mu}(c,i,d)\mathcal{S}^{\nu}(c,j,d),
\ea
and so the two configurations are only different by products of single soft currents, which can be expressed as sums
of soft factors,
\ba
\lefteqn{\mathcal{S}_{\mu}(a,i,b)\mathcal{S}_{\nu}(a,j,b)\mathcal{S}^{\mu}(c,i,d)\mathcal{S}^{\nu}(c,j,d)=}\nn\\
&&\frac{1}{4}\big(S_{aid}+S_{bic}-S_{aic}-S_{bid}\big) \big(S_{ajd}+S_{bjc}-S_{ajc}-S_{bjd}\big).
\ea

The different identities derived in this section and in \sec{SingleSoft} are sufficient to re-express the double soft limit of the R-term into a suitable form for antenna subtraction:
\ba \label{eq:Rtermsoftlim}
\bs{R}_4^0 \xrightarrow{k,l\to 0}
\big(
&-& S_{1k2} S_{1li}
+ S_{1k2} S_{1lj}
+ S_{1k2} S_{2li}
- S_{1k2} S_{2lj}
-2S_{1k2} S_{ilj} \nn\\
&-& S_{1ki} S_{1l2}
+ S_{1ki} S_{1lj}
+ S_{1ki} S_{2lj} \nn\\
&+& S_{1kj} S_{1l2}
+ S_{1kj} S_{1li}
-2S_{1kj} S_{1lj}
- S_{1kj} S_{2li}
+2S_{1kj} S_{ilj} \nn\\
&+& S_{2ki} S_{1l2}
- S_{2ki} S_{1lj}
-2S_{2ki} S_{2li}
+ S_{2ki} S_{2lj}
+2S_{2ki} S_{ilj} \nn\\
&-& S_{2kj} S_{1l2}
+ S_{2kj} S_{1li}
+ S_{2kj} S_{2li} \nn\\
&-&2S_{ikj} S_{1l2}
+2S_{ikj} S_{1lj}
+2S_{ikj} S_{2li}
-2S_{ikj} S_{ilj}
\big) B_2^0(1,i,j,2)  + (i \leftrightarrow j) \nn\\
+ \big(
&-&2 S_{1kl2}
-2 S_{1lk2} \nn\\
&+&2 S_{1k2} S_{1l2}
-2 S_{1k2} S_{ilj} \nn\\
&-& S_{1ki} S_{2li}
+ S_{1ki} S_{2lj}
+ S_{1ki} S_{ilj} \nn\\
&+& S_{1kj} S_{2li}
- S_{1kj} S_{2lj}
+ S_{1kj} S_{ilj} \nn\\
&-& S_{2ki} S_{1li}
+ S_{2ki} S_{1lj}
+ S_{2ki} S_{ilj} \nn\\
&+& S_{2kj} S_{1li}
- S_{2kj} S_{1lj}
+ S_{2kj} S_{ilj} \nn\\
&-& 2 S_{ikj} S_{1l2}
+ S_{ikj} S_{1li}
+ S_{ikj} S_{1lj}
+ S_{ikj} S_{2li}
+ S_{ikj} S_{2lj}
-2 S_{ikj} S_{ilj}
\big) \nn\\
&& \times \t{B}_{2,R}^0(1,i,j,2),
\ea
where we see that all the products of the single and double soft currents at amplitude level have been re-expressed in terms of only iterated single eikonal factors and double soft factors at the squared matrix element level. The reduced amplitudes have been squared to leading colour functions $B_2^0$ or are present as the interference matrix element $\t{B}_{2,R}^0$. Each soft factor can now be promoted to a corresponding antenna function to obtain the first candidate subtraction for the R-term. Focusing only at the double soft factors, we see that we get the following $X_4^0$ terms:
\ba
-2 A_4^0(1,k,l,2) \t{B}_{2,R}^0([1],i,j,[2]) \quad \mbox{and} \quad
-2 A_4^0(1,l,k,2) \t{B}_{2,R}^0([1],i,j,[2]).
\ea 
If we also promote the $2\,S_{1k2} S_{1l2} \t{B}_{2,R}^0(1,i,j,2)$ term to $ 2\,\t{A}_4^0(1,k,l,2) \t{B}_{2,R}^0([1],i,j,[2])$, we see that we end up with exactly the correct structure of $X_4^0$ terms we needed to subtract all the triple collinear limits of the R-term. The construction of $\df \si^{S,b_1}$ has been completed, and having found all the functions to account for the soft limits, the construction for the rest of $\df \si^S$ can continue in the same manner as for the other B-type matrix elements. Dressing the $\df \si^{S,b_1}$ terms with $X_3^0 X_3^0$, adding $\df \si^{S,a}$ and the newly $\df \si^{S,c}$ and $\df \si^{S,d}$ terms obtained from \eq{Rtermsoftlim}, all the single collinear limits of the R-term are also correctly reproduced. Due to the cross talk of the several terms with mapped momenta, the soft limits are however not correctly subtracted anymore, but this is fixed by the usual process of adding LAS terms~\cite{Currie:2013vh}, thereby completing the construction of $\df \si^S$ for the R-term.

\subsubsection{Tests of the RR subtraction term}
\begin{figure}[t]
	\centering
	\includegraphics[width=0.49\textwidth]{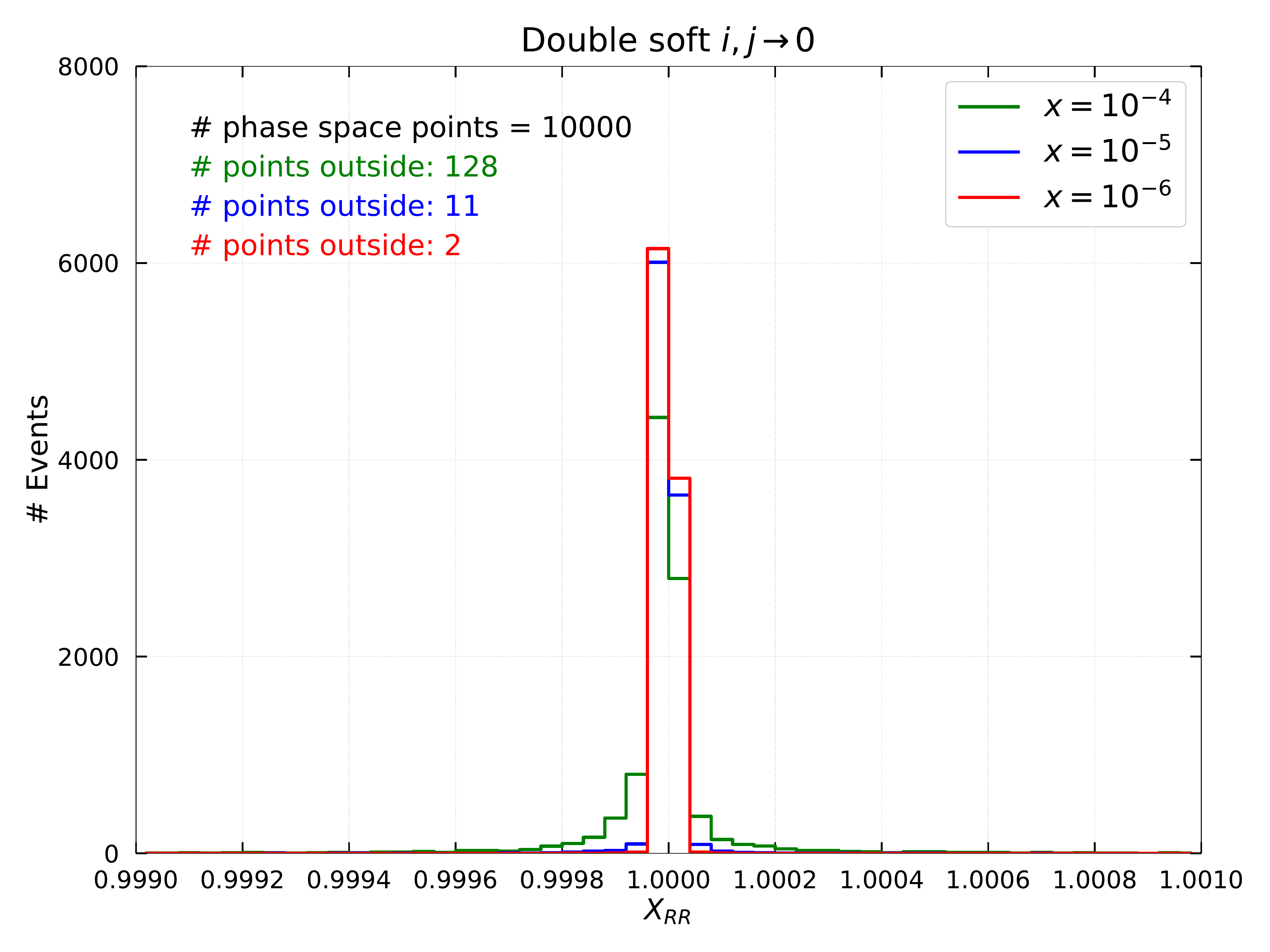} 
	\includegraphics[width=0.49\textwidth]{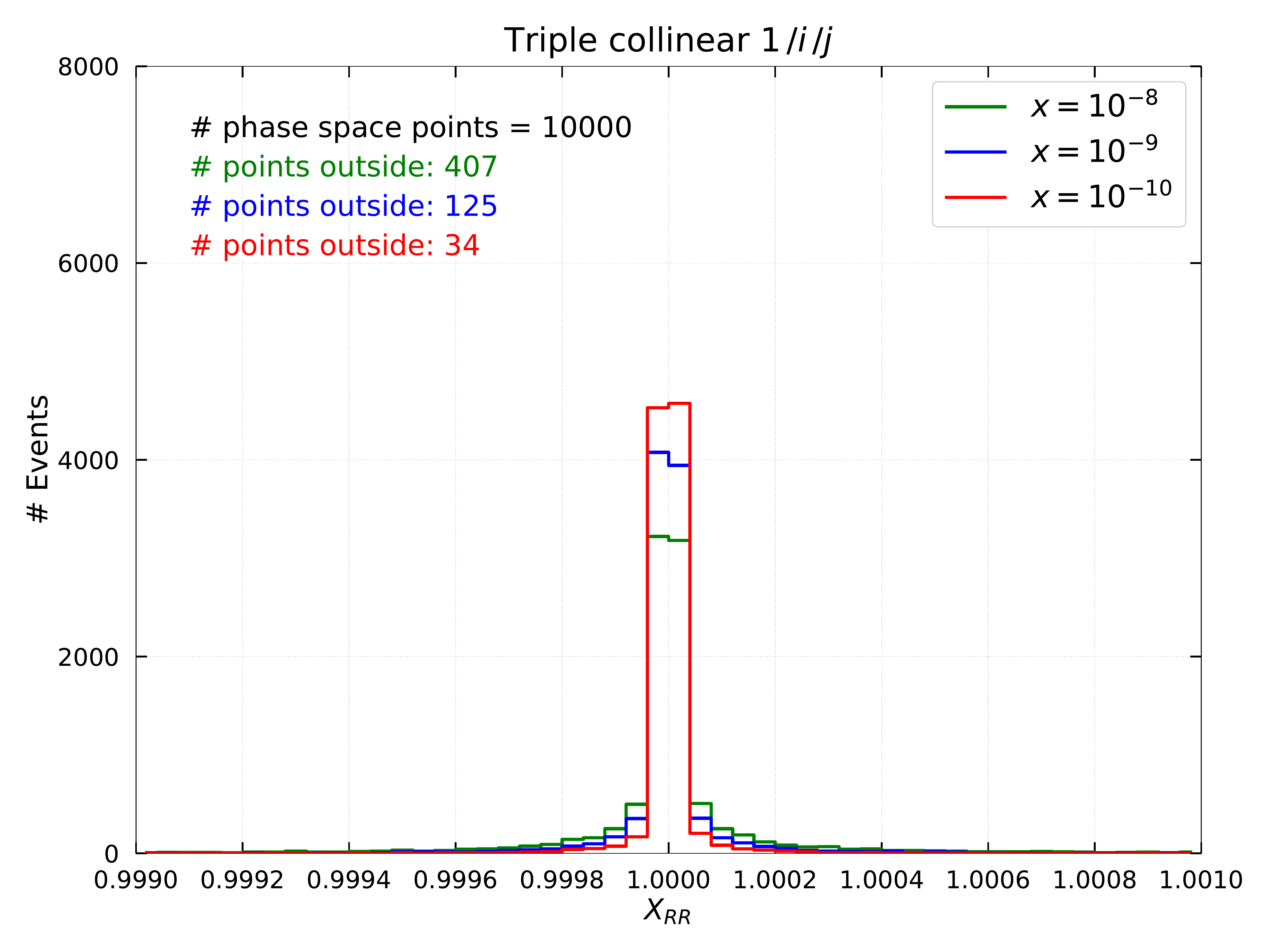}
	\caption{Spike plots of the ratio $\df \si^{\mathrm{RR}} / \df \si^S$ verifying the correct behaviour of the subtraction terms for the R-term in the double soft gluons (left) and triple collinear $q\|g\|g$ (right) limits.}
	\label{fig:qqbR4g0_spikeplots}
\end{figure}

To assess how well the subtraction term $\df \si^S$ approaches the matrix element $\df \si^{RR}$ that it is supposed to subtract in unresolved limits, we take the double soft gluons and $q\|g\|g$ triple collinear limits for the R-term as examples. The validity of the subtraction term is tested as described in~\cite{NigelGlover:2010kwr} by generating a number of phase space points approaching the considered IR limit at $\sqrt{s} = 1000$ GeV, and for each point the following ratio is calculated:
\ba
X_{RR} = \frac{\df \si^{RR}}{\df \si^S}.
\ea
This ratio is expected to converge to unity as the points go deeper into the IR limit. The depth of the IR divergence is controlled by a variable $x$, which is defined in terms of momenta invariants in a suitable way for each considered unresolved limit. For the exemplary double soft $i,j \to 0$ and triple collinear $1\|i\|j$ limits in the $1,2\rightarrow i,j,k,l$ scattering processes, $x$ is defined as $x = 1-s_{kl}/s$ and $x = s_{1ij}/s$ respectively. Three different values for $x$ are considered for each limit, with smaller $x$ representing larger IR depth. The results for $X_{RR}$ are histogrammed and shown in \fig{qqbR4g0_spikeplots} where the number of points which fall outside the plot range are also denoted. For both limits we observe narrower spikes around unity and less outliers as $x$ becomes smaller, meaning the subtraction term mimics the matrix element better as the points become closer to the IR singularity. After the full subtraction term $\df \si^{S}$ successfully produces sufficiently narrow spikes in the spike plots for each possible unresolved limit, $\df \si^{S}$ can be considered a working subtraction term.    
%

\section{Real-virtual subtraction} \label{sec:RVsub}
All five-parton one-loop matrix elements that were introduced in \sec{MEs} enter the real-virtual contribution to the dijet cross section. The antenna subtraction terms that are required for their numerical implementation are discussed in the following, focusing in particular on novel aspects that first appear for processes with four external partons at Born level. 

Since the RV corrections are $(n+1)$-parton matrix elements, we now only need to consider single unresolved limits. For the one-loop matrix elements we have the following factorization in soft limits~\cite{Bern:1999ry,Catani:2000pi}:
\ba
M_{n+1}^{1}(\dots,a,b,c,\dots) \xrightarrow{b\to 0} S_{abc} M_{n}^1(\dots,a,c,\dots) + S_{abc}^{(1)}(\ep) M_{n}^0(\dots,a,c,\dots),
\ea
and similarly for collinear limits~\cite{Kosower:1999rx,Bern:1999ry}:
\ba
M_{n+1}^{1}(\dots,i,j,\dots) \xrightarrow{i\|j} \frac{P_{ij\to K}}{s_{ij}} M_{n}^1(\dots,K,\dots) + \frac{P_{ij\to K}^{(1)}}{s_{ij}} M_{n}^0(\dots,K,\dots),
\ea
where we have the same unresolved factors from the tree-level factorization, but now factoring onto a reduced one-loop matrix element, and additional terms which have the form of a one-loop soft factor~\cite{Bern:1998sc} or one-loop splitting function~\cite{Bern:1999ry,Bern:1994zx} times a reduced tree-level matrix element. In the antenna subtraction method, these unresolved limits of the RV corrections are subtracted in the same way by ``tree times loop'' ($T \times L$) $X_3^0 M_n^1$ terms and ``loop times tree'' ($L\times T$) $X_3^1 M_n^0$ terms, where $X_3^1$ is a one-loop antenna function~\cite{Gehrmann-DeRidder:2005btv}. Besides mimicking the behaviour of the matrix element in unresolved limits, the subtraction term also needs to cancel the explicit $\ep$-poles of the one-loop matrix element, which can be obtained from the relevant one-loop amplitude expressions in \sec{MEs}. 

Similar as for the RR level, we partition the whole subtraction term into the following parts~\cite{Currie:2013vh}:
\ba
\df \si^T = \df \si^{T,a} + \df \si^{T,b} + \df \si^{T,c} + \df \si^{T,e},
\ea
and briefly describe each part:
\begin{itemize}
	\item $\df \si^{T,a}$: this part consists of the integrated $\df \si^{S,a}$ terms:
	\ba
	\df \si^{T,a} \supset \mathcal{N}_{RV} \int \frac{\df x_1}{x_1} \frac{\df x_2}{x_2} \df \Phi_{n+1} \, \mathcal{X}_3^0(s_{ij}) M_{n+1}^0(\{p_{n+1}\}) J_n^{(n+1)} (\{p_{n+1}\}),
	\ea
	and removes the explicit $\ep$-poles of the RV squared matrix element. The calligraphic $\mathcal{X}_3^0$ denotes an integrated antenna and the reduced matrix elements here are $(n+1)$-parton squared matrix elements which can 
	develop single unresolved limits. These limits, not associated with the RV matrix element, have to be taken care of by subtraction terms in the other parts. $\df \si^{T,a}$ combines with $\df \si^{\MF,1}$ to yield the single integrated dipole factors $\mathcal{J}_2^{(1)}$ that are listed in \app{J21}, multiplying the reduced squared matrix elements. 
	\item $\df \si^{T,b}$: this part is responsible for subtracting the single unresolved limits of the RV matrix element. The one-loop antennae and reduced matrix elements can contain explicit $\ep$-poles which are not associated to the RV matrix element. These poles are removed by terms of the form $\mathcal{X}_3^0 X_3^0 M_n^0$. We thus further partition $\df \si^{T,b}$ into $\df \si^{T,b_1}$ and $\df \si^{T,b_2}$ parts, corresponding to the $L \times T$, $T \times L$ terms and the terms cancelling the poles of the one-loop antennae and reduced matrix elements:
	\ba
	\df \si^{T,b_1} &\supset& \mathcal{N}_{RV} \int \frac{\df x_1}{x_1} \frac{\df x_2}{x_2} \df \Phi_{n+1} \, X_3^1(\{p_{n+1}\}) \de(1-x_1) \de(1-x_2) M_n^0(\{\t{p}_n\}) J_n^{(n)}(\{\t{p}_n\}) \nn\\
	&+&  \mathcal{N}_{RV} \int \frac{\df x_1}{x_1} \frac{\df x_2}{x_2} \df \Phi_{n+1} \, X_3^0(\{p_{n+1}\}) \de(1-x_1) \de(1-x_2) M_n^1(\{\t{p}_n\}) J_n^{(n)}(\{\t{p}_n\}), \nn\\
	\df \si^{T,b_2} &\supset& \mathcal{N}_{RV} \int \frac{\df x_1}{x_1} \frac{\df x_2}{x_2} \df \Phi_{n+1} \, \mathcal{X}_3^0(s_{ij}) X_3^0(\{p_{n+1}\}) M_n^0(\{\t{p}_n\}) J_n^{(n)}(\{\t{p}_n\}).
	\ea 
	The one-loop antennae in the $\df \si^{T,b_1}$ subtraction are renormalized at the scale of the invariant mass of their antenna partons $s_{abc}$, while the one-loop matrix elements are renormalized at the renormalization scale $\mu_r^2$. This mismatch is corrected by the following replacement:
	\ba
	X_3^1(a,b,c) \to X_3^1(a,b,c) + \frac{b_0}{\ep} X_3^0(a,b,c) \l( \l( \frac{|s_{abc}|}{\mu_r^2} \r)^{-\ep} - 1 \r).
	\ea
	\item $\df \si^{T,c}$: this part cancels the single unresolved limits of the reduced squared matrix elements in $\df \si^{T,a}$ and consists of terms of the form:
	\ba
	\df \si^{T,c} \supset \mathcal{N}_{RV} \int \frac{\df x_1}{x_1} \frac{\df x_2}{x_2} \df \Phi_{n+1} \,  \mathcal{X}_3^0(s_{ij}) X_3^0(\{p_{n+1}\}) M_n^0(\{\t{p}_n\}) J_n^{(n)}(\{\t{p}_n\}).
	\ea	
	These terms partly originate from integration of  $\df \si^{S,c}$ terms over the antenna phase space of the first $X_3^0$, and partly of terms newly introduced here, which will consequently be subtracted again at the double virtual level.
	\item $\df \si^{T,e}$: the soft functions in the LAS terms are integrated and added back at the RV level:	
	\ba
	\df \si^{T,e} \supset \mathcal{N}_{RV} \int \frac{\df x_1}{x_1} \frac{\df x_2}{x_2} \df \Phi_{n+1} \,  \mathcal{S}^{FF} X_3^0(\{p_{n+1}\}) M_n^0(\{\t{p}_n\}) J_n^{(n)}(\{\t{p}_n\}).
	\ea
\end{itemize}
Note the absence of a $\df \si^{T,d}$ part corresponding to integration of $\df \si^{S,d}$, since the colour-disconnected terms are usually added back at the VV level. 

Novel features of the RV subtraction terms for dijet processes are discussed on two examples in the following. 

\subsection{RV subtraction for the subsubleading colour C-type matrix element} \label{sec:CtypeRVsub}
We first discuss the antenna subtraction for the subsubleading colour (SSLC) C-type matrix element $\t{\t{C}}_1^1(q,\qb,Q,\Qb,g)$ defined in \eq{FullC1g1MEsimple}, as this matrix element still admits a relatively compact form of its subtraction terms while 
already exposing some of the key features of the RV subtraction. For definiteness, we take the crossing where $q,\Qb$ are in the initial state and use the notational abbreviation $(1,i,j,2,k) = (q,\qb,Q,\Qb,g)$ in this section.

The first step is to cancel the explicit $\ep$-poles of the matrix element by $\df \si^{T,a}$. The pole structure for the full SSLC level is given by
\ba \label{eq:Ctt1g1poles}
\mathrm{Poles}\big[\t{\t{C}}_1^1(1,i,j,2,k)\big] &=& 2\, \big[I_{q\qb}^{(1)}(\ep,s_{1i}) + I_{q\qb}^{(1)}(\ep,s_{2j}) + 2 I_{q\qb}^{(1)}(\ep,s_{12}) + 2 I_{q\qb}^{(1)}(\ep,s_{ij}) \nn\\
&-& 2 I_{q\qb}^{(1)}(\ep,s_{1j}) - 2 I_{q\qb}^{(1)}(\ep,s_{2i}) \big] C_{1,\ga}^0(1,i,j,2,k) \nn\\
&-& 2\, \big[ I_{q\qb}^{(1)}(\ep,s_{1i}) + I_{q\qb}^{(1)}(\ep,s_{2j}) + I_{q\qb}^{(1)}(\ep,s_{12}) + I_{q\qb}^{(1)}(\ep,s_{ij}) \nn\\
&-& I_{q\qb}^{(1)}(\ep,s_{1j}) - I_{q\qb}^{(1)}(\ep,s_{2i}) \big] \big( C_{1,c}^0(1,i,j,2,k)+C_{1,d}^0(1,i,j,2,k) \big). \nn\\
\ea
This expression can directly be used to construct $\df \si^{T,a}$ by using the known pole structures of the integrated three-parton antenna functions~\cite{Gehrmann-DeRidder:2005btv} and matching the poles. Taking also the mass factorization terms into account, we obtain the general expression:
\ba
\df \si^{T,a} = \sum_{i,j} c_{ij} \mathcal{J}_2^{(1)}(i,j) M_5^0,
\ea
where $c_{ij}$ denote numerical coefficients and $\mathcal{J}_2^{(1)}$ are the single integrated dipoles given in \app{J21} which contain integrated $X_3^0$ antennae and mass factorization kernels. One could then also unintegrate $\df \si^{T,a}$ to obtain $\df \si^{S,a}$ which subtracts all single unresolved limits of the corresponding matrix element which forms the double real correction ($\t{\t{C}}_2^0$ in this example). This approach from using the infrared singularity structure of the virtual corrections to construct the virtual subtraction term first, and subsequently unintegrating to obtain the real subtraction term~\cite{Chen:2022ktf} works well at NLO due to its relative simplicity. 

After achieving the explicit pole cancellation, we look at the single unresolved limits of the RV matrix element. The single collinear quark-antiquark limits of $\t{\t{C}}_1^1(q,\qb,Q,\Qb,g)$ are easily reproduced by the following terms:
\ba
\df \si^{T,b_1} \supset
&-& \frac{1}{2} E_{3,qq'}^0(1,j,2)  \t{\t{B}}_2^1([1],k,[2],i) \nn\\
&-&\frac{1}{2} E_{3,q'}^0(i,j,2)  \t{\t{B}}_2^1(1,k,[2],[i,j]) \nn\\
&-&\frac{1}{2} E_{3,q'}^0(j,1,i)  \t{\t{B}}_2^1([j,i],k,[1],2) \nn\\
&-&\frac{1}{2} E_{3,qq'}^0(2,1,i)  \t{\t{B}}_2^1(j,k,[1],[2]) \nn\\
\nn\\
&-&\frac{1}{2} \t{E}_{3,qq'}^1(1,j,2)  \bar{B}_2^0([1],k,[2],i) \nn\\
&-&\frac{1}{2} \t{E}_{3,q'}^1(i,j,2)  \bar{B}_2^0(1,k,[2],[i,j]) \nn\\
&-&\frac{1}{2} \t{E}_{3,q'}^1(j,1,i)  \bar{B}_2^0([j,i],k,[1],2) \nn\\
&-&\frac{1}{2} \t{E}_{3,qq'}^1(2,1,i)  \bar{B}_2^0(j,k,[1],[2]).
\ea
The unresolved limits involving the gluon, especially the soft limit, are less straightforward due to similar reasons as for the double real subtraction. Here we make use of the known pole structures of the antenna functions and matrix element to find the correct subtraction terms for the soft gluon limit. To construct these subtraction terms we take the soft gluon $k$ limit of \eq{Ctt1g1poles}. For a generic five-parton one-loop colour-ordered matrix element, one then obtains an expression of a sum of products of infrared singularity operators $I^{(1)}$, eikonal factors $S$ and four-parton reduced tree-level matrix elements:
\ba
M_5^1(1,i,j,2,k) \xrightarrow[k \,\mathrm{soft}]{\mathrm{sing.}} \sum_{a,b \in \{1,i,j,2\}} \sum_{c,d \in \{1,i,j,k,2\}} \sum_e C_{abcde} S_{akb} I_{pp' \in\{q,g\}}^{(1)}(\ep,s_{cd}) M^0_{4,e}, \nn\\
\ea
where $C_{abcde}$ contains the relevant colour factors and integer multiplicity factors, depending on the involved partons and reduced matrix element. For the case of the SSLC C-type matrix element in this example, this expression can still be written in a relatively compact form:
\ba \label{eq:Ctt1g1softsing}
&& \mathrm{Poles}\big[\t{\t{C}}_1^1(1,i,j,2,k)\big]\big|_{k\to 0} = \nn\\
&& + \big[ 2 I_{q\qb}^{(1)}(\ep,s_{1i}) + 2 I_{q\qb}^{(1)}(\ep,s_{2j}) + 4 I_{q\qb}^{(1)}(\ep,s_{12}) + 4 I_{q\qb}^{(1)}(\ep,s_{ij}) - 4 I_{q\qb}^{(1)}(\ep,s_{1j}) - 4 I_{q\qb}^{(1)}(\ep,s_{2i}) \big] \nn\\
&& \times \big( S_{1ki} + S_{2kj} \big) C_0^0(1,2,j,i) \nn\\
&& + \big[ 4 I_{q\qb}^{(1)}(\ep,s_{1i}) + 4 I_{q\qb}^{(1)}(\ep,s_{2j}) + 6 I_{q\qb}^{(1)}(\ep,s_{12}) + 6 I_{q\qb}^{(1)}(\ep,s_{ij}) - 6 I_{q\qb}^{(1)}(\ep,s_{1j}) - 6 I_{q\qb}^{(1)}(\ep,s_{2i}) \big] \nn\\
&& \times \big( S_{1k2} + S_{ikj} - S_{1kj} - S_{2ki} \big)  C_0^0(1,2,j,i).
\ea
The above expression has to be reproduced by the $L\times T$ and $T\times L$ subtraction terms in the same soft limit: eikonal factors by the antenna functions and the five-parton reduced matrix elements and the $I^{(1)}$ functions by the one-loop antennae in the $L\times T$ terms as well as the one-loop matrix elements in the $T\times L$ terms. The known pole structures of the one-loop antennae~\cite{Gehrmann-DeRidder:2005btv} and matrix elements in \sec{Cpoles}, as well as their known behaviour in the single soft limit then guides us in constructing the correct terms to match the behaviour in \eq{Ctt1g1softsing}. In this example the $L \times T$ terms are given by
\ba
\df \si^{T,b_1} \supset 
&+&\t{A}_{3,q}^1(1,k,i)  C_0^0([1],2,j,[i,k]) \nn\\
&+&\t{A}_{3,q}^1(2,k,j)  C_0^0(1,[2],[j,k],i) \nn\\
&+&2 \t{A}_{3,q\qb}^1(1,k,2)  C_0^0([1],[2],j,i) \nn\\
&+&2 \t{A}_3^1(i,k,j)  C_0^0(1,2,[j,k],[i,k]) \nn\\
&-&2 \t{A}_{3,q}^1(1,k,j)  C_0^0([1],2,[j,k],i) \nn\\
&-&2 \t{A}_{3,q}^1(2,k,i)  C_0^0(1,[2],j,[i,k]),
\ea
which do not contribute in the soft $k$ limit. The terms in \eq{Ctt1g1softsing} thus have to be fully accounted by the $T \times L$ terms. Here, we observe a similar pattern in the pole structures multiplying the soft factors and reduced matrix element as for the one-loop matrix element $\t{C}_0^1$ in \eq{Ct0g1poles}. For the first line in the right-hand side of \eq{Ctt1g1softsing} containing the soft factors which are even in the $2 \leftrightarrow j$ interchange, we can promote the soft factors to $A_3^0$ antennae and match the poles and reduced matrix element with $-\t{C}_0^1(1,2,j,i)$. However, for the second line containing a combination of soft factors odd under the $2 \leftrightarrow j$ interchange, the coefficients of the infrared operators do not match up with the one-loop four-parton matrix element. The reduced tree-level matrix element $C_0^0(1,2,j,i)$ does match up and we recall that it is invariant under a $2 \leftrightarrow j$ interchange. Thus, we can resolve this mismatch by using an antisymmetrized combination of the $\t{C}_0^1$ functions as follows:
\ba
\df \si^{T,b_1} \supset 
&-&A_{3,q}^0(1,k,i)  \t{C}_0^1([1],2,j,[i,k]) \nn\\
&-&A_{3,q}^0(2,k,j)  \t{C}_0^1(1,[2],[j,k],i) \nn\\
&-&\frac{7}{4} A_{3,q\qb}^0(1,k,2)  \t{C}_0^1([1],[2],j,i) \nn\\
&-&\frac{7}{4} A_3^0(i,k,j)  \t{C}_0^1(1,2,[j,k],[i,k]) \nn\\
&+&\frac{7}{4} A_{3,q}^0(1,k,j)  \t{C}_0^1([1],2,[j,k],i) \nn\\
&+&\frac{7}{4} A_{3,q}^0(2,k,i)  \t{C}_0^1(1,[2],j,[i,k]) \nn\\
\nn\\
&-&\frac{1}{4} A_{3,q\qb}^0(1,k,2)  \t{C}_0^1([1],j,[2],i) \nn\\
&-&\frac{1}{4} A_3^0(i,k,j)  \t{C}_0^1(1,[j,k],2,[i,k]) \nn\\
&+&\frac{1}{4} A_{3,q}^0(1,k,j)  \t{C}_0^1([1],[j,k],2,i) \nn\\
&+&\frac{1}{4} A_{3,q}^0(2,k,i)  \t{C}_0^1(1,j,[2],[i,k]),
\ea
where the $\t{C}_0^1$ in the second block have their secondary quarks interchanged with respect to the first block. Adding all $L \times T$ and $T \times L$ terms together completes the $\df \si^{T,b_1}$ part which reproduces all the collinear and soft limits of the real-virtual matrix element.

$\df \si^{T,b_1}$ is not free of $\ep$-poles, as it consists of one-loop antenna functions in the $L\times T$ terms and one-loop matrix elements in the $T\times L$ terms. The poles of the one-loop antenna functions are known~\cite{Gehrmann-DeRidder:2005btv} and the poles of the one-loop reduced matrix elements are given in \sec{B2g1Funcs}. Knowing the full pole structure, these poles are then subtracted by $\mathcal{X}_3^0 X_3^0 M_n$ terms, which come from integrated iterated $X_3^0 X_3^0 M_n$ terms from the RR subtraction or are newly introduced here at the RV level, until $\df \si^{T,b_1} + \si^{T,b_2}$ is free of $\ep$-poles. Similarly, $\df \si^{T,a}$ is not free of unresolved limits, as we can have single unresolved limits in the reduced five-parton tree-level matrix elements. These unresolved limits can again be subtracted by terms of the form $\mathcal{X}_3^0 X_3^0 M_n$. The remaining $\df \si^{T,b_2}$, $\df \si^{T,c}$ and $\df \si^{T,e}$ terms which account for these unresolved limits and poles not associated to the RV matrix elements are obtained using standard procedures~\cite{Currie:2013vh} and are not discussed here. 

We note that for one-loop matrix elements at subleading colour, the factorization in the soft gluon limits also factors onto interference terms. It is therefore usually necessary to partition the one-loop reduced matrix element into subfunctions encapsulating these interferences. For the C-type matrix elements in this example the (anti)-symmetrized combinations of $\t{C}_0^1$ coincide with the required subfunctions for the soft limit. For the D-type RV subtraction terms, we encounter a similar issue where we require the use of the subfunctions $D_{0,i}^1$ with $i \in \{a,b,c,d\}$ as defined in \eq{D0g1subfuncs}. In the B-type RV subtraction example we discuss next we will also need reduced one-loop interference matrix elements. 

\subsection{RV subtraction for the subleading colour B-type matrix element} \label{sec:BtypeRVsub}
In the following, we discuss the antenna subtraction for the first subleading colour level B-type RV squared matrix element $\t{B}_3^1$, which has as double real counter part the $\t{\bs{B}}_4^0$, $\bar{\bs{B}}_4^0$ in \eq{FullB4g0} as well as the interference terms $\bs{R}_4^0$ discussed in \secs{Rtermsub1}{Rtermsub2}. The pole structure of the RV correction can simply be obtained from the expressions in \sec{Btype1LoopPoles}. For the leading colour functions $B_3^1$ and $\hat{B}_3^1$ which consist of coherent one-loop primitive amplitudes interfered with tree-level amplitudes, the pole structure is particularly simple and transparent. This transparency in the pole structure and unresolved limits guides the construction of the LC subtraction terms. At the SLC levels however, the pole structure is much less transparent due to the subleading partial amplitudes and incoherent interferences, which complicates the construction of the SLC subtraction terms. The $\t{B}_3^1$ matrix element receives contributions from all the subleading colour partial amplitudes $\mathcal{B}_{3;3}$ and $\mathcal{B}_{3;4}$, as well as incoherent interferences of the leading partial amplitude $\mathcal{B}_{3;1}^1$. For definiteness, we take the quark pair in the initial state and use the shorthand notation $(1,i,j,k,2)$ for $(q,g_1,g_2,g_3,\qb)$. All other crossings follow the same construction principles.

The first step to cancel the explicit $\ep$-poles is achieved by integrating the corresponding $\df \si^{S,a}$ terms and adding them back here. For the $\t{B}_3^1(1,i,j,k,2)$, this implies contributions from the subtraction associated with $\t{\bs{B}}_4^0$, $\bar{\bs{B}}_4^0$ and the remainder interference terms $\bs{R}_4^0$. While these $\df \si^{T,a}$ terms at SLC are thus numerous integrated antenna functions factoring onto tree-level interference $B_{3,\mathrm{int}}^0$ terms, this procedure at SLC is no different than at LC, given that the 
RR subtraction term has been constructed previously. 

Next, we look at the unresolved limits of $\t{B}_3^1(1,i,j,k,2)$. The $L\times T$ terms can still be written in a compact form, as there are only few distinct four-parton reduced tree-level matrix elements. The following terms have the required one-loop soft and collinear behaviour: 
\ba
\df \si^{T,b} \supset \sum_{P(i,j,k)} \frac{1}{2} \Big[ && \nn\\
&+& A_3^1(1,k,2) \t{B}_{2,R}^0([1],i,j,[2]) \nn\\
&+& \t{A}_3^1(1,k,2) \t{B}_{2,R}^0([1],i,j,[2]) \nn\\
&+& \t{A}_3^1(1,k,2) \bar{B}_2^0([1],i,j,[2]) \nn\\
&-& d_{3,q}^1(1,k,i) \bar{B}_2^0([1],[i,k],j,2) \nn\\
&-& d_{3,q}^1(1,k,j) \bar{B}_2^0([1],i,[j,k],2) \nn\\
&-& d_{3,q}^1(2,k,i) \bar{B}_2^0(1,[i,k],j,[2]) \nn\\
&-& d_{3,q}^1(2,k,j) \bar{B}_2^0(1,i,[j,k],[2]) \Big].
\ea
In the crossings where at least one  quark are in the final state, it is necessary to also subtract the quark-antiquark limits with $-\t{G}_3^1(i,1,2) \bs{A}_4^0$.

The $T\times L$ terms are less compact. In the collinear limits, there is again a compact factorization of splitting functions times full four-parton one-loop squared matrix elements, e.g.
\ba
\t{B}_3^1(1,i,j,k,2) &\xrightarrow[T \times L]{1 \| i}& P_{qg\to Q} \big[ \t{B}_2^1((1i),j,k,2) - B_2^1((1i),j,k,2) - B_2^1(1i,k,j,2) \big], \nn\\
\t{B}_3^1(1,i,j,k,2) &\xrightarrow[T \times L]{i \| j}& P_{gg\to G} \big[ 2 \t{B}_2^1(1,(ij),k,2) \big].
\ea
On the other hand, the soft limit factors onto many different interferences and it becomes necessary to make use of all the subfunctions defined in \sec{B2g1Funcs}. In the soft limit we have the following $T\times L$ terms:
\ba
\t{B}_3^1(1,i,j,k,2) \xrightarrow[T \times L]{k\to 0} \frac{1}{2} \Big[
&& (- S_{1kj} - S_{2ki} + S_{ikj} ) B_2^1(1,i,j,2) \nn\\
&+& (- S_{1ki} - S_{2kj} + S_{ikj} ) B_2^1(1,j,i,2) \nn\\
&+& (S_{1k2} + S_{1ki} - S_{1kj} - S_{2ki} + S_{2kj} + 2 S_{ikj} ) \t{B}_{2,b}^1(1,i,j,2) \nn\\
&+& (S_{1k2} - S_{1ki} + S_{1kj} + S_{2ki} - S_{2kj} + 2 S_{ikj} ) \t{B}_{2,c}^1(1,i,j,2) \nn\\
&+& (S_{1ki} + S_{2kj} + S_{ikj}) \t{B}_{2,d}^1(1,i,j,2) \nn\\
&+& (-S_{1k2} + S_{1ki} + S_{1kj}+ S_{2ki} + S_{2kj} ) \t{B}_{2,e}^1(1,i,j,2) \nn\\
&+& (S_{1kj} + S_{2ki} + S_{ikj} ) \t{B}_{2,f}^1(1,i,j,2) \Big] \nn\\
&+& (i \leftrightarrow j).
\ea
Promoting the eikonal factors to tree-level antenna functions and adding the $L\times T$ terms then subtracts all the unresolved limits of $\t{B}_3^1(1,i,j,k,2)$, which concludes $\df \si^{T,b_1}$.

Although the expressions for $\df \si^{T,a}$ and $\df \si^{T,b}$ are less compact than what one usually finds at LC, knowing the behaviour and pole structure of each interference matrix element, the interplay between the different terms to cancel the unresolved limits and $\ep$-poles can be achieved in the same manner as at LC. The 
main novel feature here is the appearance of the subfunctions $\t{B}_{2,x}^1(1,i,j,2)$ in the reduced squared matrix elements of the $\df \si^{T,b}$ subtraction term.

\subsubsection{Tests of the RV subtraction term}
\begin{figure}[t]
	\centering
	\includegraphics[width=0.49\textwidth]{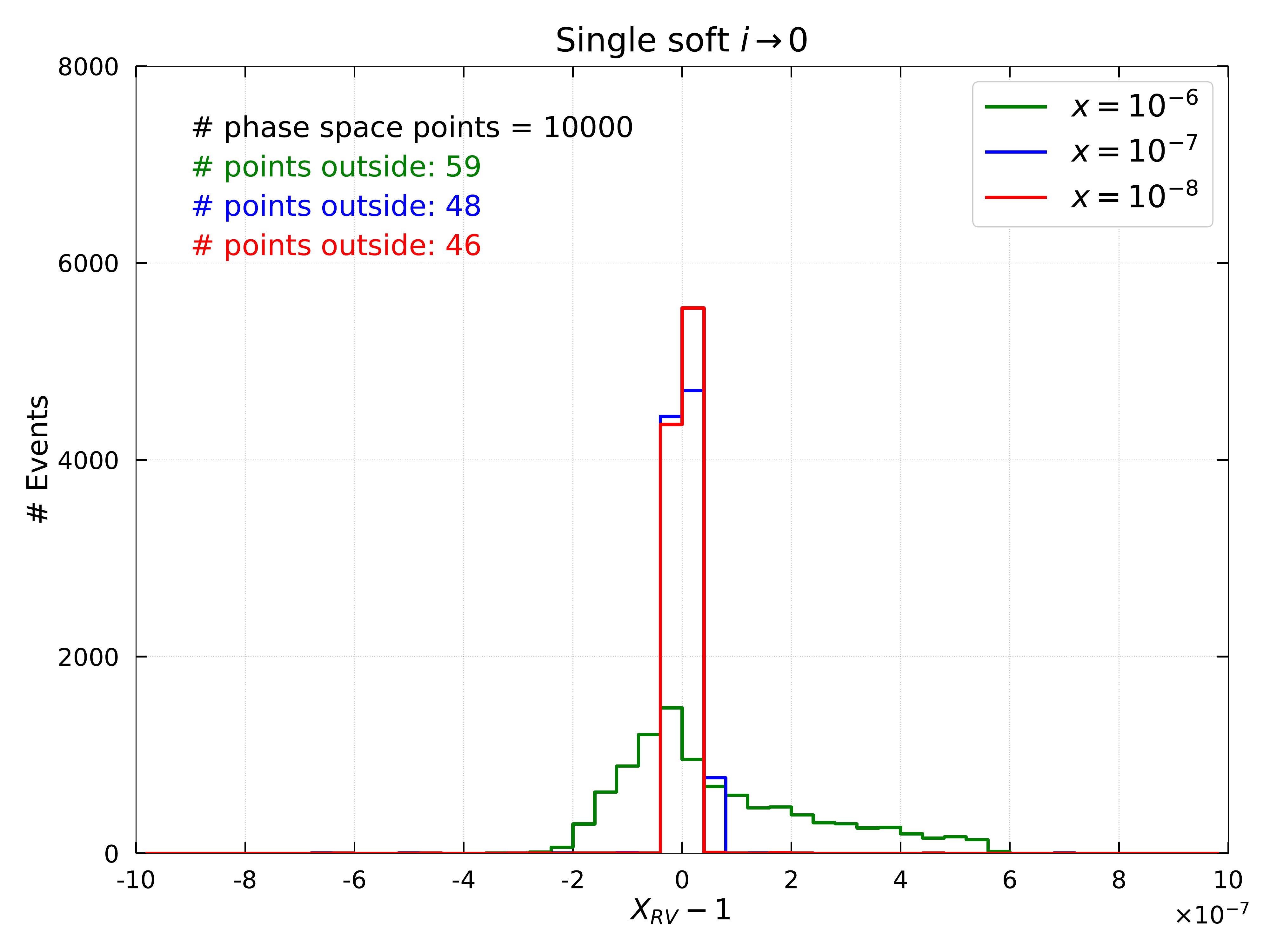} 
	\includegraphics[width=0.49\textwidth]{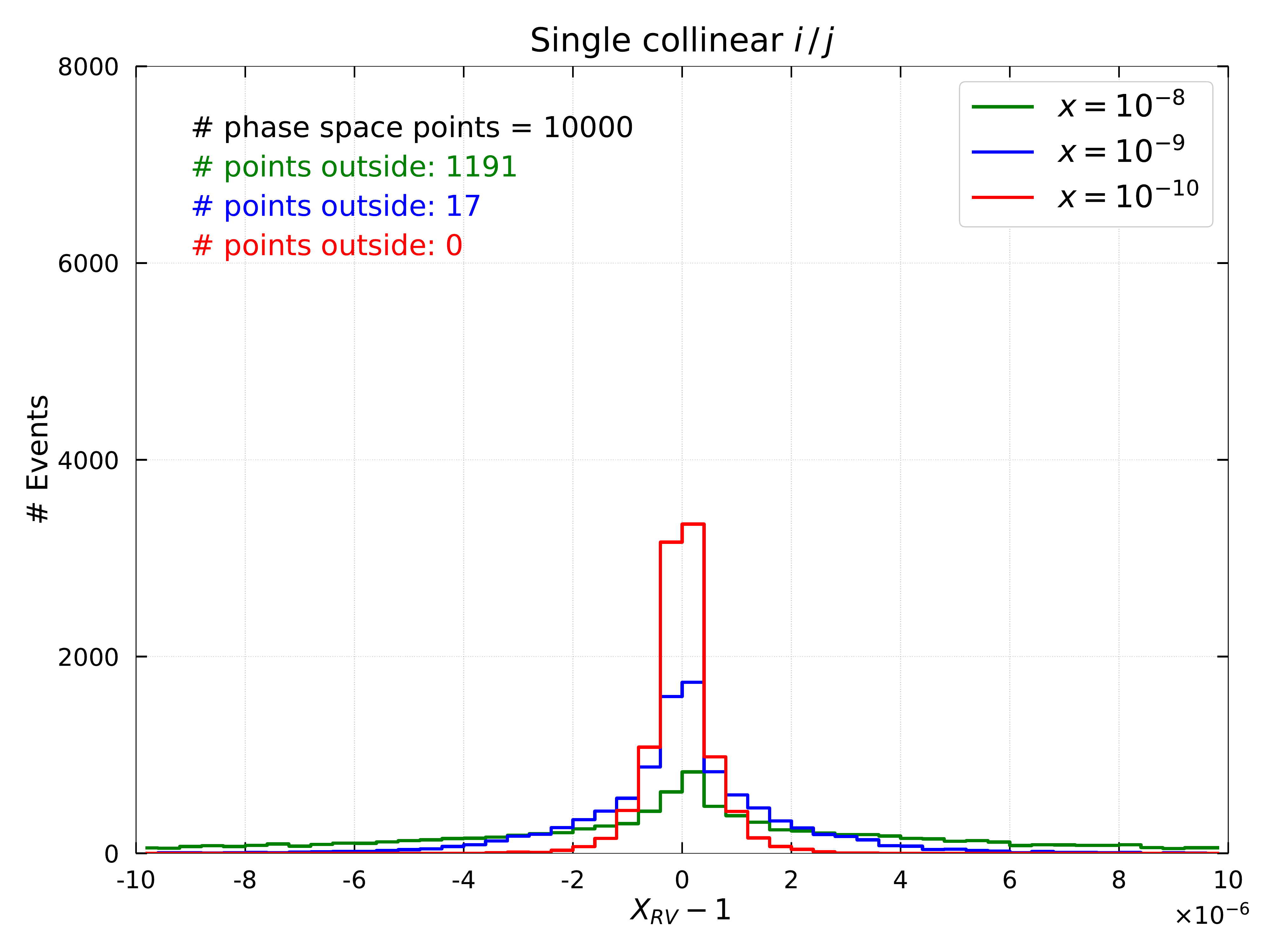}
	\caption{Spike plots of the ratio $\df \si^{\mathrm{RV}} / \df \si^T$ verifying the correct behaviour of the subtraction terms for $\t{B}_3^1$ in the single soft gluon (left) and single collinear $q\|g$ (right) limits.}
	\label{fig:qqbBt3g1_spikeplots}
\end{figure}

The correctness of the $\ep$-pole structure of the real-virtual subtraction is checked analytically, while the behaviour of $\df \si^T$ in unresolved limits is tested numerically, in the same way as was done for the double real subtraction. The results for the single soft and single collinear $g\|g$ limits for $\t{B}_3^1$ are shown in \fig{qqbBt3g1_spikeplots} as examples. The parameter controlling the depth of the IR tests is given by $x = 1-s_{jk}/s$ and $x = s_{ij}/s$ for the single soft $i\to 0$ and single collinear $i\|j$ limits in $1,2\rightarrow i,j,k$ scattering processes.
%

\section{Double virtual subtraction} \label{sec:VVsub}
The double virtual correction has no unresolved limits but contains $\ep$-poles up to order four, which must be cancelled by the subtraction term. The singularity structure of the matrix element can be obtained from Catani's factorization formula~\cite{Catani:1998bh} and after expanding in the colour levels, we can express the pole structure of each colour-ordered double virtual matrix element in terms of the colour-ordered infrared operators $I^{(1)}$, $I^{(2)}$ listed in \app{Iop_colord}, and single poles:
\ba \label{eq:2loopPoles}
\mathrm{Poles}\big[M_n^2(\{p_n\})\big] &=& \sum_{ij} \Big[ c_{1,ij} I^{(1)}(\ep,s_{ij}) \Big( M_n^1(\{p_n\}) - \frac{\bt_0}{\ep} M_n^0(\{p_n\}) \Big) \nn\\
&+& \sum_{kl} c_{ijkl} I^{(1)}(\ep,s_{ij}) I^{(1)}(\ep,s_{kl}) M_n^0(\{p_n\}) \nn\\
&+& c_{2,ij} I^{(2)}(\ep,s_{ij}) M_n^0(\{p_n\}) \nn\\
&+& c_{3,ij} H^{(2)}(\ep) M_n^0(\{p_n\}) \Big],
\ea
where the $c_x$ denote numerical coefficients and explicit expressions of the $I^{(2)}$ operators can be found in \app{Iop_colord}. The  $H^{(2)}(\ep)$ terms in \eq{2loopPoles} contain poles of maximal order one and are process dependent. We give their expressions~\cite{Glover:2001af,Anastasiou:2001sv,Anastasiou:2000kg} for dijet processes (A-type, B-type and C-/D-type matrix elements) also in colour-ordered form:
\ba
H_{A}^{(2)} &=& 4 \Big( N_c^2 H_{g,N_c^2}^{(2)} + N_c n_f H_{g,n_f}^{(2)} + n_f^2 H_{g,n_f^2}^{(2)} + + \frac{n_f}{N_c} H_{g,n_f/N_c}^{(2)} \Big), \nn\\
H_{B}^{(2)} &=& 2 \Big( N_c^2 \big( H_{g,N_c^2}^{(2)} + H_{q,N_c^2}^{(2)} \big) + N_c n_f \big( H_{g,n_f}^{(2)}+H_{q,n_f}^{(2)} \big) \nn\\
&+& n_f^2 H_{g,n_f^2}^{(2)} + H_{q}^{(2)} + \frac{n_f}{N_c} \big( H_{g,n_f/N_c}^{(2)} + H_{q,n_f/N_c}^{(2)} \big) + \frac{1}{N_c^2} H_{q,1/N_c^2}^{(2)} \Big), \nn\\
H_{C}^{(2)} &=& 4 \Big( N_c^2 H_{q,N_c^2}^{(2)} + H_{q}^{(2)} + \frac{1}{N_c^2} H_{q,1/N_c^2}^{(2)} + N_c n_f H_{q,n_f}^{(2)} + \frac{n_f}{N_c} H_{q,n_f/N_c}^{(2)} \Big),
\ea
where the individual colour level contributions are given by
\begin{alignat}{3}
& H_{q,N_c^2}^{(2)} && = \frac{409}{864} - \frac{11\pi^2}{96} + \frac{7\zeta_3}{4}, && \qquad\qquad H_{g,N_c^2}^{(2)} = \frac{5}{12} + \frac{11\pi^2}{144} + \frac{\zeta_3}{2}, \nn\\
& H_{q,n_f}^{(2)} && = -\frac{25}{216} + \frac{\pi^2}{48}, && \qquad\qquad H_{g,n_f}^{(2)} = -\frac{89}{108} - \frac{\pi^2}{72}, \nn\\
& && && \qquad\qquad H_{g,n_f^2}^{(2)} = \frac{5}{27}, \nn\\
& H_{q}^{(2)} && = -\frac{41}{108} - \frac{\pi^2}{96} - \frac{\zeta_3}{4},  && \nn\\
& H_{q,n_f/N_c}^{(2)} && = \frac{25}{216} - \frac{\pi^2}{48}, && \qquad\qquad H_{g,n_f/N_c}^{(2)} = -\frac{1}{4}, \nn\\
& H_{q,1/N_c^2}^{(2)} && = -\frac{3}{32} + \frac{\pi^2}{8} -\frac{3\zeta_3}{2}. &&
\end{alignat}

All these $\ep$-poles should be cancelled by the appropriate integrated subtraction terms from the double real and real-virtual contributions, which is checked analytically. With the process of constructing the subtraction terms starting at the double real level, continuing to the real-virtual level, and lastly the double virtual level, there is no more freedom left to add any new subtraction terms at the double virtual level. All subtraction terms of the RR or RV type which have not been integrated and added back must reappear at the VV level. Once this has been checked and the poles of the VV matrix element are correctly cancelled, a complete and consistent subtraction at NNLO has been performed.

The full double virtual subtraction term is obtained by combining all integrated RR and RV subtraction terms with the NNLO mass factorization term $\df \si^{\MF,2}$, as outlined in \eq{sigsub}. It takes the general form:
\ba \label{eq:sigUform}
\df \si^{U} &=& \mathcal{N}_{VV} \int \frac{\df x_1}{x_1} \frac{\df x_2}{x_2} \df \Phi_n \, J_n^{(n)}(\{p_n\}) \Big[ \nn\\
&& \sum_{i,j} c_{ij}^{(1)} \mathcal{J}_2^{(1)}(p_i,p_j) M_n^1(\{p_n\})  \nn\\
&+& \sum_{i,j,k,l} c_{ijkl}^{(1 \otimes 1)} \mathcal{J}_2^{(1)}(p_i,p_j) \otimes \mathcal{J}_2^{(1)}(p_k,p_l) M_n^0(\{p_n\}) \nn\\
&+& \sum_{i,j} c_{ij}^{(2)} \mathcal{J}_2^{(2)}(p_i,p_j) M_n^0(\{p_n\}) \Big].
\ea
where $c_{ij}^{(1)}$, $c_{ijkl}^{(1 \otimes 1)}$ and $c_{ij}^{(2)}$ denote numerical coefficients. The first line of \eq{sigUform} consists of single integrated dipoles multiplied by a one-loop reduced matrix element. They are the integrated $T \times L$ terms in $\df \si^{T,b_1}$ of the real-virtual subtraction which have been added back here. The second line has its origin in the $\mathcal{X}_3^0 X_3^0 M$ terms in $\df \si^{T,b_2}$ and $\df \si^{T,c}$. The last line consists of the new double integrated dipoles $\mathcal{J}_2^{(2)}$, which are defined such that we have a similar structure as for the NLO virtual subtraction. They contain the integrated four-parton antennae from $\df \si^{S,b}$, integrated one-loop antennae from the $L \times T$ terms in $\df \si^{T,b}$, $\mathcal{X}_3^0 \mathcal{X}_3^0$ terms which have not yet been accounted for, as well as mass factorization kernels for initial-state configurations. The general form of the $\mathcal{J}_2^{(2)}$ is given by
\ba
\mathcal{J}_2^{(2)}(p_i,p_j) &=& \sum_{k} c_{1,k} \mathcal{X}_{4,k}^0(s_{ij}) + c_{2,k} \mathcal{X}_{3,k}^1(s_{ij}) + c_{3,k} \frac{\bt_0}{\ep} \l( \frac{s_{ij}}{\mu^2} \r)^{-\ep} \mathcal{X}_{3,k}^0(s_{ij}) \nn\\
&+& \sum_{l} \big( c_{4,kl} \big[\mathcal{X}_{3,k}^0 \otimes \mathcal{X}_{3,l}^0\big] + c_{5,kl} \big[\mathcal{X}_{3,k}^0 \otimes \Ga_{l}^{(1)}\big] + c_{6,kl}\big[\Ga_{k}^{(1)} \otimes \Ga_{l}^{(1)}\big] \big) + c_{7,k} \bar{\Ga}_{k}^{(2)}, \nn\\
\ea
where the sum goes over the appropriate antennae and mass factorization kernels and the $c_x$ denote numerical coefficients. Written in this form, one can immediately establish the cancellation of $\ep$-poles between the subtraction term and the double virtual matrix element. The full list of the $\mathcal{J}_2^{(2)}$ integrated dipole 
factors is given in \app{J22}. 

As simple examples, we discuss the pole structure of the two-loop virtual contribution and the dipole structure of the associated double virtual subtraction term for selected colour levels of the B-type process. We select those contributions that have a particularly compact form, thereby illustrating the essential features of the pole cancellation at the double virtual level. 
 
The pole structure of the leading $N_c$ colour level function $B_2^2(1,2,3,4)$ for a given colour ordering is given by
\ba
&& \mathrm{Poles}\big[B_2^2(1,2,3,4)\big] = \nn\\
&+& 2 \big[ I_{qg}^{(1)}(\ep,s_{12}) + I_{gg}^{(1)}(\ep,s_{23}) + I_{qg}^{(1)}(\ep,s_{34}) \big] \big(B_2^1(1,2,3,4)-\frac{b_0}{\ep} B_2^0(1,2,3,4) \big) \nn\\
&-& 2 \big[ I_{qg}^{(1)}(\ep,s_{12}) + I_{gg}^{(1)}(\ep,s_{23}) + I_{qg}^{(1)}(\ep,s_{34}) \big] \nn\\
&& \times \big[ I_{qg}^{(1)}(\ep,s_{12}) + I_{gg}^{(1)}(\ep,s_{23}) + I_{qg}^{(1)}(\ep,s_{34}) \big] B_2^0(1,2,3,4) \nn\\
&+& 2 \big[ I_{qg}^{(2)}(\ep,s_{12}) + I_{gg}^{(2)}(\ep,s_{23}) + I_{qg}^{(2)}(\ep,s_{34}) \big] B_2^0(1,2,3,4) \nn\\
&+& 2 \big[ 2 H_{q,N_c^2}^{(2)} + 2 H_{g,N_c^2}^{(2)} \big] B_2^0(1,2,3,4),
\ea
and the leading $n_f$ by
\ba
&& \mathrm{Poles}\big[\hat{B}_2^2(1,2,3,4)\big] = \nn\\
&+& 2 \big[ I_{qg}^{(1)}(\ep,s_{12}) + I_{gg}^{(1)}(\ep,s_{23}) + I_{qg}^{(1)}(\ep,s_{34}) \big] \big(\hat{B}_2^1(1,2,3,4)-\frac{b_{0,F}}{\ep} B_2^0(1,2,3,4) \big) \nn\\
&+& 2 \big[ I_{qg,F}^{(1)}(\ep,s_{12}) + I_{gg,F}^{(1)}(\ep,s_{23}) + I_{qg,F}^{(1)}(\ep,s_{34}) \big] \big(B_2^1(1,2,3,4)-\frac{b_0}{\ep} B_2^0(1,2,3,4) \big) \nn\\
&-& 4 \big[ I_{qg}^{(1)}(\ep,s_{12}) + I_{gg}^{(1)}(\ep,s_{23}) + I_{qg}^{(1)}(\ep,s_{34}) \big] \nn\\
&& \times \big[ I_{qg,F}^{(1)}(\ep,s_{12}) + I_{gg,F}^{(1)}(\ep,s_{23}) + I_{qg,F}^{(1)}(\ep,s_{34}) \big] B_2^0(1,2,3,4) \nn\\
&+& 2 \big[ I_{qg,F}^{(2)}(\ep,s_{12}) + I_{gg,F}^{(2)}(\ep,s_{23}) + I_{qg,F}^{(2)}(\ep,s_{34}) \big] B_2^0(1,2,3,4) \nn\\
&+& 2 \big[2 H_{q,n_f}^{(2)} + 2 H_{g,n_f}^{(2)} \big] B_2^0(1,2,3,4).
\ea
Here, one can clearly see that the poles have the expected structure associated with the colour-connected singularities and that all the squared matrix elements on which the infrared singularity functions factor on are the colour-coherent squared leading colour matrix elements. The expressions for the subleading colour levels are less compact and somewhat lack these nice structures and naturally involve interferences of the colour-ordered amplitudes. The pole structure of the most subleading $N_c$ colour level of the two-loop B-type matrix element is however particularly simple and we take it as an example again in the $q\qb$-crossing, 
\ba \label{eq:Bttt2g2poles}
\mathrm{Poles}\big[\bar{B}_2^2(1,3,4,2)\big] &=& - 2 I_{qq}^{(1)}(\ep,s_{12}) \t{\t{B}}_2^1(1,3,4,2) \nn\\
&+& 2 I_{qq}^{(1)}(\ep,s_{12}) I_{qq}^{(1)}(\ep,s_{12}) \bar{B}_2^0(1,3,4,2) \nn\\
&-& 2 \big( 2 H_{q,1/N_c^2}^{(2)} \big) \bar{B}_2^0(1,3,4,2).
\ea
We observe no infrared operators involving quark-gluon or gluon-gluon momenta invariants, following the colour-disconnected behaviour of the abelian gluons. The integrated subtraction terms for $\bar{B}_2^2(1,3,4,2)$ are given by
\ba
\df \si^U &=& \mathcal{J}_2^{(1)}(\bar{1}_q,\bar{2}_{\qb}) \t{\t{B}}_2^1(1,3,4,2) \nn\\
&+& \frac{1}{2} \mathcal{J}_2^{(1)}(\bar{1}_q,\bar{2}_{\qb}) \otimes \mathcal{J}_2^{(1)}(\bar{1}_q,\bar{2}_{\qb}) \bar{B}_2^0(1,3,4,2) \nn\\
&+& \bar{\mathcal{J}}_2^{(2)}(\bar{1}_q,\bar{2}_{\qb}) \bar{B}_2^0(1,3,4,2),
\ea 
thereby illustrating the integrated dipole structure in \eq{sigUform} in practice. 

In our approach of constructing the antenna subtraction terms from the known IR divergent behaviour of the RR and RV squared matrix elements, the analytic pole cancellation in the VV contribution provides a strong cross-check on the correctness of the subtraction terms at the RR and RV levels. This argument could be turned around by starting the construction of the subtraction terms at the VV level, by matching the known pole structure of two-loop squared matrix elements and mass factorization terms onto a combination of integrated dipole factors. A systematic unintegration of these dipole factors~\cite{Chen:2022ktf} could then yield the RV and RR antenna subtraction terms in a systematic and algorithmic manner. 

\section{Conclusion} \label{sec:conc}
In this paper, we documented the details of the NNLO QCD calculation in full colour for jet production observables at the LHC~\cite{Chen:2022tpk} using the antenna subtraction method. All the relevant matrix elements were listed in a colour-decomposed manner, along with several properties such as their colour connections and pole structures. The use of the traditional colour-ordered antenna subtraction method to construct the NNLO subtraction terms starting from the double real and ending at the double virtual level was summarized. Several complications within this approach that arise due to the more involved structure of the high parton multiplicity matrix elements at the subleading colour levels were highlighted and explicit examples of the RR, RV and VV subtractions were discussed to show how to subtract all unresolved limits and cancel all the $\ep$-poles without needing new antenna functions or performing new analytic integrals. Building upon earlier developments at leading colour~\cite{Currie:2013vh}, we introduced the full set of integrated dipole terms for full colour, which allow the infrared pole cancellation between integrated subtraction terms, mass factorization terms and virtual matrix elements to be structured in a compact and modular manner. 

Although successful, with phenomenological results reported in~\cite{Chen:2022tpk}, this paper also illustrates the need for an automated method to apply the antenna subtraction at NNLO in full colour, considering both the more involved structure of subleading colour contributions and that the complexity only increases as the multiplicity of coloured particles in the process grows. This has motivated the development of the colourful antenna subtraction method, a reformulation of the antenna subtraction where the predictability of the pole structures of loop-amplitudes in colour space is used to generate the subtraction terms in an automated procedure starting at the double virtual and cascading to the double real level. Progress in this colourful antenna subtraction method has already been made and applied for gluonic three-jet production from proton-proton collisions and continues to be developed~\cite{Chen:2022ktf}.

\section*{Acknowledgements}
The authors thank Juan Cruz-Martinez, James Currie, Rhorry Gauld, Aude Gehrmann-De Ridder,
Marius H\"ofer, Alexander Huss, Imre Majer, Matteo Marcoli, Thomas Morgan, Jan Niehues, Jo\~ao Pires, Adrian Rodriguez Garcia, 
Robin Sch\"urmann, Giovanni Stagnitto, Duncan Walker, Steven Wells and James Whitehead for useful 
discussions and their many contributions to the \NNLOJET code.
This work has received funding from the Swiss National Science Foundation (SNF) under contract 200020-204200, from the European Research Council (ERC) under the European Union's Horizon 2020 research and innovation programme grant agreement 101019620 (ERC Advanced Grant TOPUP), from the UK Science and Technology Facilities Council (STFC) through grant ST/T001011/1 and from the Deutsche Forschungsgemeinschaft (DFG, German Research Foundation) under grant 396021762-TRR 257.

\begin{appendix}

\section{Colour-ordered infrared singularity operators} \label{app:Iop_colord}
The infrared pole structure of renormalized one-loop and two-loop QCD scattering amplitudes can be described in terms of infrared pole operators~\cite{Catani:1998bh}, which are tensors in colour space. The antenna subtraction terms are constructed for colour-ordered subprocesses, thus requiring a rearrangement of the infrared singularity operators into a colour-ordered form~\cite{Gehrmann-DeRidder:2005btv}, which was used throughout this paper. The colour-ordered singularity operators are listed for reference in the following. 

The $I^{(1)}$ infrared singularity operators for different partonic emitters are:
\ba
I_{q\qb}^{(1)}(\ep,s_{q\qb}) &=& - \frac{e^{\ep \ga_E}}{2\Ga(1-\ep)} \l( \frac{1}{\ep^2}+\frac{3}{2\ep} \r) \mathrm{Re}(-s_{q\qb})^{-\ep}, \nn\\
I_{qg}^{(1)}(\ep,s_{qg}) &=& - \frac{e^{\ep \ga_E}}{2\Ga(1-\ep)} \l( \frac{1}{\ep^2}+\frac{5}{3\ep} \r) \mathrm{Re}(-s_{qg})^{-\ep}, \nn\\
I_{gg}^{(1)}(\ep,s_{gg}) &=& - \frac{e^{\ep \ga_E}}{2\Ga(1-\ep)} \l( \frac{1}{\ep^2}+\frac{11}{6\ep} \r) \mathrm{Re}(-s_{gg})^{-\ep}, \nn\\
I_{qg,F}^{(1)}(\ep,s_{qg}) &=& 0, \nn\\
I_{qg,F}^{(1)}(\ep,s_{qg}) &=& - \frac{e^{\ep \ga_E}}{2\Ga(1-\ep)} \frac{1}{6\ep}\mathrm{Re}(-s_{qg})^{-\ep}, \nn\\
I_{gg,F}^{(1)}(\ep,s_{gg}) &=& - \frac{e^{\ep \ga_E}}{2\Ga(1-\ep)} \frac{1}{3\ep}\mathrm{Re}(-s_{gg})^{-\ep},
\ea
and the corresponding $I^{(2)}$ infrared singularity operators are given by:
\ba
I_{q\qb}^{(2)}(\ep,s_{q\qb}) &=& \frac{\Ga(1-2\ep)}{e^{\ep \ga_E} \Ga(1-\ep)} \l( \frac{b_0}{\ep} + k_0 \r) I_{q\qb}^{(1)}(2\ep,s_{q\qb}), \nn\\
I_{qg}^{(2)}(\ep,s_{qg}) &=& \frac{\Ga(1-2\ep)}{e^{\ep \ga_E} \Ga(1-\ep)} \l( \frac{b_0}{\ep} + k_0 \r) I_{qg}^{(1)}(2\ep,s_{qg}), \nn\\
I_{gg}^{(2)}(\ep,s_{gg}) &=& \frac{\Ga(1-2\ep)}{e^{\ep \ga_E} \Ga(1-\ep)} \l( \frac{b_0}{\ep} + k_0 \r) I_{gg}^{(1)}(2\ep,s_{gg}), \nn\\
\nn\\
I_{q\qb,F}^{(2)}(\ep,s_{q\qb}) &=& \frac{\Ga(1-2\ep)}{e^{\ep \ga_E} \Ga(1-\ep)} \bigg[ \l( \frac{b_{0,F}}{\ep} + k_{0,F} \r) I_{q\qb}^{(1)}(2\ep,s_{q\qb}) + \l( \frac{b_0}{\ep} + k_0 \r) I_{q\qb,F}^{(1)}(2\ep,s_{q\qb}) \bigg], \nn\\
I_{qg,F}^{(2)}(\ep,s_{qg}) &=& \frac{\Ga(1-2\ep)}{e^{\ep \ga_E} \Ga(1-\ep)} \bigg[ \l( \frac{b_{0,F}}{\ep} + k_{0,F} \r) I_{qg}^{(1)}(2\ep,s_{qg}) + \l( \frac{b_0}{\ep} + k_0 \r) I_{qg,F}^{(1)}(2\ep,s_{qg}) \bigg], \nn\\
I_{gg,F}^{(2)}(\ep,s_{gg}) &=& \frac{\Ga(1-2\ep)}{e^{\ep \ga_E} \Ga(1-\ep)} \bigg[ \l( \frac{b_{0,F}}{\ep} + k_{0,F} \r) I_{gg}^{(1)}(2\ep,s_{gg}) + \l( \frac{b_0}{\ep} + k_0 \r) I_{gg,F}^{(1)}(2\ep,s_{gg}) \bigg], \nn\\
\nn\\
I_{q\qb,F^2}^{(2)}(\ep,s_{q\qb}) &=& \frac{\Ga(1-2\ep)}{e^{\ep \ga_E} \Ga(1-\ep)} \l( \frac{b_{0,F}}{\ep} + k_{0,F} \r) I_{q\qb,F}^{(1)}(2\ep,s_{qg}), \nn\\
I_{qg,F^2}^{(2)}(\ep,s_{qg}) &=& \frac{\Ga(1-2\ep)}{e^{\ep \ga_E} \Ga(1-\ep)} \l( \frac{b_{0,F}}{\ep} + k_{0,F} \r) I_{qg,F}^{(1)}(2\ep,s_{qg}), \nn\\
I_{gg,F^2}^{(2)}(\ep,s_{gg}) &=& \frac{\Ga(1-2\ep)}{e^{\ep \ga_E} \Ga(1-\ep)} \l( \frac{b_{0,F}}{\ep} + k_{0,F} \r) I_{gg,F}^{(1)}(2\ep,s_{gg}),
\ea
where $b_0, b_{0,F}$ and $k_0, k_{0,F}$ are the coefficients of the colour-ordered $\bt$-function and collinear coefficient:
\begin{alignat}{4}
& \bt_0 && = b_0 N_c + b_{0,F} \, n_f, \qquad && b_0 = \frac{11}{6}, \qquad && b_{0,F} = - \frac{1}{3}, \nn\\
& K && = k_0 N_c + k_{0,F} \, n_f, \qquad && k_0 = \frac{67}{18} - \frac{\pi^2}{6}, \qquad && k_{0,F} = - \frac{5}{9}.
\end{alignat}

\section{Single integrated dipoles $\mathcal{J}_2^{(1)}$} \label{app:J21}
All the final-final, initial-final and initial-initial single integrated dipoles $\mathcal{J}_2^{(1)}$ used in the dijet subtraction terms are collected here. Initial states are identified by using an overhead bar and subscript $q,g$ labels are added to denote the parton species. The mass factorization kernels $\Ga^{(1)}$ can be found in~\cite{Currie:2013vh}. Identity-changing antennae are accompanied by the factors $S_{g\to q}$ or $S_{q\to g}$, which correct for the fact that the degrees of freedom in $d$-dimensions are different for a gluon and quark. Explicitly they are
\ba
S_{g\to q} &=& \frac{S_g}{S_q} = \frac{2-2\ep}{2} = 1 - \ep, \nn\\
S_{q\to g} &=& \frac{S_q}{S_g} = \frac{2}{2-2\ep} = \frac{1}{1 - \ep}. 
\ea 

\subsection{Final-final}
The final-final single integrated dipoles are given by
\ba
\mathcal{J}_2^{(1)}(i_q,j_{\qb}) &=& \mathcal{A}_3^0(s_{ij}), \nn\\
\hat{\mathcal{J}}_2^{(1)}(i_q,j_{\qb}) &=& 0, \nn\\
\mathcal{J}_2^{(1)}(i_q,j_g) &=& \frac{1}{2}\mathcal{D}_3^0(s_{ij}), \nn\\
\hat{\mathcal{J}}_2^{(1)}(i_q,j_g) &=& \frac{1}{2}\mathcal{E}_3^0(s_{ij}), \nn\\
\mathcal{J}_2^{(1)}(i_g,j_g) &=& \frac{1}{3}\mathcal{F}_3^0(s_{ij}), \nn\\
\hat{\mathcal{J}}_2^{(1)}(i_g,j_g) &=& \mathcal{G}_3^0(s_{ij}).
\ea

\subsection{Initial-final}
The initial-final single integrated dipoles are given by
\ba
\mathcal{J}_2^{(1)}(\bar{1}_q,i_{\qb}) &=& \mathcal{A}_{3,q}^0(s_{\bar{1}i}) - \Ga_{qq}^{(1)}(x_1)\de(1-x_2), \nn\\
\hat{\mathcal{J}}_2^{(1)}(\bar{1}_q,i_{\qb}) &=& 0, \nn\\
\mathcal{J}_2^{(1)}(\bar{1}_q,i_g) &=& \frac{1}{2}\mathcal{D}_{3,q}^0(s_{\bar{1}i}) - \Ga_{qq}^{(1)}(x_1)\de(1-x_2), \nn\\
\hat{\mathcal{J}}_2^{(1)}(\bar{1}_q,i_g) &=& \frac{1}{2}\mathcal{E}_{3,q'}^0(s_{\bar{1}i}), \nn\\
\mathcal{J}_2^{(1)}(i_q,\bar{1}_g) &=& \mathcal{D}_{3,g\to g}^0(s_{\bar{1}i}) - \frac{1}{2}\Ga_{gg}^{(1)}(x_1)\de(1-x_2), \nn\\
\hat{\mathcal{J}}_2^{(1)}(i_q,\bar{1}_g) &=& - \frac{1}{2}\hat{\Ga}_{gg}^{(1)}(x_1)\de(1-x_2), \nn\\
\mathcal{J}_2^{(1)}(\bar{1}_g,i_g) &=& \frac{1}{2}\mathcal{F}_{3,g}^0(s_{\bar{1}i}) - \frac{1}{2}\Ga_{gg}^{(1)}(x_1)\de(1-x_2), \nn\\
\hat{\mathcal{J}}_2^{(1)}(\bar{1}_g,i_g) &=& \frac{1}{2}\mathcal{G}_{3,g}^0(s_{\bar{1}i}) - \frac{1}{2} \hat{\Ga}_{gg}^{(1)}(x_1)\de(1-x_2), \nn\\
\mathcal{J}_{2,g\to q}^{(1)}(\bar{1}_q, i_{\qb}) &=& -\frac{1}{2} \mathcal{A}_{3,g}^0(s_{\bar{1}i}) - S_{g\to q}\Ga_{qg}^{(1)}(x_1)\de(1-x_2), \nn\\
\mathcal{J}_{2,g\to q}^{(1)}(\bar{1}_q, i_g) &=& -\mathcal{D}_{3,g}^0(s_{\bar{1}i}) - S_{g\to q}\Ga_{qg}^{(1)}(x_1)\de(1-x_2), \nn\\
\mathcal{J}_{2,q\to g}^{(1)}(i_q, \bar{1}_g) &=& - \mathcal{E}_{3,q'}^0(s_{\bar{1}i}) - S_{q\to g}\Ga_{gq}^{(1)}(x_1)\de(1-x_2), \nn\\
\mathcal{J}_{2,q\to g}^{(1)}(i_g, \bar{1}_g) &=& - \mathcal{G}_{3,q}^0(s_{\bar{1}i}) - S_{q\to g}\Ga_{gq}^{(1)}(x_1)\de(1-x_2).
\ea

\subsection{Initial-initial}
The initial-initial single integrated dipoles are given by
\ba
\mathcal{J}_2^{(1)}(\bar{1}_q,\bar{2}_{\qb}) &=& \mathcal{A}_{3,q\qb}^0(s_{\bar{1}\bar{2}}) - \Ga_{qq}^{(1)}(x_1)\de(1-x_2) - \Ga_{qq}^{(1)}(x_2)\de(1-x_1), \nn\\
\hat{\mathcal{J}}_2^{(1)}(\bar{1}_q,\bar{2}_{\qb}) &=& 0, \nn\\
\mathcal{J}_2^{(1)}(\bar{1}_q,\bar{2}_{g}) &=& \mathcal{D}_{3,qg}^0(s_{\bar{1}\bar{2}}) - \Ga_{qq}^{(1)}(x_1)\de(1-x_2) - \frac{1}{2}\Ga_{gg}^{(1)}(x_2)\de(1-x_1), \nn\\
\hat{\mathcal{J}}_2^{(1)}(\bar{1}_q,\bar{2}_g) &=& -\frac{1}{2}\hat{\Ga}_{gg}^{(1)}(x_2)\de(1-x_1), \nn\\
\mathcal{J}_2^{(1)}(\bar{1}_g,\bar{2}_{g}) &=& \mathcal{F}_{3,gg}^0(s_{\bar{1}\bar{2}}) - \frac{1}{2}\Ga_{gg}^{(1)}(x_1)\de(1-x_2) - \frac{1}{2}\Ga_{gg}^{(1)}(x_2)\de(1-x_1), \nn\\
\hat{\mathcal{J}}_2^{(1)}(\bar{1}_g,\bar{2}_{g}) &=& - \frac{1}{2}\hat{\Ga}_{gg}^{(1)}(x_1)\de(1-x_2) - \frac{1}{2}\hat{\Ga}_{gg}^{(1)}(x_2)\de(1-x_1), \nn\\
\mathcal{J}_{2,g\to q}^{(1)}(\bar{1}_q,\bar{2}_{\qb}) &=& -\mathcal{A}_{3,qg}^0(s_{\bar{1}\bar{2}}) - S_{g\to q}\Ga_{qg}^{(1)}(x_2)\de(1-x_1), \nn\\
\mathcal{J}_{2,g\to q}^{(1)}(\bar{1}_q,\bar{2}_g) &=& -\mathcal{D}_{3,gg}^0(s_{\bar{1}\bar{2}}) - S_{g\to q}\Ga_{qg}^{(1)}(x_1)\de(1-x_2), \nn\\
\mathcal{J}_{2,q\to g}^{(1)}(\bar{1}_q,\bar{2}_g) &=& -\mathcal{E}_{3,q'q}^0(s_{\bar{1}\bar{2}}) - S_{q\to g}\Ga_{gq}^{(1)}(x_2)\de(1-x_1), \nn\\
\mathcal{J}_{2,q\to g}^{(1)}(\bar{1}_g,\bar{2}_g) &=& -\mathcal{G}_{3,gq}^0(s_{\bar{1}\bar{2}}) - S_{q\to g}\Ga_{gq}^{(1)}(x_2)\de(1-x_1).
\ea

\section{Double integrated dipoles $\mathcal{J}_2^{(2)}$} \label{app:J22}
All the final-final, initial-final and initial-initial double integrated dipoles $\mathcal{J}_2^{(2)}$ used in the dijet subtraction terms are collected here. We use the notation $\de_i = \de(1-x_i)$ and $\Ga_{ab,i}^{(l)} = \Ga_{ab}^{(l)}(x_i)$ for $i \in \{1,2\}$. The expressions of the reduced two-loop mass factorization kernels $\bar{\Ga}^{(2)}$ can be found in \cite{Currie:2013vh}.

\subsection{Identity preserving: quark-antiquark}

\subsubsection{Final-final}
\ba
\mathcal{J}_2^{(2)}(1_q,2_{\qb}) &=& \mathcal{A}_4^0(s_{12}) + \mathcal{A}_3^1(s_{12}) + \frac{b_0}{\ep} \l( \frac{s_{12}}{\mu^2} \r)^{-\ep} \mathcal{A}_3^0(s_{12}) - \frac{1}{2} \big[\mathcal{A}_3^0 \otimes \mathcal{A}_3^0\big] (s_{12}), \nn\\
\t{\mathcal{J}}_2^{(2)}(1_q,2_{\qb}) &=& \frac{1}{2} \t{\mathcal{A}}_4^0(s_{12}) + 2 \mathcal{C}_4^0(s_{12}) + \t{\mathcal{A}}_3^1(s_{12}) - \frac{1}{2} \big[\mathcal{A}_3^0 \otimes \mathcal{A}_3^0\big] (s_{12}), \nn\\
\hat{\mathcal{J}}_2^{(2)}(1_q,2_{\qb}) &=& \mathcal{B}_4^0(s_{12}) + \hat{\mathcal{A}}_3^1(s_{12}) + \frac{b_{0,F}}{\ep} \l( \frac{s_{12}}{\mu^2} \r)^{-\ep} \mathcal{A}_3^0(s_{12}), \nn\\
\bar{\mathcal{J}}_2^{(2)}(1_q,2_{\qb}) &=& \frac{1}{2} \t{\mathcal{A}}_4^0(s_{12}) + \t{\mathcal{A}}_3^1(s_{12}) - \frac{1}{2} \big[\mathcal{A}_3^0 \otimes \mathcal{A}_3^0\big] (s_{12}).
\ea

\subsubsection{Initial-final}
\ba
\mathcal{J}_2^{(2)}(\bar{1}_q,2_{\qb}) &=& \mathcal{A}_{4,q}^0(s_{\bar{1}2}) + \mathcal{A}_{3,q}^1(s_{\bar{1}2}) + \frac{b_0}{\ep} \l( \frac{s_{\bar{1}2}}{\mu^2} \r)^{-\ep} \mathcal{A}_{3,q}^0(s_{\bar{1}2}) \nn\\ 
&-& \frac{1}{2} \big[\mathcal{A}_{3,q}^0 \otimes \mathcal{A}_{3,q}^0\big] (s_{\bar{1}2}) - \bar{\Ga}_{qq,1}^{(2)} \de_2, \nn\\
\t{\mathcal{J}}_2^{(2)}(\bar{1}_q,2_{\qb}) &=& \t{\mathcal{A}}_{4,q}^0(s_{\bar{1}2}) + 2\t{\mathcal{C}}_{4,q}^0(s_{\bar{1}2}) + \t{\mathcal{C}}_{4,\qb}^0(s_{\bar{1}2}) + \t{\mathcal{A}}_{3,q}^1(s_{\bar{1}2}) \nn\\
&-& \frac{1}{2} \big[\mathcal{A}_{3,q}^0 \otimes \mathcal{A}_{3,q}^0\big] (s_{\bar{1}2}) + \t{\t{\bar{\Ga}}}_{qq,1}^{(2)} \de_2, \nn\\
\hat{\mathcal{J}}_2^{(2)}(\bar{1}_q,2_{\qb}) &=& \mathcal{B}_{4,q}^0(s_{\bar{1}2}) + \hat{\mathcal{A}}_{3,q}^1(s_{\bar{1}2}) + \frac{b_{0,F}}{\ep} \l( \frac{s_{\bar{1}2}}{\mu^2} \r)^{-\ep} \mathcal{A}_{3,q}^0(s_{\bar{1}2}) - \hat{\bar{\Ga}}_{qq,1}^{(2)} \de_2, \nn\\
\bar{\mathcal{J}}_2^{(2)}(\bar{1}_q,2_{\qb}) &=& \frac{1}{2} \mathcal{A}_{4,q}^0(s_{\bar{1}2}) + \t{\mathcal{A}}_{3,q}^1(s_{\bar{1}2}) - \frac{1}{2}\big[\mathcal{A}_{3,q}^0 \otimes \mathcal{A}_{3,q}^0\big](s_{\bar{1}2}).
\ea

\subsubsection{Initial-initial}
\ba
\mathcal{J}_2^{(2)}(\bar{1}_q,\bar{2}_{\qb}) &=& \mathcal{A}_{4,q\qb}^0(s_{\bar{1}\bar{2}}) + \mathcal{A}_{3,q\qb}^1(s_{\bar{1}\bar{2}}) + \frac{b_0}{\ep} \l( \frac{s_{\bar{1}\bar{2}}}{\mu^2} \r)^{-\ep} \mathcal{A}_{3,q\qb}^0(s_{\bar{1}\bar{2}}) \nn\\
&-& \frac{1}{2} \big[\mathcal{A}_{3,q\qb}^0 \otimes \mathcal{A}_{3,q\qb}^0\big] (s_{\bar{1}\bar{2}}) - \bar{\Ga}_{qq,1}^{(2)} \de_2 - \bar{\Ga}_{qq,2}^{(2)} \de_1, \nn\\
\t{\mathcal{J}}_2^{(2)}(\bar{1}_q,\bar{2}_{\qb}) &=& \frac{1}{2}\t{\mathcal{A}}_{4,q\qb}^0(s_{\bar{1}\bar{2}}) + 2\mathcal{C}_{4,q\qb}^0(s_{\bar{1}\bar{2}}) + 2\mathcal{C}_{4,\qb q}^0(s_{\bar{1}\bar{2}}) + \t{\mathcal{A}}_{3,q\qb}^1(s_{\bar{1}\bar{2}}) \nn\\
&-& \frac{1}{2} \big[\mathcal{A}_{3,q\qb}^0 \otimes \mathcal{A}_{3,q\qb}^0\big] (s_{\bar{1}\bar{2}}) + \t{\t{\bar{\Ga}}}_{qq,1}^{(2)} \de_2 + \t{\t{\bar{\Ga}}}_{qq,2}^{(2)} \de_1, \nn\\
\hat{\mathcal{J}}_2^{(2)}(\bar{1}_q,\bar{2}_{\qb}) &=& \mathcal{B}_{4,q\qb}^0(s_{\bar{1}\bar{2}}) + \hat{\mathcal{A}}_{3,q\qb}^1(s_{\bar{1}\bar{2}}) + \frac{b_{0,F}}{\ep} \l( \frac{s_{\bar{1}\bar{2}}}{\mu^2} \r)^{-\ep} \mathcal{A}_{3,q\qb}^0(s_{\bar{1}\bar{2}}) - \hat{\bar{\Ga}}_{qq,1}^{(2)} \de_2 - \hat{\bar{\Ga}}_{qq,2}^{(2)} \de_1, \nn\\
\bar{\mathcal{J}}_2^{(2)}(\bar{1}_q,\bar{2}_{\qb}) &=& \frac{1}{2}\t{\mathcal{A}}_{4,q\qb}^0(s_{\bar{1}\bar{2}}) + \t{\mathcal{A}}_{3,q\qb}^1(s_{\bar{1}\bar{2}}) - \frac{1}{2} \big[\mathcal{A}_{3,q\qb}^0 \otimes \mathcal{A}_{3,q\qb}^0\big] (s_{\bar{1}\bar{2}}).
\ea

\subsection{Identity preserving: quark-gluon}

\subsubsection{Final-final}
\ba
\mathcal{J}_2^{(2)}(1_q,2_g) &=& \frac{1}{2}\mathcal{D}_4^0(s_{12}) + \frac{1}{2}\mathcal{D}_3^1(s_{12}) + \frac{1}{2}\frac{b_0}{\ep} \l( \frac{s_{12}}{\mu^2} \r)^{-\ep} \mathcal{D}_3^0(s_{12}) - \frac{1}{4}\big[\mathcal{D}_3^0 \otimes \mathcal{D}_3^0\big] (s_{12}), \nn\\
\hat{\mathcal{J}}_2^{(2)}(1_q,2_g) &=& \frac{1}{2}\mathcal{E}_4^0(s_{12}) + \frac{1}{2}\mathcal{E}_3^1(s_{12}) + \frac{1}{2}\hat{\mathcal{D}}_3^1(s_{12}) + \frac{1}{2}\frac{b_0}{\ep} \l( \frac{s_{12}}{\mu^2} \r)^{-\ep} \mathcal{E}_3^0(s_{12}) \nn\\
&+& \frac{1}{2}\frac{b_{0,F}}{\ep} \l( \frac{s_{12}}{\mu^2} \r)^{-\ep} \mathcal{D}_3^0(s_{12}) - \frac{1}{2}\big[\mathcal{E}_3^0 \otimes \mathcal{D}_3^0\big] (s_{12}), \nn\\
\hat{\t{\mathcal{J}}}_2^{(2)}(1_q,2_g) &=& \frac{1}{2}\t{\mathcal{E}}_4^0(s_{12}) + \frac{1}{2}\t{\mathcal{E}}_3^1(s_{12}), \nn\\
\hat{\hat{\t{\mathcal{J}}}}_2^{(2)}(1_q,2_g) &=& \frac{1}{2}\hat{\mathcal{E}}_3^1(s_{12}) + \frac{1}{2}\frac{b_{0,F}}{\ep} \l( \frac{s_{12}}{\mu^2} \r)^{-\ep} \mathcal{E}_3^0(s_{12}) - \frac{1}{4}\big[\mathcal{E}_3^0 \otimes \mathcal{E}_3^0\big] (s_{12}).
\ea

\subsubsection{Initial-final: quark-initiated}
\ba
\mathcal{J}_2^{(2)}(\bar{1}_q,2_g) &=& \frac{1}{2}\mathcal{D}_{4,q}^0(s_{\bar{1}2}) + \frac{1}{2}\mathcal{D}_{3,q}^1(s_{\bar{1}2}) + \frac{1}{2}\frac{b_0}{\ep} \l( \frac{s_{\bar{1}2}}{\mu^2} \r)^{-\ep} \mathcal{D}_{3,q}^0(s_{\bar{1}2}) \nn\\
&-& \frac{1}{4} \big[\mathcal{D}_{3,q}^0 \otimes \mathcal{D}_{3,q}^0\big] (s_{\bar{1}2}) - \bar{\Ga}_{qq,1}^{(2)} \de_2, \nn\\
\hat{\mathcal{J}}_2^{(2)}(\bar{1}_q,2_g) &=& \mathcal{E}_{4,q}^0(s_{\bar{1}2}) + \frac{1}{2}\mathcal{E}_{3,q}^1(s_{\bar{1}2}) + \frac{1}{2}\hat{\mathcal{D}}_{3,q}^1(s_{\bar{1}2}) + \frac{1}{2}\frac{b_{0}}{\ep} \l( \frac{s_{\bar{1}2}}{\mu^2} \r)^{-\ep} \mathcal{E}_{3,q}^0(s_{\bar{1}2}) \nn\\ 
&+& \frac{1}{2}\frac{b_{0,F}}{\ep} \l( \frac{s_{\bar{1}2}}{\mu^2} \r)^{-\ep} \mathcal{D}_{3,q}^0(s_{\bar{1}2}) - \frac{1}{2}\big[\mathcal{E}_{3,q}^0 \otimes \mathcal{D}_{3,q}^0\big](s_{\bar{1}2}) - \hat{\bar{\Ga}}_{qq,1}^{(2)} \de_2, \nn\\
\hat{\t{\mathcal{J}}}_2^{(2)}(\bar{1}_q,2_g) &=& \frac{1}{2}\t{\mathcal{E}}_{4,q}^0(s_{\bar{1}2}) + \frac{1}{2}\t{\mathcal{E}}_{3,q}^1(s_{\bar{1}2}), \nn\\
\hat{\hat{\t{\mathcal{J}}}}_2^{(2)}(\bar{1}_q,2_g) &=& \frac{1}{2}\hat{\mathcal{E}}_{3,q}^1(s_{\bar{1}2}) + \frac{1}{2}\frac{b_{0,F}}{\ep} \l( \frac{s_{\bar{1}2}}{\mu^2} \r)^{-\ep} \mathcal{E}_{3,q}^0(s_{\bar{1}2}) - \frac{1}{4}\big[\mathcal{E}_{3,q}^0 \otimes \mathcal{E}_{3,q}^0\big](s_{\bar{1}2}).
\ea

\subsubsection{Initial-final: gluon-initiated}
\ba
\mathcal{J}_2^{(2)}(\bar{1}_g,2_q) &=& \mathcal{D}_{4,g}^0(s_{\bar{1}2}) + \frac{1}{2}\mathcal{D}_{4,g'}^0(s_{\bar{1}2}) + \mathcal{D}_{3,g}^1(s_{\bar{1}2}) + \frac{b_0}{\ep} \l( \frac{s_{\bar{1}2}}{\mu^2} \r)^{-\ep} \mathcal{D}_{3,g}^0(s_{\bar{1}2}) \nn\\
&-& \big[\mathcal{D}_{3,g\to g}^0 \otimes \mathcal{D}_{3,g\to g}^0\big] (s_{\bar{1}2}) - \frac{1}{2}\bar{\Ga}_{gg,1}^{(2)} \de_2 - \big[\mathcal{D}_{3,g\to q}^0 \otimes \mathcal{D}_{3,q}^0\big] (s_{\bar{1}2}) \nn\\
&-& \big[\mathcal{D}_{3,g\to q}^0 \otimes \Ga_{gg,1}^{(1)}\big] + 2\big[\mathcal{D}_{3,g\to q}^0 \otimes \Ga_{qq,1}^{(1)}\big] - \mathcal{A}_{4,g}^0(s_{\bar{1}2}) \nn\\
&-& \frac{1}{2}\t{\mathcal{A}}_{4,g}^0(s_{\bar{1}2}) - \frac{1}{2}\mathcal{A}_{3,g}^1(s_{\bar{1}2}) - \frac{1}{2}\t{\mathcal{A}}_{3,g}^1(s_{\bar{1}2}) - \frac{1}{2}\frac{b_0}{\ep} \l( \frac{s_{\bar{1}2}}{\mu^2} \r)^{-\ep} \mathcal{A}_{3,g}^0(s_{\bar{1}2}) \nn\\
&+& \big[\mathcal{A}_{3,q}^0 \otimes \mathcal{A}_{3,g}^0\big] (s_{\bar{1}2}) + \frac{1}{2}\big[\Ga_{gg,1}^{(1)} \otimes \mathcal{A}_{3,g}^0\big] - \big[\Ga_{qq,1}^{(1)} \otimes \mathcal{A}_{3,g}^0\big], \nn\\
\hat{\mathcal{J}}_2^{(2)}(\bar{1}_g,2_q) &=& \mathcal{E}_{4,g}^0(s_{\bar{1}2}) + \hat{\mathcal{D}}_{3,g}^1(s_{\bar{1}2}) + \frac{b_{0,F}}{\ep} \l( \frac{s_{\bar{1}2}}{\mu^2} \r)^{-\ep} \mathcal{D}_{3,g}^0(s_{\bar{1}2}) - \big[\mathcal{E}_{3,q}^0 \otimes \mathcal{D}_{3,g\to q}^0\big](s_{\bar{1}2}) \nn\\
&-& \big[\hat{\Ga}_{gg,1}^{(1)} \otimes \mathcal{D}_{3,g\to q}^0\big] - \frac{1}{2}\hat{\bar{\Ga}}_{gg,1}^{(2)} \de_2 + S_{g\to q}\big[\Ga_{qg,1}^{(1)} \otimes \mathcal{E}_{3,q'}^0\big] + \frac{1}{2}\big[\Ga_{qg,1}^{(1)} \otimes \Ga_{gq,1}^{(1)}\big] \nn\\
&-& \frac{1}{2} \hat{\mathcal{A}}_{3,g}^1(s_{\bar{1}2}) - \frac{1}{2}\frac{b_{0,F}}{\ep} \l( \frac{s_{\bar{1}2}}{\mu^2} \r)^{-\ep} \mathcal{A}_{3,g}^0(s_{\bar{1}2}) + \frac{1}{2}\big[\hat{\Ga}_{gg,1}^{(1)} \otimes \mathcal{A}_{3,g}^0\big], \nn\\
\hat{\t{\mathcal{J}}}_2^{(2)}(\bar{1}_g,2_q) &=& \frac{1}{2}\t{\mathcal{E}}_{4,g}^0(s_{\bar{1}2}) + \frac{1}{2}\hat{\t{\bar{\Ga}}}_{gg,1}^{(2)} \de_2 + S_{g\to q}\big[\Ga_{qg,1}^{(1)} \otimes \mathcal{E}_{3,q'}^0\big] + \frac{1}{2}\big[\Ga_{qg,1}^{(1)} \otimes \Ga_{gq,1}^{(1)}\big], \nn\\
\hat{\hat{\mathcal{J}}}_2^{(2)}(\bar{1}_g,2_q) &=& -\frac{1}{2} \hat{\hat{\bar{\Ga}}}_{gg,1}^{(2)} \de_2.
\ea

\subsubsection{Initial-initial}
\ba
\mathcal{J}_2^{(2)}(\bar{1}_q,\bar{2}_g) &=& \mathcal{D}_{4,qg}^{0,adj}(s_{\bar{1}\bar{2}}) + \frac{1}{2}\mathcal{D}_{4,qg}^{0,nadj}(s_{\bar{1}\bar{2}}) + \mathcal{D}_{3,qg}^1(s_{\bar{1}\bar{2}}) + \frac{b_0}{\ep} \l( \frac{s_{\bar{1}\bar{2}}}{\mu^2} \r)^{-\ep} \mathcal{D}_{3,qg}^0(s_{\bar{1}\bar{2}}) \nn\\
&-& \big[\mathcal{D}_{3,qg}^0 \otimes \mathcal{D}_{3,qg}^0\big] (s_{\bar{1}\bar{2}}) - \bar{\Ga}_{qq,1}^{(2)} \de_2 - \frac{1}{2}\bar{\Ga}_{gg,2}^{(2)} \de_1, \nn\\
\hat{\mathcal{J}}_2^{(2)}(\bar{1}_q,\bar{2}_g) &=& \mathcal{E}_{4,qg}^{0}(s_{\bar{1}\bar{2}}) + \hat{\mathcal{D}}_{3,qg}^{1}(s_{\bar{1}\bar{2}}) + \frac{b_{0,F}}{\ep} \l( \frac{s_{\bar{1}\bar{2}}}{\mu^2} \r)^{-\ep} \mathcal{D}_{3,qg}^0(s_{\bar{1}\bar{2}}) - \hat{\bar{\Ga}}_{qq,1}^{(2)} \de_2 \nn\\
&-& \frac{1}{2}\hat{\bar{\Ga}}_{gg,2}^{(2)} \de_1 + S_{g\to q} \big[\Ga_{qg,2}^{(1)} \otimes \mathcal{E}_{3,qq'}^0\big] + \frac{1}{2} \big[\Ga_{qg,2}^{(1)} \otimes \Ga_{gq,2}^{(1)}\big], \nn\\
\hat{\t{\mathcal{J}}}_2^{(2)}(\bar{1}_q,\bar{2}_g) &=& \frac{1}{2}\t{\mathcal{E}}_{4,qg}^{0}(s_{\bar{1}\bar{2}}) + \frac{1}{2} \hat{\t{\bar{\Ga}}}_{gg,2}^{(2)} \de_1 + S_{g\to q} \big[\Ga_{qg,2}^{(1)} \otimes \mathcal{E}_{3,qq'}^0\big] + \frac{1}{2} \big[\Ga_{qg,2}^{(1)} \otimes \Ga_{gq,2}^{(1)}\big], \nn\\
\hat{\hat{\mathcal{J}}}_2^{(2)}(\bar{1}_q,\bar{2}_g) &=& -\frac{1}{2} \hat{\hat{\bar{\Ga}}}_{gg,2}^{(2)} \de_1.
\ea

\subsection{Identity preserving: gluon-gluon}

\subsubsection{Final-final}
\ba
\mathcal{J}_2^{(2)}(1_g,2_g) &=& \frac{1}{4}\mathcal{F}_4^0(s_{12}) + \frac{1}{3}\mathcal{F}_3^1(s_{12})  + \frac{1}{3}\frac{b_0}{\ep} \l( \frac{s_{12}}{\mu^2} \r)^{-\ep} \mathcal{F}_3^0(s_{12}) - \frac{1}{9} \big[\mathcal{F}_3^0 \otimes \mathcal{F}_3^0\big] (s_{12}), \nn\\
\hat{\mathcal{J}}_2^{(2)}(1_g,2_g) &=& \mathcal{G}_4^0(s_{12}) + \mathcal{G}_3^1(s_{12}) + \frac{1}{3}\hat{\mathcal{F}}_3^1(s_{12}) + \frac{b_{0}}{\ep} \l( \frac{s_{12}}{\mu^2} \r)^{-\ep} \mathcal{G}_3^0(s_{12}) \nn\\
&+& \frac{1}{3}\frac{b_{0,F}}{\ep} \l( \frac{s_{12}}{\mu^2} \r)^{-\ep} \mathcal{F}_3^0(s_{12}) - \frac{2}{3}\big[\mathcal{G}_3^0 \otimes \mathcal{F}_3^0\big](s_{12}), \nn\\
\hat{\t{\mathcal{J}}}_2^{(2)}(1_g,2_g) &=& \frac{1}{2}\t{\mathcal{G}}_4^0(s_{12}) + \frac{1}{2}\t{\mathcal{G}}_3^1(s_{12}), \nn\\
\hat{\hat{\t{\mathcal{J}}}}_2^{(2)}(1_g,2_g) &=& \frac{1}{2}\mathcal{H}_4^0(s_{12}) + \hat{\mathcal{G}}_3^1(s_{12}) + \frac{b_{0,F}}{\ep} \l( \frac{s_{12}}{\mu^2} \r)^{-\ep} \mathcal{G}_3^0(s_{12}) - \big[\mathcal{G}_3^0 \otimes \mathcal{G}_3^0\big] (s_{12}).
\ea

\subsubsection{Initial-final}
\ba
\mathcal{J}_2^{(2)}(\bar{1}_g,2_g) &=& \frac{1}{2}\mathcal{F}_{4,g}^0(s_{\bar{1}2}) + \frac{1}{2}\mathcal{F}_{3,g}^1(s_{\bar{1}2}) + \frac{1}{2}\frac{b_0}{\ep} \l( \frac{s_{\bar{1}2}}{\mu^2} \r)^{-\ep} \mathcal{F}_{3,g}^0(s_{\bar{1}2}) \nn\\
&-& \frac{1}{4} \big[\mathcal{F}_{3,g}^0 \otimes \mathcal{F}_{3,g}^0\big] (s_{\bar{1}2}) - \frac{1}{2}\bar{\Ga}_{gg,1}^{(2)} \de_2, \nn\\
\hat{\mathcal{J}}_2^{(2)}(\bar{1}_g,2_g) &=& \mathcal{G}_{4,g}^0(s_{\bar{1}2}) + \frac{1}{2}\mathcal{G}_{3,g}^1(s_{\bar{1}2}) + \frac{1}{2}\hat{\mathcal{F}}_{3,g}^1(s_{\bar{1}2}) + \frac{1}{2}\frac{b_{0}}{\ep} \l( \frac{s_{\bar{1}2}}{\mu^2} \r)^{-\ep} \mathcal{G}_{3,g}^0(s_{\bar{1}2}) \nn\\
&+& \frac{1}{2}\frac{b_{0,F}}{\ep} \l( \frac{s_{\bar{1}2}}{\mu^2} \r)^{-\ep} \mathcal{F}_{3,g}^0(s_{\bar{1}2}) - \frac{1}{2}\big[\mathcal{G}_{3,g}^0 \otimes \mathcal{F}_{3,g}^0\big](s_{\bar{1}2}) - \frac{1}{2} \hat{\bar{\Ga}}_{gg,1}^{(2)} \de_2 \nn\\
&+& S_{g\to q} \big[\Ga_{qg,1}^{(1)} \otimes \mathcal{G}_{3,q'}^0\big] + \frac{1}{2}\big[\Ga_{qg,1}^{(1)} \otimes \Ga_{gq,1}^{(1)}\big], \nn\\
\hat{\t{\mathcal{J}}}_2^{(2)}(\bar{1}_g,2_g) &=& \frac{1}{2}\t{\mathcal{G}}_{4,g}^0(s_{\bar{1}2}) + \frac{1}{2}\t{\mathcal{G}}_{3,g}^1(s_{\bar{1}2}) + \frac{1}{2}\hat{\t{\bar{\Ga}}}_{gg,1}^{(2)} \de_2 \nn\\
&+& S_{g\to q} \big[\Ga_{qg,1}^{(1)} \otimes \mathcal{G}_{3,q'}^0\big] + \frac{1}{2}\big[\Ga_{qg,1}^{(1)} \otimes \Ga_{gq,1}^{(1)}\big], \nn\\
\hat{\hat{\mathcal{J}}}_2^{(2)}(\bar{1}_g,2_g) &=& \frac{1}{2}\hat{\mathcal{G}}_3^1(s_{\bar{1}2}) + \frac{1}{2}\frac{b_{0,F}}{\ep} \l( \frac{s_{\bar{1}2}}{\mu^2} \r)^{-\ep} \mathcal{G}_{3,g}^0(s_{\bar{1}2}) - \frac{1}{4}\big[\mathcal{G}_{3,g}^0 \otimes \mathcal{G}_{3,g}^0\big] (s_{\bar{1}2}) - \frac{1}{2}\hat{\hat{\bar{\Ga}}}_{gg,1}^{(2)} \de_2. \nn\\
\ea

\subsubsection{Initial-initial}
\ba
\mathcal{J}_2^{(2)}(\bar{1}_g,\bar{2}_g) &=& \mathcal{F}_{4,gg}^{0,adj}(s_{\bar{1}\bar{2}}) + \frac{1}{2}\mathcal{F}_{4,gg}^{0,nadj}(s_{\bar{1}\bar{2}}) + \mathcal{F}_{3,gg}^1(s_{\bar{1}\bar{2}}) + \frac{b_0}{\ep} \l( \frac{s_{\bar{1}\bar{2}}}{\mu^2} \r)^{-\ep} \mathcal{F}_{3,gg}^0(s_{\bar{1}\bar{2}}) \nn\\
&-& \big[\mathcal{F}_{3,gg}^0 \otimes \mathcal{F}_{3,gg}^0\big] (s_{\bar{1}\bar{2}}) - \frac{1}{2}\bar{\Ga}_{gg,1}^{(2)} \de_2 - \frac{1}{2}\bar{\Ga}_{gg,2}^{(2)} \de_1, \nn\\
\hat{\mathcal{J}}_2^{(2)}(\bar{1}_g,\bar{2}_g) &=& \mathcal{G}_{4,gg}^0(s_{\bar{1}\bar{2}}) + \hat{\mathcal{F}}_{3,gg}^1(s_{\bar{1}\bar{2}}) + \frac{b_{0,F}}{\ep} \l( \frac{s_{\bar{1}\bar{2}}}{\mu^2} \r)^{-\ep} \mathcal{F}_{3,gg}^0(s_{\bar{1}\bar{2}}) \nn\\
&-& \frac{1}{2}\hat{\bar{\Ga}}_{gg,1}^{(2)} \de_2 + S_{g\to q} \big[\Ga_{qg,1}^{(1)} \otimes \mathcal{G}_{3,qg}^0\big] + \frac{1}{2} \big[\Ga_{qg,1}^{(1)} \otimes \Ga_{gq,1}^0\big] \nn\\
&-& \frac{1}{2}\hat{\bar{\Ga}}_{gg,2}^{(2)} \de_1 + S_{g\to q} \big[\Ga_{qg,2}^{(1)} \otimes \mathcal{G}_{3,qg}^0\big] + \frac{1}{2} \big[\Ga_{qg,2}^{(1)} \otimes \Ga_{gq,2}^0\big], \nn\\
\hat{\t{\mathcal{J}}}_2^{(2)}(\bar{1}_g,\bar{2}_g) &=& \t{\mathcal{G}}_{4,gg}^0(s_{\bar{1}\bar{2}}) \nn\\
&+& \hat{\t{\bar{\Ga}}}_{gg,1}^{(2)} \de_2 + 2 S_{g\to q} \big[\Ga_{qg,1}^{(1)} \otimes \mathcal{G}_{3,qg}^0\big] + \big[\Ga_{qg,1}^{(1)} \otimes \Ga_{gq,1}^0\big] \nn\\
&+& \hat{\t{\bar{\Ga}}}_{gg,2}^{(2)} \de_1 + 2 S_{g\to q} \big[\Ga_{qg,2}^{(1)} \otimes \mathcal{G}_{3,qg}^0\big] + \big[\Ga_{qg,2}^{(1)} \otimes \Ga_{gq,2}^0\big], \nn\\
\hat{\hat{\t{\mathcal{J}}}}_2^{(2)}(\bar{1}_g,\bar{2}_g) &=& -\frac{1}{2}\hat{\hat{\bar{\Ga}}}_{gg,1}^{(2)} \de_2 - \frac{1}{2}\hat{\hat{\bar{\Ga}}}_{gg,2}^{(2)} \de_1.
\ea

\subsection{Identity changing: $g \to q$}

\subsubsection{Initial-final}
\ba
\mathcal{J}_{2,g\to q}^{(2)}(\bar{1}_{g\to q},2_{\qb}) &=& -\mathcal{A}_{4,g}^0(s_{\bar{1}2}) - \frac{1}{2}\mathcal{A}_{3,g}^1(s_{\bar{1}2}) - \frac{1}{2} \frac{b_0}{\ep} \l( \frac{s_{\bar{1}2}}{\mu^2} \r)^{-\ep} \mathcal{A}_{3,g}^0(s_{\bar{1}2}) \nn\\
&+& \frac{1}{2}\big[\mathcal{A}_{3,g}^0 \otimes \mathcal{A}_{3,q}^0\big](s_{\bar{1}2}) - S_{g\to q} \bar{\Ga}_{qg,1}^{(2)} \de_2 \nn\\
&+& \frac{1}{2}\big[\mathcal{A}_{3,g}^0 \otimes \Ga_{gg,1}^{(1)}\big] + \frac{1}{2} S_{g\to q}\big[\Ga_{qg,1}^{(1)} \otimes \Ga_{gg,1}^{(1)}\big] \nn\\
&-& \frac{1}{2}\big[\mathcal{A}_{3,g}^0 \otimes \Ga_{qq,1}^{(1)}\big] - \frac{1}{2} S_{g\to q}\big[\Ga_{qg,1}^{(1)} \otimes \Ga_{qq,1}^{(1)}\big], \nn\\
\hat{\mathcal{J}}_{2,g\to q}^{(2)}(\bar{1}_{g\to q},2_{\qb}) &=& -\frac{1}{2} \hat{A}_{3,g}^1(s_{\bar{1}2}) - \frac{1}{2} \frac{b_{0,F}}{\ep} \l( \frac{s_{\bar{1}2}}{\mu^2}\r)^{-\ep} \mathcal{A}_{3,g}^0(s_{\bar{1}2}) \nn\\
&+& \frac{1}{2}\big[\mathcal{A}_{3,g}^0 \otimes \hat{\Ga}_{gg,1}^{(1)}\big] + \frac{1}{2} S_{g\to q} \big[\Ga_{qg,1}^{(1)} \otimes \hat{\Ga}_{gg,1}^{(1)}\big] - S_{g\to q} \hat{\bar{\Ga}}_{qg,1}^{(2)} \de_2, \nn\\
\t{\mathcal{J}}_{2,g\to q}^{(2)}(\bar{1}_{g\to q},2_{\qb}) &=& -\t{\mathcal{A}}_{4,g}^0(s_{\bar{1}2}) - \t{\mathcal{A}}_{3,g}^1(s_{\bar{1}2}) + \big[\mathcal{A}_{3,g}^0 \otimes \mathcal{A}_{3,q}^0\big](s_{\bar{1}2}) - \big[\Ga_{qq,1}^{(1)} \otimes \mathcal{A}_{3,g}^0\big] \nn\\
&-& S_{g\to q} \big[\Ga_{qg,1}^{(1)} \otimes \Ga_{qq,1}^{(1)}\big] + 2 S_{g\to q} \t{\bar{\Ga}}_{qg,1}^{(2)} \de_2.
\ea

\subsubsection{Initial-initial}
\ba
\mathcal{J}_{2,g\to q}^{(2)}(\bar{1}_{g\to q},\bar{2}_{\qb}) &=& - \mathcal{A}_{4,qg}^0(s_{\bar{1}\bar{2}}) - \mathcal{A}_{4,qg}^{0,nadj}(s_{\bar{1}\bar{2}}) - \mathcal{A}_{3,qg}^1(s_{\bar{1}\bar{2}}) - \frac{b_0}{\ep} \l( \frac{s_{\bar{1}\bar{2}}}{\mu^2} \r)^{-\ep} \mathcal{A}_{3,qg}^0(s_{\bar{1}\bar{2}}) \nn\\
&+& \big[\mathcal{A}_{3,qg}^0 \otimes \mathcal{A}_{3,qq}^0\big](s_{\bar{1}\bar{2}}) - \big[\Ga_{qq,1}^{(1)} \otimes \mathcal{A}_{3,qg}^0\big] + \big[\Ga_{gg,1}^{(1)} \otimes \mathcal{A}_{3,qg}^0\big] \nn\\
&-& \frac{1}{2} S_{g\to q} \big[\Ga_{qg,1}^{(1)} \otimes \Ga_{qq,1}^{(1)}\big] + \frac{1}{2} S_{g\to q} \big[\Ga_{qg,1}^{(1)} \otimes \Ga_{gg,1}^{(1)}\big] - S_{g\to q} \bar{\Ga}_{qg,1}^{(2)} \de_2, \nn\\
\mathcal{J}_{2,g\to q}^{(2)}(\bar{1}_{g\to q},\bar{2}_{\qb}) &=& - \hat{\mathcal{A}}_{3,qg}^1(s_{\bar{1}\bar{2}}) - \frac{b_{0,F}}{\ep} \l( \frac{s_{\bar{1}\bar{2}}}{\mu^2} \r)^{-\ep} \mathcal{A}_{3,qg}^0(s_{\bar{1}\bar{2}}) \nn\\
&+& \big[\hat{\Ga}_{gg,1}^{(1)} \otimes \mathcal{A}_{3,qg}^0\big] + \frac{1}{2} S_{g\to q} \big[\Ga_{qg,1}^{(1)} \otimes \hat{\Ga}_{gg,1}^{(1)}\big] - S_{g\to q} \hat{\bar{\Ga}}_{qg,1}^{(2)} \de_2, \nn\\
\t{\mathcal{J}}_{2,g\to q}^{(2)}(\bar{1}_{g\to q},\bar{2}_{\qb}) &=& - \t{\mathcal{A}}_{4,qg}^0(s_{\bar{1}\bar{2}}) - \t{\mathcal{A}}_{3,qg}^1(s_{\bar{1}\bar{2}}) + \big[\mathcal{A}_{3,qg}^0 \otimes \mathcal{A}_{3,qq}^0\big](s_{\bar{1}\bar{2}}) \nn\\
&-& \big[\Ga_{qq,1}^{(1)} \otimes \mathcal{A}_{3,qg}^0\big] - \frac{1}{2} S_{g\to q} \big[\Ga_{qg,1}^{(1)} \otimes \Ga_{qq,1}^{(1)}\big] + S_{g\to q} \t{\bar{\Ga}}_{qg,1}^{(2)} \de_2.
\ea

\subsection{Identity changing: $q \to g$}

\subsubsection{Initial-final}
\ba
\t{\mathcal{J}}_{2,q\to g}^{(2)}(\bar{1}_{q\to g},2_g) &=& -\t{\mathcal{E}}_{4,q'}^0(s_{\bar{1}2}) - \t{\mathcal{E}}_{3,q'}^1(s_{\bar{1}2}) + \big[\mathcal{E}_{3,q'}^0 \otimes \Ga_{qq,1}^{(1)}\big] \nn\\
&+& \frac{1}{2} S_{q\to g} \big[\Ga_{qq,1}^{(1)} \otimes \Ga_{gq,1}^{(1)}\big] + S_{q\to g} \t{\bar{\Ga}}_{gq,1}^{(2)} \de_2, \nn\\
\hat{\mathcal{J}}_{2,q \to g}^{(2)}(\bar{1}_{q\to g},2_g) &=& - \mathcal{H}_{4,q}^0(s_{\bar{1}2}) - \hat{\mathcal{G}}_{3,q'}^1(s_{\bar{1}2}) - \frac{b_{0,F}}{\ep} \l( \frac{s_{\bar{1}2}}{\mu^2} \r)^{-\ep} \mathcal{G}_{3,q'}^0(s_{\bar{1}2}) \nn\\
&+& S_{q\to g} \big[\Ga_{gq,1}^{(1)} \otimes \mathcal{G}_{3,g}^0\big] - \frac{1}{2} S_{q\to g} \big[\Ga_{gq,1}^{(1)} \otimes \hat{\Ga}_{gg,1}^{(1)}\big] + S_{q\to g}  \hat{\bar{\Ga}}_{gq,1}^{(2)} \de_2. \nn\\
\ea

\subsubsection{Initial-initial: quark-gluon}
\ba
\mathcal{J}_{2,q\to g}^{(2)}(\bar{1}_{q\to g},\bar{2}_g) &=& -\mathcal{G}_{4,qg}^{0,adj}(s_{\bar{1}\bar{2}}) -\mathcal{G}_{4,qg}^{0,nadj}(s_{\bar{1}\bar{2}}) - \mathcal{G}_{3,qg}^1(s_{\bar{1}\bar{2}}) - \frac{b_0}{\ep} \l( \frac{s_{\bar{1}\bar{2}}}{\mu^2} \r)^{-\ep} \mathcal{G}_{3,qg}^0(s_{\bar{1}2}) \nn\\
&+& 2 \big[\mathcal{G}_{3,qg}^0 \otimes \mathcal{F}_{3,gg}^0\big](s_{\bar{1}2}) - S_{q\to g} \bar{\Ga}_{gq,1}^{(2)} \de_2 \nn\\
&+& \big[\mathcal{A}_{3,qg}^0 \otimes \mathcal{G}_{3,qq}\big](s_{\bar{1}2}) + \big[\Ga_{qq,1}^{(1)} \otimes \mathcal{G}_{3,qg}^0\big] - \big[\Ga_{gg,1}^{(1)} \otimes \mathcal{G}_{3,qg}^0\big] \nn\\
&+& \frac{1}{2} S_{q\to g} \big[\Ga_{qq,1}^{(1)} \otimes \Ga_{gq,1}^{(1)}\big] - \frac{1}{2} S_{q\to g} \big[\Ga_{gg,1}^{(1)} \otimes \Ga_{gq,1}^{(1)}\big], \nn\\
\hat{\mathcal{J}}_{2,q\to g}^{(2)}(\bar{1}_{q\to g},\bar{2}_g) &=& - \hat{\mathcal{G}}_{3,qg}^1(s_{\bar{1}\bar{2}}) - \frac{b_{0,F}}{\ep} \l( \frac{s_{\bar{1}\bar{2}}}{\mu^2} \r)^{-\ep} \mathcal{G}_{3,qg}^0(s_{\bar{1}2}) - \big[\mathcal{G}_{3,qg}^0 \otimes \hat{\Ga}_{gg,1}^{(1)}\big] \nn\\ 
&-&\frac{1}{2} S_{q\to g} \big[\Ga_{gq,1}^{(1)} \otimes \hat{\Ga}_{gg,1}^{(1)}\big] - S_{q\to g} \hat{\bar{\Ga}}_{gq,1}^{(2)} \de_2, \nn\\
\t{\mathcal{J}}_{2,q\to g}^{(2)}(\bar{1}_{q\to g},\bar{2}_g) &=& -\t{\mathcal{G}}_{4,qg}^0(s_{\bar{1}\bar{2}}) - \t{\mathcal{G}}_{3,qg}^1(s_{\bar{1}\bar{2}}) + \big[\mathcal{G}_{3,qg}^0 \otimes \Ga_{qq,1}^{(1)}\big] + \frac{1}{2} S_{q\to g} \big[\Ga_{qq,1}^{(1)} \otimes \Ga_{gq,1}^{(1)}\big] \nn\\
&+& S_{q\to g} \t{\bar{\Ga}}_{gq,1}^{(2)} \de_2 - S_{g\to q} \big[\mathcal{G}_{3,qq}^0 \otimes \Ga_{qg,2}^{(1)}\big].
\ea

\subsubsection{Initial-initial: quark-antiquark}
\ba
\mathcal{J}_{2,q\to g}^{(2)}(\bar{1}_{q\to g},\bar{2}_{\qb'}) &=& - \mathcal{E}_{4,\qb'q}^0(s_{\bar{1}\bar{2}}) - \mathcal{E}_{4,q'q}^0(s_{\bar{1}\bar{2}}) - \mathcal{E}_{3,q'q}^1(s_{\bar{1}\bar{2}}) - \frac{b_0}{\ep} \l( \frac{s_{\bar{1}\bar{2}}}{\mu^2} \r)^{-\ep} \mathcal{E}_{3,q'q}^0(s_{\bar{1}\bar{2}}) \nn\\
&+& 2 \big[\mathcal{D}_{3,gq}^0 \otimes \mathcal{E}_{3,q'q}^0\big](s_{\bar{1}\bar{2}}) + \big[\Ga_{qq,1}^{(1)} \otimes \mathcal{E}_{3,q'q}^0\big] - \big[\Ga_{gg,1}^{(1)} \otimes \mathcal{E}_{3,q'q}^0\big] \nn\\
&+& \frac{1}{2} S_{q\to g} \big[\Ga_{qq,1}^{(1)} \otimes \Ga_{gq,1}^{(1)}\big] - \frac{1}{2} S_{q\to g} \big[\Ga_{gg,1}^{(1)} \otimes \Ga_{gq,1}^{(1)}\big] - S_{q\to g} \bar{\Ga}_{gq,1}^{(2)} \de_2, \nn\\
\hat{\mathcal{J}}_{2,q\to g}^{(2)}(\bar{1}_{q\to g},\bar{2}_{\qb'}) &=& - \hat{\mathcal{E}}_{3,q'q}^1(s_{\bar{1}\bar{2}}) - \frac{b_{0,F}}{\ep} \l( \frac{s_{\bar{1}\bar{2}}}{\mu^2} \r)^{-\ep} \mathcal{E}_{3,q'q}^0(s_{\bar{1}\bar{2}}) - \big[\hat{\Ga}_{gg,1}^{(1)} \otimes \mathcal{E}_{3,q'q}^0\big]  \nn\\
&-& \frac{1}{2} S_{q\to g} \big[\Ga_{gq,1}^{(1)} \otimes \hat{\Ga}_{gg,1}^{(1)}\big] - S_{q\to g} \hat{\bar{\Ga}}_{gq,1} \de_2, \nn\\ 
\t{\mathcal{J}}_{2,q\to g}^{(2)}(\bar{1}_{q\to g},\bar{2}_{\qb'}) &=& - \t{\mathcal{E}}_{4,q'q}^0(s_{\bar{1}\bar{2}}) - \t{\mathcal{E}}_{3,q'q}^1(s_{\bar{1}\bar{2}}) + \big[\Ga_{qq,1}^{(1)} \otimes \mathcal{E}_{3,q'q}^0\big] \nn\\
&+& \frac{1}{2} S_{q\to g} \big[\Ga_{qq,1}^{(1)} \otimes \Ga_{gq,1}^{(1)}\big] + S_{q\to g} \t{\bar{\Ga}}_{gq,1}^{(2)} \de_2.
\ea

\subsection{Identity changing: $q \to \qb$}

\subsubsection{Initial-initial}
\ba
\t{\mathcal{J}}_{2,q\to \qb}^{(2)}(\bar{1}_{q\to \qb},\bar{2}_q) &=& \mathcal{C}_{4,qq'}^0(s_{\bar{1}\bar{2}}) + \t{\bar{\Ga}}_{q\qb,1}^{(2)} \de_2.
\ea

\subsection{Identity changing: $q \to q'$}

\subsubsection{Initial-initial}
\ba
\mathcal{J}_{2,q\to q'}^{(2)}(\bar{1}_{q\to q'},\bar{2}_{\qb}) &=& \mathcal{B}_{4,q'q}^0(s_{\bar{1}\bar{2}}) + S_{q\to g} \big[\Ga_{gq,1}^{(1)} \otimes \mathcal{A}_{3,gq}^0\big] + \frac{1}{2} S_{q\to g} \big[\Ga_{gq,1}^{(1)} \otimes \Ga_{qg,1}^{(1)}\big] - \bar{\Ga}_{q\qb',1}^{(2)} \de_1. \nn\\ 
\ea

\subsection{Identity changing: $gg \to q\qb$}

\subsubsection{Initial-initial}
\ba
\mathcal{J}_{2,gg\to q\qb}^{(2)}(\bar{1}_{g\to q},\bar{2}_{g\to \qb}) &=& \mathcal{A}_{4,gg}^0(s_{\bar{1}\bar{2}}) + S_{g\to q} \big[\mathcal{A}_{3,qg}^0 \otimes \Ga_{qg,1}^{(1)}\big] \nn\\
&+& S_{g\to q} \big[\mathcal{A}_{3,gq}^0 \otimes \Ga_{qg,2}^{(1)}\big] + S_{g\to q} \Ga_{qg,1}^{(1)} S_{g\to q}  \Ga_{qg,2}^{(1)}, \nn\\
\t{\mathcal{J}}_{2,gg\to q\qb}^{(2)}(\bar{1}_{g\to q},\bar{2}_{g\to \qb}) &=& \t{\mathcal{A}}_{4,gg}^0(s_{\bar{1}\bar{2}}) + 2 S_{g\to q} \big[\mathcal{A}_{3,qg}^0 \otimes \Ga_{qg,1}^{(1)}\big] \nn\\
&+& 2 S_{g\to q} \big[\mathcal{A}_{3,gq}^0 \otimes \Ga_{qg,2}^{(1)}\big] + 2 S_{g\to q} \Ga_{qg,1}^{(1)} S_{g\to q} \Ga_{qg,2}^{(1)}.
\ea

\subsection{Identity changing: $q\qb' \to gg$}

\subsubsection{Initial-initial}
\ba
\mathcal{J}_{2,q\qb' \to gg}^{(2)}(\bar{1}_{q\to g},\bar{2}_{\qb'\to g}) &=& \mathcal{H}_{4,q\qb'}^0(s_{\bar{1}\bar{2}}) + S_{q\to g} \big[\Ga_{gq,1}^{(1)} \otimes \mathcal{G}_{3,gq}^0\big] + S_{q\to g} \big[\Ga_{gq,2}^{(1)} \otimes \mathcal{G}_{3,qg}^0\big] \nn\\
&+& S_{q\to g} \Ga_{gq,1}^{(1)} S_{q\to g} \Ga_{gq,2}^{(1)}.
\ea

\subsection{Identity changing: $q \to g \to q$}

\subsubsection{Initial-final}
\ba
\mathcal{J}_{2,q\to g\to q}^{(2)}(\bar{1}_{q\to g\to q},2_g) &=& \mathcal{B}_{4,q'}^0(s_{\bar{1}2}) + S_{q\to g} \big[\Ga_{gq,1}^{(1)} \otimes \mathcal{A}_{3,g}^0\big] + \big[\Ga_{gq,1}^{(1)} \otimes \Ga_{qg,1}^{(1)}\big] - 2 \t{\bar{\Ga}}_{qq,1}^{(2)} \de_2. \nn\\
\ea

\end{appendix}


\bibliographystyle{JHEP}
\bibliography{jetfc_thy}

\end{document}